\documentclass[psamsfonts,reqno]{amsart}

\usepackage{amsmath}
\usepackage{amssymb}
\usepackage{amsxtra}

\newcommand {\la} {\left \langle}
\newcommand {\ra} {\right \rangle}
\newcommand {\lb} {\left (}
\newcommand {\rb} {\right )}

\newcommand{\vect}[1]{\overrightarrow{ #1 }}

\newcommand {\CalD} {\mathcal D}
\newcommand {\CalE} {\mathcal E}

\newcommand {\CalG} {\mathcal G}

\newcommand {\CalO} {\mathcal O}

\newcommand {\CalN} {\mathcal N}

\newcommand {\CalL} {\mathcal L}
\newcommand {\CalS} {\mathcal S}

\newcommand {\CalM} {\mathcal M}

\newcommand {\BH}   {\mathbb H}
\newcommand {\BR}   {\mathbb R}
\newcommand {\BZ}   {\mathbb Z}
\newcommand {\BC}   {\mathbb C}

\newcommand {\CP}   {\mathbb C \mathbb P}

\newcommand {\de} {\delta}
\newcommand {\ve}  {\varepsilon}
\newcommand {\ep}  {\epsilon}

\newcommand{\g}{\mathfrak{g}}
\newcommand{\h}{\mathfrak{h}}

\renewcommand{\Re} {\mathrm{Re}}

\newcommand {\p} {\partial}
\newcommand{\Dslash}{\ensuremath \raisebox{0.025cm}{\slash}\hspace{-0.25cm} D}

\DeclareMathOperator{\coker}{coker}

\DeclareMathOperator{\tr} {tr}

\DeclareMathOperator{\str} {str}

\DeclareMathOperator{\vol}{vol}

\DeclareMathOperator{\Pexp} {Pexp}
\DeclareMathOperator{\ad} {ad}
\DeclareMathOperator{\ind} {ind}

\DeclareMathOperator{\diag}{diag}

\newcommand{\SU}{SU}
\newcommand{\U}{U}
\newcommand{\SO}{SO}
\newcommand{\Spin}{Spin}
\newcommand{\Sp}{Sp}
\newcommand{\OSp}{OSp}
\newcommand{\GL}{GL}
\newcommand{\SL}{SL}
\newcommand{\PSL}{PSL}

\newcommand{\Cl}{\mathrm Cl}

\numberwithin{equation}{section}

\usepackage[pagebackref=true]{hyperref}

\hypersetup{
    colorlinks = false,
    linkcolor = blue,
    anchorcolor = red,
    citecolor = blue,
    filecolor = red,
    pagecolor = red,
    urlcolor = blue
}

\usepackage[numbers,sort&compress]{natbib}

\usepackage[dvips]{color}

\usepackage{graphicx}

\newcommand{\gfive}{\Gamma^{(\overline{14})}}
\newcommand{\geight}{\Gamma^{(\overline{58})}}
\newcommand{\ti}{0}
\newcommand{\tic}{9}
\newcommand{\ZN}{Z_{\text{inst}}}
\newcommand{\qq}{q}

\newcommand{\gym}{g_{\text{\tiny \textsc{ym}}}}

\newcommand{\MyGreen}{\color [rgb]{0,0.7,0}}

\newcommand{\GREEN}[1]{{\MyGreen{#1}}}

\begin{document}

\title[Localization of gauge theory on $S^4$]{Localization of gauge theory on a
  four-sphere and supersymmetric Wilson loops } \author{Vasily Pestun}
\address{Physics Department, Princeton University, Princeton NJ 08544}
\email{pestun@princeton.edu} \thanks{On leave of absence from ITEP, Moscow,
  117259, Russia.}  \date{December, 2007}

\begin{abstract}
 We prove conjecture due to Erickson-Semenoff-Zarembo and Drukker-Gross  which relates supersymmetric circular Wilson
  loop operators in the $\CalN=4$ supersymmetric Yang-Mills theory with a Gaussian matrix
  model.  We also compute the partition function and give a new matrix model
  formula for the expectation value of a supersymmetric circular Wilson loop
  operator for the pure $\CalN=2$ and the $\CalN=2^*$  supersymmetric Yang-Mills theory on a
  four-sphere. A four-dimensional $\CalN=2$ superconformal gauge theory is treated similarly.
\end{abstract}

\begin{flushright}
  ITEP-TH-41/07 \\
  PUTP-2248
\end{flushright}
\maketitle
\setcounter{tocdepth}{1} 
\tableofcontents
\section{Introduction}

Topological gauge theory can be constructed by twisting $\CalN=2$ supersymmetric
Yang-Mills theory~\cite{Witten:1988ze}. The path integral of the twisted theory
localizes to the moduli space of instantons and computes the Donaldson-Witten
invariants of four-manifolds~\cite{Witten:1988ze,MR1094734,MR710056}.

In the flat space the twist does not change the Lagrangian.
In~\cite{Nekrasov:2002qd} Nekrasov used a $\U(1)^2$ subgroup of the $\SO(4)$
Lorentz symmetry group of $\BR^4$ to define a $\U(1)^2$-equivariant version of the
topological partition function, or, equivalently, the partition function of the
$\CalN=2$ supersymmetric gauge theory in the $\Omega$-deformed
background~\cite{Nekrasov:2003rj}.  The integral over the moduli space of instantons
$\CalM_{inst}$ localizes at the fixed point set of the group
$\U(1)^2 \times T$, where $T$ is the maximal torus of the gauge
group $G$. The group $\U(1)^2 \times T$ acts on
$\CalM_{inst}$ by Lorentz rotations of the space-time and by gauge transformations
at infinity. The partition function $\ZN(a,\ep_1,\ep_2,q)$ depends on the
parameters $(\ep_1,\ep_2,a)  \in Lie(U(1)^2 \times T)$ and the coupling
constant $q = \exp(2 \pi i \tau)$. 
 The partition function is finite because the
$\Omega$-background effectively confines the dynamics to the finite volume
$V_{\text{eff}} = \frac {1}{\ep_1 \ep_2}$.  In the limit $\ep_1
\ep_2 \to 0$, the effective volume $V_{\text{eff}}$ diverges as well as
the free energy $F=-\log \ZN$. The specific free energy
$F/V_{\text{eff}}$ does not diverge  and  coincides with the Seiberg-Witten
low-energy effective prepotenial  of the $\CalN=2$ supersymmetric
Yang-Mills theory~\cite{Seiberg:1994aj,Seiberg:1994rs}. Hence, the instanton 
counting derives the Seiberg-Witten prepotential from the first principles.

Here we  consider another interesting situation when the
 gauge theory partition function  can be exactly computed. 
  We consider the $\CalN=2$, the $\CalN=2^{*}$ and the $\CalN=4$
  Yang-Mills theory on the standard four-sphere $S^4$. Our
  definition 
of   $\CalN=2$ supersymmetry on $S^4$ is explained in
  section~\ref{se:fields-action-symmetries}.\footnote{
  It would be interesting to extend the analysis to more general backgrounds~\cite{Karlhede:1988ax}.}  

Around the trivial background, no fields in the path integral have
zero modes. There are no zero modes for the gauge fields, because the first cohomology group of $S^4$ vanishes.  
There are
no zero modes for the fermions. This follows from the fact that the
Laplacian operator on a compact space is semipositive and the
formula $\Dslash^2 = \Delta + \frac R 4$, where $\Dslash$ is the Dirac operator,
 $\Delta$ is the Laplacian, and $R$ is the scalar curvature, which is positive on $S^4$.
There are no zero modes for the scalar fields, 
because the conformal
coupling in the kinetic term effectively generates mass for scalars.

Since there are no zero modes, we want to integrate
over all fields in the path integral and to compute the total partition function of the
theory and expectation values of certain 
 observables.

\begin{figure}
\includegraphics[width=4cm]{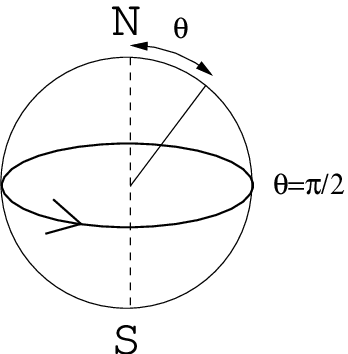}
\label{fig:Wilson-loop}
\caption{Wilson loop on the equator of $S^4$}
\end{figure}

We are most interested in the
supersymmetric circular Wilson loop operator (see Fig.~\ref{fig:Wilson-loop})
\begin{equation}
  \begin{aligned}
    \label{eq:Wilson-loop-defined}
    W_R(C) = \tr_{R} \Pexp \oint_{C} (A_{\mu} dx^{\mu} + i \Phi_\ti^{E} ds),
  \end{aligned}
\end{equation}
where  $R$ is a representation of the gauge group, $\Pexp$ is the path-ordered
exponent, $C$ is a circular loop located at the equator of $S^4$, $A_{\mu}$ is
the gauge field and $i \Phi_{\ti}^{E}$ is one of the scalar fields of the
$\CalN=2$ vector multiplet. We use notation $\Phi_{\ti}^E$ for 
the scalar field in the theory obtained by dimensional reduction
of the Euclidean six-dimensional or ten-dimensional $\CalN=1$
Yang-Mills theory. 
In our conventions, all fields take value in 
the real Lie algebra of the gauge group.

In~\cite{Erickson:2000af} Erickson, Semenoff and Zarembo conjectured that the
expectation value $ \langle W_R(C) \rangle$ of the Wilson loop operator~\eqref{eq:Wilson-loop-defined} 
in the four-dimensional $\CalN=4$ $\SU(N)$ gauge theory in the large $N$ limit
can be exactly
computed summing all rainbow diagrams in Feynman gauge. The combinatorics of rainbow
diagrams can be represented by a Gaussian matrix
model. In~\cite{Erickson:2000af} the conjecture was tested at the one-loop level in
gauge theory.  In~\cite{Drukker:2000rr} Drukker and Gross conjectured that the
exact relation to the Gaussian matrix model holds for any $N$ and argued
that the expectation value of the Wilson loop operator~\eqref{eq:Wilson-loop-defined} can
 be computed by a matrix model. However, 
Drukker-Gross argument does not show that the matrix model is
Gaussian.

The conjecture was used in many studies dealing with the $AdS/CFT$
correspondence~\cite{Maldacena:1997re,Witten:1998qj,Gubser:1998bc}; see for example~\cite{Maldacena:1998im,Berenstein:1998ij,Rey:1998ik,Drukker:1999zq,Semenoff:2001xp,Zarembo:2002an,
  Zarembo:2002ph,Tseytlin:2002tr, Bianchi:2002gz,Semenoff:2002kk,
  Pestun:2002mr,Drukker:2005kx,Semenoff:2006am,
  Drukker:2006zk,Drukker:2007yx,Yamaguchi:2007ps,Okuyama:2006jc,Okuyama:2006ir,
  Gomis:2006im,Tai:2006bt,Giombi:2006de,Chen:2006iu,Hartnoll:2006ib,
  Drukker:2006ga,Drukker:2007dw,Chu:2007pb} and references there in.  
Still there has been no direct gauge theory verification of the conjecture beyond the two-loop
level~\cite{Plefka:2001bu,Arutyunov:2001hs} or leading instanton corrections~\cite{Bianchi:2002gz}.

In this paper, we prove the Erickson-Semenoff-Zarembo/Drukker-Gross conjecture for the $\CalN=4$ supersymmetric
Yang-Mills theory and generalize the result in several
directions.
   Let $r$ be the radius of $S^4$. 
The conjecture states that 
\begin{equation}
  \label{eq:main-result-op}
  \langle W_{R}(C) \rangle  = \frac { \int_{\g} [da] \, e^{-\frac { 8 \pi^2 r^2} {\gym^2}(a,a)}  \tr_R e^{2 \pi r i a} }
  {  \int_{\g} [da] \, e^{-\frac { 8 \pi^2 r^2} {\gym^2} (a,a)} }.
\end{equation}

The finite dimensional integrals in this formula are taken over the Lie algebra
$\g$ of the gauge group.   In  
our conventions the kinetic term in the gauge theory is normalized 
as $\frac {1} {2 \gym^2} \int d^4 x \sqrt{g}  (F_{\mu \nu},F^{\mu \nu})$.
The formula~(\ref{eq:main-result-op}) can be
rewritten in terms of the integral over the Cartan subalgebra $\h
\subset \g$ with 
insertion of the usual Weyl measure $\Delta(a) = \prod_{\alpha \in \text{roots of $\g$}} \alpha \cdot a$.

For the $\CalN=2$ and the $\CalN=2^*$ supersymmetric
Yang-Mills theory we get a new formula for the $\langle W_R(C) \rangle$.
 As in the $\CalN=4$ theory, the result can be written in terms of a
matrix model. However, this matrix model is much more complicated than a
Gaussian matrix model. We derive the exact matrix model action
effectively describing all orders of the
perturbation theory. Then we consider the non-perturbative corrections.  

Our  main result is
\begin{equation}
  \label{eq:main-result}
  \boxed{ \langle W_R(C) \rangle =\frac{1}{ Z_{S^4}} \frac {1}
    {\vol(G)} \int_{\g} [da] \, e^{-\frac { 8 \pi^2 r^2} {\gym^2}
      (a,a) } Z_{\text{1-loop}}(ia)|\ZN(ia, r^{-1},r^{-1},q)|^2 \tr_R e^{2\pi r i a}}.
\end{equation}

Here $Z_{S^4}$ is the partition function of the $\CalN=2$, the $\CalN=2^*$ or
the $\CalN=4$ supersymmetric Yang-Mills theory on $S^4$, defined by the path
integral over all fields in the theory, and $\langle W_R(C) \rangle_{\CalN}$ is the
expectation value of $W_R(C)$ in the corresponding theory.  In particular, if
 $R$ is the trivial one-dimensional representation, the formula
 reads
\begin{equation}
  \label{eq:main-result2}
  Z_{S^4} = \frac {1} {\vol (G)}\int [da] e^{-\frac { 8 \pi^2
      r^2} {\gym^2} (a,a) }
  Z_{\text{1-loop}}(ia)|\ZN(ia,r^{-1},r^{-1},q)|^2.
\end{equation}

In other words, we show that the Wilson loop
observable~\eqref{eq:Wilson-loop-defined} is compatible with the localization of
the path integral to the finite dimensional integral~\eqref{eq:main-result} and
that
\begin{equation}
  \boxed{ \langle W_{R}(C) \rangle_{\text{4d theory}}  = \langle \tr_R e^{2 \pi r i a} \rangle_{\text{matrix model}}},
\end{equation}
where the matrix model measure  $\langle \dots \rangle_{\text{matrix model}}$ is given by the
integrand in~\eqref{eq:main-result2}.

The factor $Z_{\text{1-loop}}(ia)$ is a certain infinite dimensional product, which appears as a
determinant in the localization computation.  It can be expressed as a product of Barnes $G$-functions~\cite{Barnes}. In the $\CalN=2^*$ case,
the factor $Z_{\text{1-loop}}(ia)$ is given by the formula~\eqref{eq:Z-Barnes-hyper}. 
The $\CalN=2$ and $\CalN=4$ cases 
can be obtained by taking respectively the limits $m=\infty$ and $m=0$, where $m$ is the hypermultiplet mass 
in the $\CalN=2^{*}$ theory.
For the $\CalN=4$ theory we get $Z_{\text{1-loop}}=1$.

The factor $\ZN(ia,\ep_1,\ep_2, q)$ is Nekrasov's instanton partition
function~\cite{Nekrasov:2003rj,Losev:1997tp,Moore:1997dj}
of the gauge theory in the $\Omega$-background 
 on $\BR^4$. In the $\CalN=2^*$ case $Z_{\text{inst}}$ is given 
by the formula~(\ref{eq:Z-inst-N-star}). In the limit $m=\infty$, one gets the $\CalN=2$ case~(\ref{eq:Z-inst}).
In the limit $m=0$, describing the $\CalN=4$ conformal theory, all instanton corrections vanish.
Vanishing of instanton corrections for the $\CalN=4$ theory contradicts to ~\cite{Bianchi:2002gz}, where 
the first instanton correction for the $\SU(2)$ gauge group was found to be non-zero.
The~\cite{Bianchi:2002gz}  introduces a certain cut-off on the instanton moduli space, 
which is not compatible with the relevant supersymmetry of the theory and the Wilson loop operator. 
Our instanton calculation is based on Nekrasov's partition function on $\BR^4$.
This partition function is regularized by a certain non-commutative deformation of $\BR^4$ compatible with the relevant supersymmetry.
Though we do not construct explicitly the non-commutative deformation of the theory on $S^4$, 
we assume that such deformation can be well defined. We also assume that
 near the North or the South pole of $S^4$ this non-commutative
deformation   agrees with the non-commutative deformation 
used by Nekrasov~\cite{Nekrasov:2002qd} on $\BR^4$.

Since both
$\ZN(ia,\ep_1,\ep_2,q)$ and its complex conjugate appear in the formula,  we count both instantons and anti-instantons. The formula is similar to
Ooguri-Strominger-Vafa relation between the black hole entropy and the
topological string partition function~\cite{Ooguri:2004zv,Beasley:2006us}
\begin{equation}
  \label{eq:OSV}
  Z_{BH} \propto |Z_{top}|^2.
\end{equation}

 The localization can compute more general observables than the Wilson
 loop  on the equator~\eqref{eq:Wilson-loop-defined}. We fix two opposite points on the $S^4$ and call them the North and the South
poles. Then we can consider a class of Wilson loops of
arbitrary radius with a common center at the North pole conjugated
 to each other by a dilation in the north-south
 direction and an  $\SU(2)_{R}$ rotation, where $\SU(2)_R \subset
 \SO(4) \subset \SO(5)$ is the group of self-dual rotations around the North pole. 
Let
$C_\theta$ be a circle located at a polar angle $\theta$. We use
conventions that $\theta = 0$ at the North Pole. We consider
\begin{equation}
  \begin{aligned}
    \label{eq:Wilson-loop-gen-defined}
 W_R(C_\theta) = \tr_{R} \Pexp \oint_{C_\theta} (A_{\mu} dx^{\mu} + \frac {1
    } {\sin \theta} ( i\Phi_\ti^{E} -  \Phi_9 \cos \theta) ds ),
  \end{aligned}
\end{equation}
where $ \Phi_0^E$ and $\Phi_9$ are the scalar fields of the $\CalN=2$ vector multiplet. 

Equivalently,
\begin{equation}
  \label{eq:Wilson-loop-gen-defined-eq}
  W_R(C_\theta) = \tr_{R} \Pexp \oint_{C_\theta} (A_{\mu} dx^{\mu} +
  (i \Phi_\ti^{E} - \Phi_9 \cos \theta) r d\alpha).
\end{equation}
where $\alpha \in [0, 2\pi)$ is the angular coordinate on the circle
$C$. Formally, as the the circle shrinks in the limit $\theta \to 0$, we get the
holomorphic observable $W_R(C_{\theta \to 0}) = \tr_{R} \exp 2\pi r
 i \Phi(\text{N})$,
where $ \Phi(\text{N})$ is the complex scalar field 
$\Phi_0^E +  i \Phi_9$ evaluated at the North pole. In the opposite 
limit ($\theta \to \pi$) we get the anti-holomorphic observable
$W_R(C_{\theta \to \pi}) = \tr_{R} \exp 2\pi r i \bar \Phi(\text{S})$,
where $\bar \Phi({S})$ is the conjugated scalar field $\Phi_0^E -i
\Phi_9$ evaluated at the South pole. 
In this paper, we assume that the size of $C$ is finite.

For any set $\{W_{R_1}(C_{\theta_1}),\dots,
W_{R_n}(C_{\theta_n})\}$ of Wilson loops in the described  class
we claim
\begin{equation}
  \boxed{ \langle W_{R_1}(C_{\theta_1} ) \dots W_{R_n}(C_{\theta_n}) \rangle_{\text{4d theory}}  = \langle \tr_{R_1} e^{2 \pi r i a} \dots 
    \tr_{R_n} e^{2 \pi r i a}  \rangle_{\text{matrix model}}}.
\end{equation}

The Drukker-Gross argument only applies to a single circular loop, because
a circle can be related 
to a straight line on $\BR^4$ by a conformal transformation. In our
approach we can consider several circular loops of certain class  simultaneously. 
Above we have described the class of observables  computable in the massive $\CalN=2^{*}$ theory. 
All these observables are invariant under the same operator $Q$ generated by a conformal Killing spinor on $S^4$ of
constant norm. This operator $Q$ is a fermionic symmetry at quantum level. 

Now we describe more general classes of circular Wilson loops which can be solved in $\CalN=4$ theory. 
Because of the conformal symmetry, the $\CalN=4$ theory contains
a family of
supersymmetry operators usable for localization. Denoting the parameter
of this family by $t$, such that  $t=1$ describes $Q$ 
used to construct the $\CalN=2^{*}$ theory, in the $\CalN=4$ theory we consider 
\begin{equation}
  \begin{aligned}
    \label{eq:Wilson-loop-gen-defined-more-general}
    W_R(C_\theta,t) = \tr_{R} \Pexp \oint_{C_\theta} \left(A_{\mu} dx^{\mu} + \frac {1
    } {t \sin \theta} \left( (\sin^2 \frac \theta 2+t^2 \cos^2 \frac \theta 2)     i\Phi_\ti^{E} + \Phi_9
( \sin^2 \frac \theta 2 - t^2 \cos^2 \frac \theta 2 )\right) ds \right).
  \end{aligned}
\end{equation}
The $\CalN=4$ theory with insertion of the operator $W_{R}(C_{\theta},t)$ still
 localizes to the Gaussian matrix model. 

The localization principle is that in some situations the integral is
exactly equal to its semiclassical approximation. For example, the
Duistermaat-Heckman formula~\cite{MR674406} is
\[ \int_M \frac {\omega^{n}}{ (2 \pi)^n n!} e^{i H(\phi)} = i^n \sum_{p \in F}
\frac {e^{i H(\phi)}} {\prod \alpha_i^{p}(\phi)},\] 
where $(M,\omega)$ is a
symplectic manifold,  $H: M \to \g^*$ is a moment map\footnote{In other words,
  $ i_{\phi} \omega = d H(\phi)$ for any $\phi \in \g$, where $i_{\phi}$ is a
  contraction with a vector field generated by $\phi$.} for a Hamiltonian action
of a torus $G=\U(1)^k$ on $M$, and $F$ is the $G$-fixed point set. 

The Duistermaat-Heckman formula is a particular case of a
more general Atiyah-Bott-Berline-Vergne localization
formula~\cite{MR721448,MR685019}.  Let a torus $G$ act on a compact
manifold $M$. We consider the complex of $G$-equivariant differential forms on $M$
valued in functions on $\g$. The differential is $Q = d - \phi^{a} i_{a}$. The
$Q^2$ is  a symmetry transformation: $Q^2 = - \phi^{a}
\CalL_{v^{a}}$. Here $\CalL_{v^{a}}$ represents the $G$-action of vector
fields $v^{a}$ by Lie derivatives. The operator
$Q^2$ annihilates $G$-invariant differential forms.  

Given a  $Q$-closed form $\alpha$,
Atiyah-Bott-Berline-Vergne localization formula claims
\[ \int_{M} \alpha = \int_{F} \frac {{i^*_F \alpha}}{e(N_{F})}, \] where \
$F \overset{i}{\hookrightarrow} M$ is the $G$-fixed point set, and $e(N_{F})$ is the equivariant Euler class
of the normal bundle of $F$ in $M$. When $F$ is a discrete set of points, the
equivariant Euler class $e(N_F)$ at each point $f \in F$ is simply the determinant of the representation in which $\g$ acts on 
the tangent bundle of $M$ at a point $f$. 

Localization can be argued in the following
way~\cite{Witten:1988ze,Witten:1991zz}.  Let $Q$ be a fermionic symmetry of a
theory. The $Q^2 = \CalL_{\phi}$ is a bosonic symmetry. Let $S$ be a
$Q$-invariant action, so that $QS = 0$.  Consider a functional $V$ which is
invariant under $\CalL_{\phi}$, so that $Q^2 V = 0$.  Deformation of the action
by a $Q$-exact term $QV$ can be written as a total derivative.
Hence, such deformation does not
change the integral up to a boundary contribution
\[ \frac {d} {d t} \int e^{S + t Q V } = \int \{Q,V\} e^{S + t Q V} = \int \{Q,
V e^{S + t Q V}\} = 0. \] As $t \to \infty$, the one-loop approximation at the
critical set of $QV$ becomes exact.  For a sufficiently nice $V$, the
integral is then computed evaluating $S$ at critical points of $QV$ and the corresponding
one-loop determinant.

We apply this strategy to the $\CalN=2$, the $\CalN=2^*$ and the $\CalN=4$
supersymmetric Yang-Mills gauge theories on $S^4$ and show that the path integral
is localized to the constant modes of the scalar field $\Phi_0$ with all other
fields vanishing. In this way we also compute exactly the expectation value of
the circular supersymmetric Wilson loop operator~\eqref{eq:Wilson-loop-defined}.

\emph{Remark.} Most of the arguments in this paper apply to $\CalN=2$ theory with an arbitrary 
matter content. Because of technical regularization issues, we limit 
our discussion to the $\CalN=2$ theory with a single $\CalN=2$ massive hypermultiplet in the adjoint representation,
also known as the $\CalN=2^{*}$. By taking the limit of zero or infinite mass
we can respectively recover the $\CalN=4$ or the $\CalN=2$ theory.

In~(\ref{eq:Z-1-loop-any-matter}) we give $Z_{\text{1-loop}}$ for an $\CalN=2$ gauge theory with a massless hypermultiplet in
such representation that the theory is conformal.
Perhaps, one could check our result by the traditional Feynman diagram computations 
directly in the gauge theory. To simplify comparison, 
we compute the supersymmetric Wilson loop expectation value in the
$\SU(2)$ theory with $N_f=4$ fundamental hypermultiplets up to
$\gym^6$, see explicitly (\ref{se:first-corrections}).

 Note that the first difference between the $\CalN=2$ superconformal 
theory and the $\CalN=4$ theory appears at the order $\gym^6$. 
In this order  the Feynman diagrams in the $\CalN=4$ theory were computed in~\cite{Plefka:2001bu,Arutyunov:2001hs}.
Therefore a direct computation of Feynman diagrams in the $\CalN=2$ theory up to $\gym^6$ seems 
to be possible and will test our results.

We notice a couple of features in this work unusual in the studies of
topological field theories: (i) the theory localizes
not on a counting problem, but on a nontrivial matrix model, 
(ii) there is a one-loop factor involving an index theorem for
transversally elliptic operators~\cite{MR0482866,MR0341538}. However, we should
not be surprised, as even though we are using cohomological field
theory methods, we study the physical untwisted theory.

In section 2 we describe the $\CalN=2$, the $\CalN=2^*$ and
the $\CalN=4$ SYM theories
on $S^4$. In section 3 we explain the localization principle for these theories. 
 Section 4 proceeds with the computation of the one-loop determinant~\cite{MR0341538,MR0482866}, 
or, mathematically speaking, of the equivariant Euler class of the infinite-dimensional 
normal bundle to the localization locus. 
In section 5 we consider instanton corrections.

There are open questions and several immediate directions to explore:
\begin{enumerate}
 \item{ One can consider more general supersymmetric Wilson
     loops~\cite{Drukker:2007qr,Drukker:2007yx,Drukker:2007dw} and try
     to prove the conjectures relating those loops with matrix models or two-dimensional super Yang-Mills theory. 
}

 \item{ Using localization, one can try to compute exactly the
 expectation value of a circular supersymmetric 
't Hooft-Wilson operator. Such operator is a generalization of Wilson loop in which the loop carries 
both electric and magnetic charges~\cite{Kapustin:2006hi,Kapustin:2005py,Kapustin:2006pk}. 
The expectation values of such operators should transform in the right way under the $S$-duality
transformation which replaces the coupling constant by its inverse and the gauge group $G$
by its Langlands dual $^LG$. 
From such computation we can learn more about the relation between the four-dimensional gauge theory 
and geometric Langlands~\cite{Kapustin:2006pk} where 't Hooft-Wilson loops play the key role.}

\item{ Is there a more precise relation between 
the gauge theory formula of this paper  
and Ooguri-Strominger-Vafa~\cite{Ooguri:2004zv}
conjecture~(\ref{eq:OSV}), or is there a four-dimensional analogue of the $tt^{*}$-fusion~\cite{Cecotti:1991me}?
}
\end{enumerate}

\bigskip {\bf{Acknowledgment.}} I would like to thank E.~Witten and N.~Nekrasov
for many inspiring discussions, important comments and suggestions.  I thank
M.~Atiyah, N.~Berkovits, A.~Dymarsky, D.~Gaiotto, S.~Gukov, J.~Maldacena,
I.~Klebanov, H.~Nakajima, A.~Neitzke, I.~Singer, M.~Rocek, K.~Zarembo for interesting discussions and
remarks.  Part of this research was done during my visit to Physics Department
of Harvard University in January 2007. The work was supported in part by Federal
Agency of Atomic Energy of Russia, grant RFBR 07-02-00645, grant for support of
scientific schools NSh-8004.2006.2 and grant of the National Science Foundation PHY-0243680.
Any opinions, findings, and conclusions or recommendations expressed in this material are those
of the authors and do not necessarily reflect the views of these
funding agencies.

\section{Fields, action and symmetries}
\label{se:fields-action-symmetries}
To construct explicitly the action of $\CalN=4$ SYM on $S^4$, we start from
the $\CalN=4$ SYM on $\BR^{4}$ obtained by  dimensional reduction
of the $\CalN=1$ SYM~\cite{Brink:1977bc} on $\BR^{9,1}$ along $5+1$ dimensions.  Let $G$ be the gauge group, which we assume 
to be  a compact Lie group. Let $\g$ be the Lie algebra of $G$.  
By $A_{M}$,  $M = 0, \dots, 9$ we denote the components
of the gauge field in ten dimensions, where we assume the Minkowski metric $ds^2 =
- dx_0^2 + dx_1^2 + \dots + dx_9^2$.  In 10d Euclidean signature
with the metric $ds^2 = dx_0^2 + dx_1^2 + \dots + dx_9^2$ we use notation 
$A_0^E$ for the zero component of the gauge field.

By $\Psi$ we denote $\Spin(9,1)$ Majorana-Weyl fermion valued in the adjoint
representation of $G$.  In the $(9,1)$ signature $\Psi$ has 16 real
components. In the $(10,0)$ signature $\Psi$ is not real, but its complex conjugate
does not appear in the theory. The ten-dimensional $\CalN=1$ SYM action is $S= \int
d^{10} x \CalL$  with the Lagrangian
\begin{align}
  \CalL = \frac {1} { \gym^2} \lb \frac 1 2 F_{MN} F^{MN} - \Psi \Gamma^{M}
  D_{M} \Psi \rb.
\end{align}
In Euclidean signature we integrate over fields taking value 
in the real Lie algebra of the gauge group. We write the covariant derivative
as $D_{\mu} = \p_{\mu} + A_{\mu}$ and the field strength as $F_{\mu \nu} = [D_{\mu}, D_{\nu}]$.
For example, for the $\U(N)$ gauge
group the fields are represented by the antihermitian matrices. 
We are not writing explicitly the color and spinor indices. We  assume
contraction of color indices in all bilinear terms 
using an invariant positive definite bilinear form $(\cdot, \cdot)$ on
$\g$. For the $\U(N)$ gauge group  we define the Killing form as $(a,b) = -\tr_{\text{F}} a b$, where
$\tr_{\text{F}}$ is the trace in the fundamental representation.

The action is invariant under the supersymmetry transformations
\begin{align*}
  & \delta_{\ve} A_{M} = \ve \Gamma_{M} \Psi \\
  & \delta_{\ve} \Psi = \frac 1 2 F_{MN} \Gamma^{MN} \ve.
\end{align*}
Here $\ve$ is a constant Majorana-Weyl spinor parametrizing the
supersymmetry transformations in ten dimensions. (See appendix~\ref{sec:Octonionic-gamma-matrices}
for our conventions on the algebra of gamma-matrices.)

We take $(x_1,\dots, x_4)$ to be the coordinates along the four-dimensional
Euclidean space-time, and we do dimensional reduction in $(x_5,\dots, x_9,x_0)$.

Now we describe the symmetries of the four-dimensional theory obtained
 from the $9+1$ dimensional $\CalN=1$ SYM. 
Notice that in the dimensional reduction 
we eliminate the time coordinate $x_0$. 
Therefore, we get negative kinetic
term for the scalar field $\Phi_0$ obtained from the $0$-th component of the 
gauge field $A_M$. To ensure convergence of
the path integral we integrate over imaginary $\Phi_0$, which we present
as $\Phi_0 = i \Phi_0^E$ where $\Phi_0^E$ is real.
This makes the path integral of the reduced $(9,1)$ theory
to be equivalent to the path integral of the reduced $(10,0)$ theory. 

The ten-dimensional $\Spin(9,1)$ Lorentz symmetry group is broken to $\Spin(4) \times \Spin(5,1)^R$,
where  $\Spin(4)$ is the four-dimensional Lorentz group acting on $(x_1,
\dots, x_4)$ and the $\Spin(5,1)^R$ is the R-symmetry group acting on $(x_5,
\dots, x_9, x_0)$. It is convenient to split the Lorentz group as $\Spin(4) =
SU(2)_L \times SU(2)_R$, and brake the $\Spin(5,1)^R$-symmetry group into
$\Spin(4)^R \times \SO(1,1)^R=\SU(2)_L^R \times \SU(2)_R^R \times
\SO(1,1)^R$.
 We denote by $A_{\mu}$,
 $\mu = 1,\dots, 4$, the four-dimensional gauge fields and by $\Phi_{A}$, $A =
0,5,\dots,9$ the four-dimensional scalar fields. 
 Here are the bosonic fields and  the symmetry groups
acting on them
\[ \overbrace{A_{1}, \dots A_{4}}^{ SU(2)_L \times SU(2)_R } \, \,
\overbrace{\Phi_5, \dots, \Phi_8}^{{ \SU(2)_L^R \times \SU(2)_R^R }} \, \,
\overbrace{\Phi_9,\Phi_0}^{{\SO(1,1)^R}}.\]

Using a certain Majorana-Weyl representation of the Clifford algebra
$\Cl(9,1)$ (see appendix~\ref{sec:Octonionic-gamma-matrices} for our
conventions), we write $\Psi$ in terms of four-dimensional chiral spinors
as \[
\Psi = \begin{pmatrix}
\psi^L \\ \chi^R \\ \psi^R \\ \chi^L
\end{pmatrix}.
\] 

Each of these spinors ($\psi^L,\chi^R,\psi^R,\chi^L$) has four
real components. Their transformation properties are summarized in the table:

\bigskip
\begin{tabular}{|c|c|c|c|c|c|c|}
  \hline
  $\ve$ &  $\Psi$   &$\SU(2)_L$& $\SU(2)_R$ & $\SU(2)_L^R$ & $\SU(2)_R^R$ & $\SO(1,1)^R$ \\
  $ * $ & $\psi^{L}$ & $1/2$ & 0  & $1/2$  & 0 & $+$ \\
  $ 0 $ & $\chi^{R}$ & $0$  & $1/2$& 0   & $1/2$ & $+$ \\
  $ * $ & $\psi^{R}$ & $0$  & $1/2$& $1/2$ &0 & $-$ \\
  $ 0 $ &  $\chi^{L}$ & $1/2$ & 0 & 0   & $1/2$ & $-$ \\
  \hline
\end{tabular}
\bigskip

Let the spinor $\ve$ be the parameter of the supersymmetry transformations. 
We restrict the $\CalN=4$ supersymmetry algebra to the $\CalN=2$ subalgebra 
by taking $\ve$ in the $+1$-eigenspace of the operator $\Gamma^{5678}$. 
Such spinor $\ve$ has the structure 
\[
\ve=
\begin{pmatrix}
  \ve^{L} \\ 0 \\ \ve^{R} \\ 0
\end{pmatrix},
\]
transforms in the spin-$\frac 1 2$ representation of the $\SU(2)^R_L$ and in the trivial representation of the $\SU(2)^R_R$.

With respect to the supersymmetry transformation generated by such $\ve$,
the $\CalN=4$ gauge multiplet splits in two parts 
\begin{itemize}
\item{ $(A_1\dots A_4,\Phi_9,\Phi_0,\, \psi^L, \psi^{R})$ is the $\CalN=2$ vector
    multiplet }
\item{ $(\Phi_5 \dots \Phi_8,\, \chi^L, \chi^R)$ is the $\CalN=2$ hypermultiplet}.
\end{itemize}

After we have reviewed the dimensional reduction from the
 $\BR^{9,1}$ to the $\BR^4$, we put the 4d theory on the round
 four-sphere $S^4$. Let $r$ be the radius of $S^4$. 

Because of the curvature, the kinetic term for the scalar fields is
deformed as $(\partial \Phi)^2 \to
(\partial \Phi)^2 + \frac R 6 \Phi^2$, where $R$ is the scalar curvature. 
This can be derived from the conformal invariance; one 
can check that $\int d^{4} x \, \sqrt{g} ( (\p \Phi)^2 + \frac R 6 \Phi^2 )$
is invariant under Weyl transformations of the metric $g_{\mu \nu} \to e^{2\Omega}g_{\mu \nu}$
and scalar fields $\Phi \to e^{-\Omega} \Phi$. 
Then the action on $S^4$ of the $\CalN=4$ SYM is
\begin{equation}
  \label{eq:SYM_S^4_N=4}
  S_{\CalN=4} = \frac 1 { \gym^2} \int_{S^4} \sqrt{g} d^4 x \left(\frac 1 2 F_{MN} F^{MN}
    - \Psi \Gamma^{M} D_{M} \Psi + \frac 2 {r^2} \Phi^A \Phi_A  \right),
\end{equation}
where we used that the scalar curvature on $S^4$ is $R= \frac {12}{r^2}$. 

The action~\eqref{eq:SYM_S^4_N=4} is invariant under the $\CalN=4$ superconformal
transformations
\begin{align}
  \label{eq:SYM_S^4_N=4-trans}
  & \delta_{\ve} A_{M} = \ve \Gamma_{M} \Psi \\
  & \delta_{\ve} \Psi = \frac 1 2 F_{MN} \Gamma^{MN} \ve + \frac 1 2 \Gamma_{\mu A}
  \Phi^A \nabla^{\mu} \ve,
\end{align}
where $\ve$ is a conformal Killing spinor solving the equations
\begin{align}
  \label{eq:conformal-Killing}
  & \nabla_{\mu} \ve = \tilde \Gamma_{\mu} \tilde \ve \\
  & \nabla_{\mu} \tilde \ve = - \frac {1} {4 r^2} \Gamma_{\mu} \ve.
\end{align}
(See e.g.~\cite{math0202008v1} for a review on conformal Killing spinors,
and for the explicit solution of these equations on $S^4$ see appendix~\ref{se:Killing-spinors}.)
The meaning of $\ve$ and $\tilde \ve$ simplifies in  the flat space limit $r \to \infty$. 
In this limit, $\tilde \ve$ becomes a constant spinor $\tilde \ve = \hat \ve_{c}$, 
while $\ve$ becomes a spinor with 
at most linear dependence on the flat coordinates $x^{\mu}$ on $\BR^4$: 
$\ve = \hat \ve_s + x^{\mu} \Gamma_{\mu} \hat \ve_{c}$.
The spinors $\hat \ve_{s}$ and $\hat \ve_{c}$  parametrize the
solution. The $\hat \ve_{s}$ generates Poincare 
supersymmetry transformations and the $\hat \ve_{c}$ generates special superconformal symmetry transformations.

The superconformal algebra closes only on-shell (see appendix~\ref{se:off-shell susy}) 
\begin{equation}
  \begin{aligned}
    \label{eq:delta2-closed-main-text}
    & \delta_{\ve}^2 A_{\mu} = - (\ve \Gamma^{\nu} \ve) F_{\nu \mu} - [(\ve \Gamma^{B}\ve) \Phi_B, D_{\mu}] \\
    & \delta_{\ve}^2 \Phi_A = - (\ve \Gamma^{\nu} \ve) D_{\nu} \Phi_{A} - [(\ve
    \Gamma^{B} \ve )\Phi_B, \Phi_A] +
    2 (\tilde \ve  \Gamma_{AB}  \ve) \Phi^{B} - 2(\ve \tilde \ve) \Phi_A \\
    & \delta_{\ve}^{2} \Psi = -(\ve \Gamma^{\nu} \ve) D_{\nu} \Psi - [(\ve \Gamma^{B}
    \ve )\Phi_B, \Psi] - \frac 1 2 (\tilde \ve \Gamma_{\mu \nu} \ve) \Gamma^{\mu
      \nu} \Psi + \frac 1 2 (\tilde \ve  \Gamma_{AB}  \ve) \Gamma^{AB} \Psi
    -3(\tilde \ve \ve) \Psi + \text{eom}[\Psi].
  \end{aligned}
\end{equation}
The term denoted by $\text{eom}[\Psi]$ is proportional to the Dirac equation of
motion
\begin{equation}
  \label{eq:opsi}
  \text{eom}[\Psi] = \frac 1 2 (\ve \Gamma_N \ve) \tilde \Gamma^{N} \Dslash \Psi - (\ve \Dslash \Psi) \ve.
\end{equation}

We present $\delta_{\ve}^2$  as 
\begin{equation}
  \label{eq:square-susy}
  \delta^{2}_{\ve}  = -\CalL_{v}  - R  - \Omega.
\end{equation}

The $\CalL_{v}$ is the gauge covariant Lie derivative in the
direction of the vector field
\begin{equation}\label{eq:v-in-terms-eps}
  v^M = \ve \gamma^{M} \ve.
\end{equation}
The gauge covariant Lie derivative $\CalL_{v}$ acts on scalar fields as follows: $\CalL_v {\Phi_A} = v^M D_{M} \Phi= 
v^{\mu} D_{\mu} \Phi_A + v^{B} [\Phi_{B}, \Phi]$, 
where $D_{\mu}$ is the usual covariant derivative $D_{\mu} = \p_{\mu} +
A_{\mu}$.

The $R$ in~\eqref{eq:square-susy} is a $\Spin(5,1)^R$-symmetry
transformation acting on scalar fields as $(R \cdot \Phi)_{A} = R_{AB}
\Phi^{B}$, and on fermions as $R \cdot \Psi = \frac 1 4 R_{AB} \Gamma^{AB}
\Psi$, where $R_{AB} = 2 \ve \tilde \Gamma_{AB} \tilde \ve$.  When
$\ve$ and $\tilde \ve$ are restricted to the $\CalN=2$ subspace of $\CalN=4$
algebra, $\Gamma^{5678} \ve = \ve$ and $\Gamma^{5678} \tilde \ve = \tilde
\ve$, the matrix $R_{AB}$, $A,B=5,\dots, 8$, is an
anti-self-dual left generator of $\SU(2)_L^R \subset \SO(4)^{R}$ rotations. 
 The fermionic fields of 
the $\CalN=2$ vector multiplet $\psi$ transform in the trivial
representation of $R$ and 
the fermionic fields of the $\CalN=2$ hypermultiplet $\chi$
transform in the spin-$\frac 1 2$ representation of $R$. 

Finally, the term denoted by $\Omega$ in~\eqref{eq:square-susy} generates 
a local dilatation with the parameter $2(\ve \tilde
\ve)$, under which the gauge fields do not transform, the scalar fields transform with weight $1$, 
and the fermions transform with weight $\frac 3 2$. In other words, if we make Weyl transformation 
$g_{\mu \nu} \to e^{2\Omega} g_{\mu \nu}$, we scale the fields as $A_{\mu} \to A_{\mu}, 
\Phi \to e^{-\Omega} \Phi, \Psi \to e^{-\frac 3 2 \Omega} \Psi$ to keep the action invariant.

Classically, it is easy to restrict the fields and the symmetries of the $\CalN=4$ SYM
to the minimal $\CalN=2$ SYM: discard all fields of the
$\CalN=2$ hypermultiplet and restrict $\ve$ to the $+1$ eigenspace of
$\Gamma^{5678}$. The resulting action is invariant under $\CalN=2$ superconformal symmetry.  
On quantum level, the minimal $\CalN=2$ SYM is not  conformally invariant.  
We will precisely define  the quantum $\CalN=2$ theory on $S^4$,
considering it as the $\CalN=4$ theory softly broken by a massive
deformation defined in a certain way on $S^4$.

In the $\CalN=2$ case, $\ve$ is a Dirac spinor on $S^4$.  The
equation~\eqref{eq:conformal-Killing} has 16 linearly independent
solutions corresponding to the fermionic generators of the $\CalN=2$
superconformal algebra. Intuitively, $8$ generators out of these 16
correspond to $8$ charges of
$\CalN=2$ Poincare supersymmetry algebra on $\BR^4$, and the other $8$ correspond to the
remaining generators of $\CalN=2$ superconformal algebra.  The full $\CalN=2$
superconformal group on $S^4$ is $\SL(1|2, \BH)$.\footnote{By $\SL(n,\BH)$ we denote
the  group of general linear transformation $\GL(n, \BH)$ over quaternions factored
  by $\BR^{*}$;  the real dimension of $\SL(n,\BH)$ is $4n^2-1$.}
Its bosonic subgroup is $\SL(1,\BH)\times \SL(2,\BH)\times \SO(1,1)$.
The $\SL(1,\BH) \simeq \SU(2)$ generates
the $R$-symmetry $\SU(2)^R_L$ transformations. 
The $\SL(2,\BH) \simeq \SU^{*}(4,\BC) \simeq \Spin(5,1)$ generates  conformal
transformations of $S^4$. 
The $\SO(1,1)^R$ generates the $\SO(1,1)^R$ symmetry transformations.  
The fermionic generators of $\SL(1,2|\BH)$ transform in the
$\mathbf{2+2'}$ of the $\SL(2,\BH)$, where $\mathbf{2}$ denotes
the quaternionic fundamental representation. 
This representation can be identified with the
complex fundamental representation $\mathbf{4}$ of $\SU^{*}(4)$, or with chiral Weyl spinor representation of the conformal group $\Spin(5,1)$.  
The other representation $\mathbf{2'}$ corresponds to the other chiral spinor representation of $\Spin(5,1)$ of
the opposite chirality.

In the $\CalN=4$ case we do not impose the chirality condition on $\ve$. 
Hence a sixteen component Majorana-Weyl spinor $\ve$ of $\Spin(9,1)$ reduces to
a pair of the four-dimensional Dirac spinors $(\ve_{\psi}, \ve_{\chi})$, 
where $\ve_{\psi}$ and $\ve_{\chi}$ are respectively 
in the $+1$ and $-1$ eigenspaces
of  $\Gamma^{5678}$. 
Each of the Dirac spinors $\ve_{\psi}$ and $\ve_{\chi}$ independently
satisfies the conformal Killing spinor equation~(\ref{eq:conformal-Killing}) 
because the operators $\Gamma_{\mu}$ commute with $\Gamma^{5678}$.
Then we get 16+16 = 32 linearly independent conformal Killing
spinors. Each 
of them  corresponds to a generator of the $\CalN=4$ superconformal symmetry. 
The full $\CalN=4$ superconformal group on $S^4$ is $\PSL(2|2, \BH)$.

Since the mass terms of the $\CalN=2^*$ theory breaks conformal invariance,
we should expect the $\CalN=2^{*}$ theory to be invariant only
under
the half of $16$ 
fermionic symmetries of the $\CalN=2$ superconformal group
$\SL(1,2|\BH)$. Therefore we restrict the spinor $\ve$.

First, let us describe the general solution for the conformal spinor
Killing equation on $S^4$. 
Let $x^{\mu}$ be the stereographic coordinates on $S^4$. The point
$x^{\mu} = 0$ is the North pole, the point $|x| = \infty$ is the South
pole.  
The metric is
\begin{equation}\label{eq:metric-S4-stereo}
  g_{\mu \nu} = \delta_{\mu \nu} e^{2\Omega},
  \quad \text{where} \quad  e^{2 \Omega} := \frac 1 { (1 + \frac {x^2} {4 r^2})^2 }.
\end{equation}
We use the vielbein $e_{\mu}^{i} = \delta^{i}_{\mu} e^{\Omega}$ where
$\delta^{i}_{\mu}$ is the Kronecker delta,  $\mu=1,\dots,4$ is the
space-time index, $i=1,\dots,4$ is the vielbein index. Solving
 the conformal Killing equation~\eqref{eq:conformal-Killing} we get (appendix~\ref{se:Killing-spinors})
\begin{align}
  \label{eq:Killing-solution-in-S^4-stereo}
  \ve =   \frac 1 {\sqrt{1+ \frac {x^2} {4 r^2}}} (\hat \ve_{s} +
  x^{i} \tilde \Gamma_{i} \hat \ve_{c})  \\
  \tilde \ve = \frac 1 {\sqrt{1+ \frac {x^2} {4 r^2}}} (\hat \ve_{c} - \frac
  {x^{i} \Gamma_{i} }{4 r^2} \hat \ve_{s}),
\end{align}
where $\hat \ve_{s}$ and $\hat \ve_{c}$ are the Dirac spinor valued parameters.

Classically, the action of $\CalN=2$ SYM on $\BR^4$ with a  massless
hypermultiplet is invariant under the $\CalN=2$ superconformal group, which has
16 fermionic generators. Massive deformation breaks 
8 superconformal fermionic symmetries, but preserves the other 8
Poincare supersymmetries. 
These 8 charges are known to be preserved on quantum level~\cite{Seiberg:1994rs}. 
The usual $\CalN=2$ supersymmetry algebra closes to the 
isometries of $\BR^4$ compatible with the massive deformation of the
action. 

For the theory on $S^4$, following the same logic, we want to find
 a subgroup  $\CalS \subset \SL(1|2,\BH)$  such that: 
(i) $\CalS$ contains $8$ fermionic generators, (ii)
the bosonic transformations of $\CalS$ are the isometries 
compatible with mass terms for the hypermultiplet. 
We will call the group $\CalS$ the $\CalN=2$ supersymmetry group on $S^4$.

The  group of conformal symmetries of  $S^4$ is $\SO(5,1)$. The
group of the isometries of $S^4$ is $\SO(5)$. 
  We require that the space-time bosonic part of $\CalS$ is a subgroup of the $\SO(5)$. 
This means that for any conformal Killing spinor $\ve$  generating
a fermionic transformation of $\CalS$, the bilinear $(\tilde \ve  \ve)$
in the $\delta^2_{\ve}$ vanishes. 

For a general $\ve$ generating the $\CalN=2$ superconformal transformation, the $\delta_{\ve}^2$ contains
$\SO(1,1)^{R}$ generator.  Since the $\SO(1,1)^R$ symmetry is broken
explicitly by hypermultiplet mass terms, 
 we require that $\CalS$ 
contains no $\SO(1,1)^{R}$ transformations: the equation
\eqref{eq:delta2-closed-main-text} implies
 $\tilde \ve
\Gamma^{09} \ve = 0$.  

Using the explicit
solution~\eqref{eq:Killing-solution-in-S^4-stereo} we rewrite the equation $(\tilde \ve
 \ve) = (\tilde \ve \Gamma^{09} \ve) = 0$ in terms of $\hat \ve_{s}$ and $\hat
\ve_{c}$
\begin{equation}\label{eq:good-Killing-spinors}
  \begin{aligned}
    \hat \ve_{s} \hat \ve_{c} = \hat \ve_{s} \Gamma^{09} \hat \ve_{c} = 0 \\
    \hat \ve_{c} \Gamma^{\mu} \hat \ve_{c} - \frac 1 {4 r^2} \hat \ve_{s}
    \Gamma^{\mu} \hat \ve_{s} =0.
  \end{aligned}
\end{equation}
To solve the second equation, we take chiral $\hat \ve_{s}$ and $\hat
\ve_{c}$ with respect to the four-dimensional chirality operator $\Gamma^{1234}$.
Since the operators $\Gamma^{\mu}$ reverse the four-dimensional chirality,
both terms in the second equation vanish automatically.  
There are two interesting cases: (i) the chiralities of $\hat \ve_{s}$ and $\hat \ve_{c}$ 
are opposite, (ii) the chiralities of $\hat \ve_{s}$ and $\hat \ve_{c}$ are the same. 
This paper focuses on the second case.

1. In the first case we can assume that 
\[ \hat \ve_{s}^{L} = 0, \quad \hat \ve_{c}^{R} = 0. \] 
Here by $\hat \ve_{s}^{L}$ and $\hat \ve_{s}^{R}$ we denote left/right four-dimensional 
chiral components, which are in the 
$+1/-1$ eigenspaces of $\Gamma^{1234}$.
The first equation in~(\ref{eq:good-Killing-spinors})
is  automatically satisfied. 
Moreover, the spinors $\ve$ and $\tilde \ve$  have
opposite chirality everywhere on  $S^4$.  The $8$ generators, components
of $\hat \ve_{c}^{L}$ and  $\hat \ve_{s}^{R}$ anticommute to pure gauge
transformations
 generated by the field $\Phi := (\ve \Gamma^{A} \ve) \Phi_{A}$.  The $\delta_{\ve}$-closed
observables are the gauge invariant functions of $\Phi$ and their descendants.
Such $\delta_{\ve}$ relates to the cohomological BRST operator $Q$ of
Donaldson-Witten theory on a punctured $S^4$; the puncture is at the point where
$\ve$ vanishes. This is unlike the twisted
theory~\cite{Witten:1988ze,Vafa:1994tf} in which twisted $\ve$ is a
nowhere vanishing constant scalar.

2. The spinors $\hat \ve_{s}$ and $\hat \ve_{c}$ have the same chirality, say
right, and the first equation restricts them to be orthogonal
\[ \hat \ve_{s}^{L} = 0, \quad \hat \ve_{c}^{L} = 0, \quad (\hat \ve_{s}^R \hat
\ve_{c}^R) =0. \] 
The Killing vector field $v^{\mu} = \ve \Gamma^{\mu} \ve$,
associated with the $\de_{\ve}^2$, generates self-dual right rotation
of $S^4$ around the North pole. In addition, $\de_{\ve}^2$ generates a
$\SU(2)_{L}^R$-symmetry transformation and a gauge symmetry transformation. The
spinor $\ve$ is chiral only at the poles of
$S^4$:   right at the North poles and left at the South pole. 
The circular Wilson loop operators~\eqref{eq:Wilson-loop-defined}  are invariant under such
$\delta_{\ve}$. 
If both spinors $\hat \ve_{s}$ and $\hat \ve_{c}$ do not vanish, then $\ve$
is a nowhere vanishing spinor on $S^4$.  Such $\de_{\ve}$ relates the
physical $\CalN=2$ gauge theory on $S^4$ to an unusual equivariant
topological theory, for which localization
methods~\cite{Witten:1988ze,MR1094734} 
can be applied.

Before we describe this unusual cohomological field theory, we will
complete our description of the supersymmetry group $\CalS$ of 
the $\CalN=2^{*}$ theory on $S^4$. 
First we find the basis of conformal Killing spinors $\{\ve^i\}$ that  simultaneously
satisfy the equations
\begin{equation}
  \label{eq:no-dilat-cond}
  \ve^{(i} \tilde \ve^{j)} =
  \ve^{(i} \Gamma^{09} \tilde \ve^{j)} = 0,
\end{equation}
and then we find the superconformal algebra generated by this basis.
The conformal Killing spinor equation
restricted
to the  $+1$ eigenspace of $\Gamma^{5678}$ 
is  
\begin{equation}
  \label{eq:Killing-spinor1}
  D_{\mu} \ve = \frac 1 {2 r}  \Gamma_{\mu} \Lambda \ve,
\end{equation}
where $\Lambda$ is a generator of $SU(2)^{R}_L$-symmetry. 
For example, for the reduced $(9,1)$ $\CalN=1$ SYM we take 
$\Lambda=\Gamma^{0}\Gamma_{ij}$, where $i,j = 5, \dots, 8$; 
for the reduced $(10,0)$ $\CalN=1$ SYM we
 take $\Lambda=-i\Gamma^{0} \Gamma_{ij}$.
The matrix  $\Lambda$ is a
real antisymmetric matrix, commuting with
$\Gamma^{5678}$ and $\Gamma^{m}$, $m=1,\dots,4,0,9$, and such that
$\Lambda^2 = -1$. 
The equation~\eqref{eq:Killing-spinor1} has 8 linearly independent 
solutions.  Let $V_{\Lambda}$ be the vector space spanned by
these solutions.  Then the
space of solutions of the conformal Killing spinor equations~\eqref{eq:conformal-Killing} 
is $V_{\Lambda} \oplus V_{-\Lambda}$ with $\tilde \ve = \frac {1} {2r} \Lambda \ve$.

The spinors in the space $V_{\Lambda}$ satisfy our requirement~\eqref{eq:no-dilat-cond},
because $\Lambda$ is antisymmetric and commutes with $\Gamma^{\tic}$. The generators
$\{\delta_{\ve} | \ve \in V_{\Lambda}\}$ anticommute to generators of $\Spin(5)
\times \SO(2)^{R}$, where $\Spin(5)$ rotates $S^4$, and $\SO(2)^{R}
\subset \SU(2)^R_L$ is generated by $\Lambda$. 
The space $V_{\Lambda}$
transforms in the fundamental representation of $\Sp(4) \simeq
\Spin(5)$.  
We conclude that restricting the fermionic generators to the space $V_{\Lambda}$
 breaks the $\CalN=2$ superconformal group
$\SL(1|2,\BH)$ to the supergroup $\OSp(2|4)$, where the choice of the $\SU(2)^{R}_L$
generator $\Lambda$ determines the embedding  $\SO(2)_{R}
\hookrightarrow \SU(2)^{R}_L$.

Besides the space $V_{\Lambda}$, obtained as solutions
of~\eqref{eq:Killing-spinor1}, there are other half-dimensional fermionic
subspaces of the $\CalN=2$ superconformal algebra which 
satisfy~\eqref{eq:no-dilat-cond}.  We obtain these spaces by
$SO(1,1)_{R}$ conjugation of $V_{\Lambda}$.  Indeed, if the spinors $\ve$ and
$\tilde \ve$ satisfy~\eqref{eq:no-dilat-cond}, then so do the spinors $\ve' =
e^{\frac 1 2 \beta \Gamma^{\ti\tic}} \ve $ and $\tilde \ve' = e^{-\frac 1 2
  \beta \Gamma^{\ti\tic} \tilde \ve}$, where $\Gamma^{\ti\tic}$ generates
$\SO(1,1)_{R'}$, and $\beta$ is a free parameter.   The $\SO(1,1)_{R}$ twisted space $V_{\Lambda, \beta}$ is
the space of solutions to the modified conformal Killing equation
\begin{equation}
\label{eq:Killing-spinor11}
 D_{\mu} \ve = \frac 1 {2 r}  \Gamma_{\mu}  e^{ -\beta \Gamma^{\ti\tic}} \Lambda \ve.
\end{equation}
The choice of $\beta$ depends on the choice of the radius $r$ of the
Wilson loop $W$ under the condition that $\delta_{\ve} \in V_{\Lambda,
  \beta}$ annihilates $W$. 
In the  $(9,1)$ conventions the Wilson loop operator is 
\begin{equation}
\label{eq:eqLoop}
\begin{aligned}
 W_R(\rho) = \tr_{R} \Pexp \oint_{C} ((A_{\mu} \frac {dx^{\mu}}{ds} +
 \Phi_\ti) ds).
\end{aligned}
\end{equation}
Let circular contour $C$  be defined in the stereographic coordinates as  $(x^{1},x^{2}, x^{3}, x^{4}) = t (\cos \alpha,
\sin \alpha, 0, 0)$, $\alpha = 0\dots 2 \pi$. The $t$ relates to the polar angle $\theta$ as 
 $t = 2 r \tan \frac {\theta} 2$.  The
field $v^{M} A_{M} = v^{\mu} A_{\mu} + v^{A} \phi_{A}$ is annihilated by
$\delta_{\ve}$, since $(\ve \Gamma^{M} \ve) (\Psi \Gamma_M {\ve})$ vanishes
because of the triality identity~\eqref{eq:triality}.  Thus the Wilson
loop~\eqref{eq:eqLoop} is $\delta_{\ve}$-closed if $(v^{\mu}, v^{\tic},
v^{\ti}) = (\frac {dx^{\mu}}{ds}, 0, 1)$.  
We get
\begin{equation}
\hat \ve_{c} =  \frac 1 {t} \Gamma^{0} \Gamma_{12} \hat \ve_{s}.
\end{equation}
To satisfy~\eqref{eq:Killing-spinor1} we must have
\begin{equation}
\hat \ve_{c} = \frac 1 {2 r} e^{-\beta \Gamma^{\ti\tic}} \Lambda \hat \ve_{s}.
\end{equation}
Assume $\hat \ve_{s}, \hat \ve_{c}$ are of the positive four-dimensional chirality.  Then $
\beta = \log \frac {t} {2r}$, and $(\Lambda - \Gamma_{12})  \hat \ve_{s}
= 0$.  This equation has a non-zero solution for $\hat \ve_{s}$ only when
$\det (\Lambda - \Gamma_{12})=0$. That determines $\Lambda$ uniquely up to a
sign.  

The location of the Wilson loop on $S^4$
determines how the $\SU(2)_R$ symmetry group breaks to $\SO(2)$, and
the Wilson loop radius determines the $\SO(1,1)$ twist parameter $\beta$.  For the Wilson loop
located at the equator $t=2 r$. In the rest of the paper we focus on
this case. 

A conformal Killing spinor $\ve$ generating a transformation of $\OSp(2|4)$ 
 has constant norm over $S^4$ like a constant spinor
 on $\BR^4$ generating Poincare supersymmetries, and the fermionic
 transformations of $\OSp(2|4)$ anticommute to the isometry
 transformations on $S^4$.  Therefore we let the group $\OSp(2|4)$ to be
 the $\CalN=2$ supersymmetry group on $S^4$.

Now we consider massive deformation of the hypermultiplet Lagrangian
preserving $\OSp(2|4)$ symmetry. 

To generate the mass term in four
dimensions we use the Scherk-Schwarz reduction of ten-dimensional
$\CalN=1$ SYM with a Wilson line twist in the $\SU(2)^{R}_{R}$ symmetry group along the
coordinate $x_0$.  The $\CalN=2$ vector multiplet fields $A_{\mu}, \Phi_0,
\Phi_9, \Psi$ are not charged under $\SU(2)^R_{R}$. Therefore their kinetic terms
are not changed. The hypermultiplet fields $\chi$ and $\Phi_{i}$,  $i =
5,\dots,8$, transform in the spin-$\frac 1 2$ representation under $\SU(2)^R_R$.
We should replace $D_0 \Phi_{i}$ by $D_{0} \Phi_{i} + M_{ij} \Phi_j$, and $D_0 \chi$ by
$D_0 \chi + \frac 1 4 M_{ij} \Gamma_{ij} \chi$, where an antisymmetric $4\times
4 $ matrix $M_{ij}$ generates the $\SU(2)^{R}_{R}$
symmetry.  Since $F_{0i}$ is replaced by $[\Phi_0, \Phi_i] + M_{ij} \Phi_j$,
the $F_{0i} F^{0i}$ term in the action generates mass for the scalars of the hypermultiplet.

In the flat space, the massive action is still invariant under the usual
$\CalN=2$ supersymmetry.  However, on $S^4$ we need to be more careful with the
$\ve$-derivative terms in the supersymmetry transformations.  Let us explicitly
compute the variation of the Scherk-Schwarz deformed $\CalN=4$ theory on $S^4$.  We
use the conformal Killing spinor $\ve$ from the $\CalN=2$ superconformal
subsector satisfying  $\Gamma^{5678} \ve = \ve$.  Then $\ve$ is not charged under
$SU(2)_{R}^{R}$, so $D_0 \ve = 0$.  Variation of~\eqref{eq:SYM_S^4_N=4}
by~\eqref{eq:SYM_S^4_N=4-trans} up to the total derivative terms is
\begin{multline*}
  \delta_{\ve}( \frac 1 2 F_{MN} F^{MN} - \Psi \Gamma^{M} D_{M} \Psi + \frac 2 {r^2}
  \Phi_{A} \Phi^{A}) = \\ = 2 D_{M} (\ve \Gamma_{N} \Psi) F^{MN} + 2 \Psi
  \Gamma^{M} D_{M} ( \frac 1 2 F_{PQ} \Gamma^{PQ} \ve -2 \Phi_{A} \tilde
  \Gamma^{A} \tilde \ve) + \frac 4 {r^2} (\ve \Gamma^{A} \psi) \Phi_{A} = \\= -
  2 (\ve \Gamma_{N} \Psi) D_{M} F^{MN} + \Psi D_{M} F_{PQ} \Gamma^{M}
  \Gamma^{PQ} \ve + \Psi \Gamma^{M} \Gamma^{PQ} F_{PQ} D_{M} \ve - 4 \Psi
  \Gamma^{M} \tilde \Gamma^{A} \tilde \ve D_{M} \Phi_{A} + \\+ \frac 1 {r^2}
  \Psi \Gamma^{\mu} \tilde \Gamma^{A} \Phi_{A} \Gamma_{\mu} \ve + \frac 4 {r^2}
  (\ve \Gamma^{A} \Psi) \Phi_A = \dots
\end{multline*}
Using
\begin{equation}
  \label{eq:GammaMPQ}
  \Gamma^M \Gamma^{PQ} = \frac 1 3 (\Gamma^{M} \Gamma^{PQ} + \Gamma^{P} \Gamma^{QM} + \Gamma^{Q} \Gamma^{MP})
  + 2 g^{M[P} \Gamma^{Q]}
\end{equation}
and the Bianchi identity, we see that the first term cancels the second, and that
the last two terms cancel each other. Then
\begin{multline*}
  \dots = \Psi \Gamma^{\mu} \Gamma^{PQ} \Gamma_{\mu} \tilde \ve F_{PQ} - 4\Psi
  \Gamma^{M} \tilde \Gamma^{A} \tilde \ve D_{M} \Phi_{A} = 4 \Psi \tilde
  \Gamma^{MA} \tilde \ve F_{MA} - 4 \Psi \Gamma^{M} \tilde \Gamma^{A} \tilde \ve
  D_{M} \Phi_{A}
\end{multline*}
where we use the index conventions $M,N,P,Q = 0,\dots,9$, $\mu=1,\dots,4$, $A
=5,\dots,9,0$.  In the absence of Scherk-Schwarz deformation we have $F_{MA} =
D_{M} \Phi_{A}$ for all $M=0,\dots, 9$ and $A =5,\dots,9,0$, hence the two terms
cancel. After the deformation, we have $F_{0i} = D_{0} \Phi_i$, but $F_{i0} =
-D_{0} \Phi_{i} = -[\Phi_0, \Phi_i] - M_{ij} \Phi_{j} = D_i \Phi_0 - M_{ij}
\Phi_j$. Therefore, the naively Scherk-Schwarz deformed $\CalN=4$ theory on
$S^4$ is not invariant under $\delta_{\ve}$:
\begin{equation}
  \delta_{\ve}( \frac 1 2 F_{MN} F^{MN} - \Psi \Gamma^{M} D_{M} \Psi + \frac 2 {r^2} \Phi_{A} \Phi^{A})  = - 4 \Psi \Gamma^{i} \tilde \Gamma^{0} \tilde \ve M_{ij} \Phi_{j}.
\end{equation}
To make the Scherk-Schwarz action of massive theory invariant on $S^4$
under the $\OSp(2|4)$ group we need to add extra terms to the action.
Assume that  $\tilde \ve = \frac {1}
{2r} \Lambda \ve$, where $\Lambda$ is a generator of $\SU(2)_{L}^R$-group
normalized as $\Lambda^2 = -1$. Concretely we can take $\Lambda = \frac 1 4 \Gamma_{kl}
R_{kl}$ where $R_{kl}$ is an anti-self-dual matrix normalized as $R_{kl}R^{kl} =
4$, where $k,l = 5, \dots, 8$.  Then we get
\begin{equation}
  \begin{aligned}
    \delta_{\ve}( \frac 1 2 F_{MN} F^{MN} - \Psi \Gamma^{M} D_{M} \Psi + \frac 2 {r^2} \Phi_{A} \Phi^{A})  =  \frac 1 {2r} \Psi \Gamma^{0} \Gamma^{i} \Gamma^{kl} \ve R_{kl} M_{ij} \Phi_{j} = \\
    = \frac {1} {2r} (\Psi \Gamma^{i} \ve) R_{ki} M_{kj} \Phi_{j} = \frac {1}
    {2r} (\delta_{\ve} \Phi^{i}) (R_{ki} M_{kj}) \Phi_j
  \end{aligned}
\end{equation}
Hence, after we add the  $\frac {-1} {4r} (R_{ki} M_{kj}) \Phi^{i}
\Phi^{j}$  to the Scherk-Schwarz deformed action, we get the action invariant under the $\OSp(2|4)$.

Finally, the complete action, invariant under the $\OSp(2|4)$
transformations generated by $\ve$ such that  
$D_{\mu} \ve = \frac {1} {8r} \Gamma_{\mu} \Gamma^{0kl} R_{kl} \ve$ and 
$\Gamma^{5678} \ve = \ve$, is
\begin{align}
  S_{\CalN=2^{*}} = -\frac {1} { \gym^2} \tr \int d^4 x \sqrt{g} \lb \frac 1 2
  F_{MN} F^{MN} - \Psi \Gamma^{M} D_{M} \Psi + \frac 2 {r^2} \Phi_{A} \Phi^{A} -
  \frac {1} {4r} (R_{ki} M_{kj}) \Phi^{i} \Phi^{j} \rb,
\end{align}
where $D_{0}\Phi^{i} = [\Phi_0, \Phi^{i}] + M_{ij} \Phi^{j}$ and $D_{0} \Psi =
[\Phi_0, \Psi] + \frac 1 4 \Gamma^{ij} M_{ij} \Psi$.

The $\delta^2_{\ve}$ generates the covariant Lie derivative along the
vector field $-v^{M} = -\ve \Gamma^M \ve$, therefore it is contributed 
by  the gauge transformation along the $0$-th direction. Because
of the Scherk-Schwarz massive deformation, $\delta^2_{\ve}$ gets
new contributions 
\begin{equation}
  \begin{aligned}
    \delta^2_{\ve} \Phi_i= \delta^2_{\ve, \text{M=0}} \Phi_i   - v^{0} M_{ij} \Phi_j \\
    \delta^2_{\ve} \chi = \delta^2_{\ve, \text{M=0}} \chi -\frac 1 4 v^{0}
    M_{ij} \Gamma^{ij} \chi.
  \end{aligned}
\end{equation}

So far we have computed $\delta^2_{\ve}$  on-shell. To use the localization  we
need to close $\delta_{\ve}$ off-shell.
The pure $\CalN=2$ SYM involving transformation only for the vector
multiplet can be easily closed by adding three auxiliary scalar fields. 
However, the  $\CalN=2$ supersymmetry of the hypermultiplet can not be
closed using a finite number of auxiliary fields. 

For our purposes, we do not need to close off-shell 
completely the whole $\OSp(2|4)$ symmetry group. Since the
localization uses only one  generator $Q_{\ve}$, it is
enough to close off-shell only the symmetry generated by the $Q_{\ve}$.

To close off-shell the relevant $Q_{\ve}$ in the $\CalN=4$ theory 
we use the Berkovits construction~\cite{Berkovits:1993zz} for 10d
$\CalN=1$ SYM (see also
~\cite{Evans:1994cb,Baulieu:2007ew}).  The number of auxiliary fields compensates
the difference between the number of fermionic and bosonic off-shell degrees of
freedom modulo gauge transformations.  For $\CalN=4$ theory we add
 $16 - (10-1)
= 7 $ auxiliary fields $K_i$ with free quadratic action and modify the
superconformal transformations to
\begin{equation}
  \begin{aligned}
    \label{eq:off-shell-susy}
    &  \delta_{\ve} A_{M} = \Psi \Gamma_{M} \ve \\
    &  \delta_{\ve} \Psi = \frac 1 2 \gamma^{MN} F_{MN} + \frac 1 2 \gamma^{\mu A} \phi_{A} D_{\mu} \ve + K^i \nu_i \\
    & \delta_{\ve} K_i = - \nu_i \gamma^{M} D_{M} \Psi,
  \end{aligned}
\end{equation}
where spinors $\nu_i$ with $i=1, \dots, 7$ are chosen to satisfy
\begin{align}
  \label{eq:nu1-relations1}
  &\ve \Gamma^M \nu_i = 0 \\
  \label{eq:nu1-relations2}
  &\frac 1 2 (\ve \Gamma_N \ve) \tilde \Gamma^{N}_{\alpha \beta} =
  \nu^i_{\alpha} \nu^i_{\beta} + \ve_{\alpha} \ve_{\beta} \\
  \label{eq:nu1-relations3}
  &\nu_i \Gamma^M \nu_j = \delta_{ij} \ve \Gamma^M \ve.
\end{align}
For any non-zero Majorana-Weyl spinor $\ve$ of $\Spin(9,1)$ there exist seven
linearly independent spinors $\nu_{i}$, which satisfy these
equations\footnote{The author thanks N.Berkovits for
  communications.}~\cite{Berkovits:1993zz}.  They are determined up to an
$\SO(7)$ transformations.  The equation~\eqref{eq:nu1-relations1} ensures
closure on $A_M$, the equation~\eqref{eq:nu1-relations2} ensures closure on
$\Psi$, and the equations~\eqref{eq:nu1-relations1}
and~\eqref{eq:nu1-relations3} ensure closure on $K$
\begin{equation}
  \label{eq:delta2K}
  \delta_{\ve}^2  K_i =
  - (\ve \gamma^{M} \ve) D_{M} K^{i} - (\nu_{[i} \gamma^{M} D_{M} \nu_{j]}) K^{j}
  - 4(\tilde \ve \ve) K_{i}.
\end{equation}

The auxiliary fields $K_i$ are sections of a  $\SO(7) \otimes
\mathrm{ad} (G) $ vector bundle $\CalE_K$ over $S^4$. 
The equation \eqref{eq:delta2K} can
be interpreted as the covariant Lie derivative along the vector field
$v^{\mu}$: a lift of the $L_{v}$  on $S^4$ to the vector bundle
$\CalE_{K} \to S^4$. 

We present the  $\OSp(2|4)$ spinor $\ve$ in the form  
 (see appendix~\ref{se:Killing-spinors} for details)
\begin{equation}
\label{eq:spin5}
  \ve (x) = \exp \lb - \frac \theta 2 n_{i }(x)  \Gamma^{i} \Gamma^{9} \rb \hat \ve_{s},
\end{equation}
where $x^i$ are the stereographic  coordinates on $S^4$, 
$n_{i}$ is the unit vector in the direction of the vector field $v_{i} =
\frac 1 r x^{i} \omega_{i j} $.  The $\ve(x)$ satisfies  
$(\ve(x), \ve(x)) = 1$ and $\Gamma^{9} \hat \ve_{s}
= \GREEN{-} \hat \ve_{s}$.
 The self-dual matrix $\omega_{ij}$ generates an  $\SU(2)_R \subset
 \SO(4)$ rotation around the North pole. 
The equation (\ref{eq:spin5}) says that the spinor $\ve(x)$ relates to $\ve(0)$ by  $\Spin(5)$ transformation.

To close off-shell $\CalN=4$ supersymmetry we need seven spinors $\nu_i$ which
satisfy~\eqref{eq:nu1-relations1}-\eqref{eq:nu1-relations3}. Following~\cite{Berkovits:1993zz},
at the origin we can take $\hat \nu_i = \Gamma^{i8} \hat \ve_{s}$ for $i =
1\dots 7$. Then we transform $\hat \nu_i$ to any point on $S^4$ by the rule
\begin{equation}
  \label{eq:nu-defined}
  \nu_i(x) = \exp (\frac \theta 2 n_{i}(x)  \Gamma^{i8}) \hat \ve_{s}.
\end{equation}

Finally, we conclude that the action
\begin{align}
  \label{eq:actionNstar-offshell}
  S_{\CalN=2^{*}} =- \frac {1} { \gym^2} \tr \int d^4 x \sqrt{g} \lb \frac 1 2
  F_{MN} F^{MN} - \Psi \Gamma^{M} D_{M} \Psi + \frac 2 {r^2} \Phi_{A} \Phi^{A} -
  \frac {1} {4r} (R_{ki} M_{kj}) \Phi^{i} \Phi^{j} - K_i K_i\rb,
\end{align}
is invariant under the off-shell supersymmetry $Q_{\ve}$ given
by~\eqref{eq:off-shell-susy} with $\nu_i$ defined
by~\eqref{eq:nu-defined}. 

\section{Localization}

To localize the path integral of a theory invariant under a fermionic symmetry 
 we deform the action by 
a $Q$-exact term \cite{Witten:1988ze}
\begin{align}
  \label{eq:deformation}
  S \to S + t Q V.
\end{align}
Since  $Q$  squares to a symmetry of the theory, and since the action
and the Wilson loop observable are $Q$-closed, we can use the localization
principle.  Given a $Q^2$-invariant functional $V$, the deformation~(\ref{eq:deformation}) does not change the
expectation value of $Q$-closed observables. When we send $t$ to
infinity, the theory localizes to a space $F$ of critical points of $QV$. 
In the end we need to integrate over $F$. The measure in the integral over $F$ comes
from the restriction of the action $S$ to $F$ and the determinant of the kinetic 
term of $QV$ which counts
fluctuations in the normal directions to $F$.

To ensure convergence of the four-dimensional path integral,
we compute it for the theory reduced from the Euclidean $(10,0)$
signature. 

To technically 
simplify the presentation 
of the supersymmetry in the previous section, 
we have used the $\CalN=1$ theory dimensionally reduced from the $(9,1)$ signature.
The path integral of the  reduced $(9,1)$ theory
with $\Phi_0 = i\Phi_0^E$ and integration over real $\Phi_0^E$
coincides with the
 path integral of the reduced $(10,0)$ theory. 

For convergence, in the reduced $(9,1)$ theory, we  integrate over the
imaginary
auxiliary fields $K_i$. We use conventions $K_i = iK_i^E$, 
where $K_i^E$ is real. 

For localization we use the following functional
\begin{align}
  \label{eq:V-definition}
  V = (\Psi, \overline{Q\Psi}),
\end{align}
where $\overline{Q \Psi}$ is complex conjugation of the $Q \Psi$ in the
$(10,0)$ Euclidean conventions.\footnote{The bar symbol in $\overline {Q\Psi}$
  literally means complex conjugation only in the Euclidean
  conventions at the real integration contour for $\Phi^E_0,
  K^E$. Generally, however, one  should assume the second line of
  (\ref{eq:QPsi}) as the definition of $\overline {Q \Psi}$ without
  referring to the operation of complex conjugation.}
Then the bosonic part of the $QV$-term is positively definite
\begin{align}
  S^{Q}|_{bos}  = (Q\Psi, \overline{Q\Psi}).
\end{align}

Explicitly, in the conventions of the reduced $(9,1)$ theory 
 we have
\begin{equation}
  \begin{aligned}
    \label{eq:QPsi}
    Q \Psi = \frac 1 2 F_{MN} \Gamma^{MN} \ve + \frac 1 2 \Phi_{A} \Gamma^{\mu A}  \nabla_{\mu} \ve + K^{i} \nu_{i} \\
    \overline{ Q \Psi} = \frac 1 2 F_{MN} \tilde \Gamma^{MN} \ve + \frac 1 2
    \Phi^{A} \tilde \Gamma^{\mu A} \nabla_{\mu} \ve - K^{i} \nu_i,
  \end{aligned}
\end{equation}
where $\tilde \Gamma^{0} = -\Gamma^{0}, \tilde \Gamma^{M} = \Gamma^{M}$ for $M=1,\dots,9$, and $\Gamma^{MN} = \tilde \Gamma^{[M} \Gamma^{N]}, \tilde
\Gamma^{MN} = \Gamma^{[M} \tilde \Gamma^{N]}$.

To proceed with the technical details,  we should explicitly 
describe the conformal Killing spinor $\ve$
and the vector field $v^{M} = \ve \Gamma^{M} \ve$ generated 
by the $\delta_{\ve}^2$. We take~$\ve$ in the form
(\ref{eq:Killing-solution-in-S^4-stereo}), where $\hat \ve_{s}$ is 
any spinor such that
\begin{enumerate}
\item{$\Gamma^{5678} \ve_s = \ve_s$},
\item{$\Gamma^{1234} \ve_s =  -\ve_s$},
\item{$\hat \ve_{s} \hat \ve_{s} = 1$}.
\end{enumerate}
The condition (1) restricts $\ve$ to the $\CalN=2$ 
supersymmetry subgroup of $\CalN=4$. The condition (2)  
 ensures that $\ve$ is right chiral at  the North pole of $S^4$. 
The condition (3) is a convenient normalization. 
In our conventions for the gamma-matrices (appendix \ref{sec:Octonionic-gamma-matrices})
we take  $\hat \ve_{s} = (0_{8},1,0_7)^{t}$.

We consider the Wilson circular loop in the $(x_1, x_2)$ plane. 
In the $(9,1)$ gamma-matrices conventions we set  $\hat \ve_{c} = \frac {1} {2r} \Gamma^{012} \hat \ve_{s}$. 
The conformal Killing spinor $\ve$ defined by such $\hat \ve_{s}$ and $\hat \ve_{c}$ has 
constant unit norm on the $S^4$. 
Now we compute  the vector field $v^{M} = \ve \Gamma^{M} \ve$. 
In the reduced $(9,1)$ conventions we get 
\begin{equation}
\begin{aligned}
& v_{t} = \sin \theta \\
& v^0 = 1 \\  
& v^{9} = -\cos \theta \\
& v^{i} = 0, \quad i=5,\dots,8,
\end{aligned}
\end{equation}
where $\theta$ is the polar angle on $S^4$ measured from the North pole, and 
$v_t = \sqrt{v_{\mu} v^{\mu}}$. The space-time part of $v$ 
is the vector field of the $\U(1) \subset \SU(2)_R \subset \SO(4)$
transformation rotating equally 
$(x_1, x_2)$ and $(x_3,x_4)$  planes. In the reduced  $(10,0)$ theory we
get $v^{0}=i$.

To simplify $S^{Q}|_{bos}$ we use the Bianchi identity for $F_{MN}$, the gamma-matrices
algebra and integration by parts. 
The principal contribution to $S^{Q}|_{bos}$
is the curvature term
\begin{equation}
  S_{FF} = \frac 1 4 (\ve \tilde \Gamma^{N} \Gamma^{M} \tilde \Gamma^{P} \Gamma^{Q} \ve) F^{MN} F^{PQ}.
\end{equation}
The cross-terms $F_{MN} K_i$  vanish because $\nu_i \Gamma^{0} \Gamma^{M} \ve
= \nu_i \Gamma^{M} \ve = 0$. The auxiliary
$KK$-term contributes 
\begin{equation}
  S_{KK} = -K_i K^i.
\end{equation}
In the flat space limit the spinor $\ve$ is covariantly
constant: $\nabla_{\mu} \ve =0$. Therefore, in the flat space we get $S^{Q}|_{bos} = S_{FF} + S_{KK}$.
Up to the total derivatives and $\nabla_{\mu} \ve$-terms,
using the Bianchi identity and the gamma-matrices algebra, we see
that $S_{FF}$ is equivalent to the usual Yang-Mills action $\frac 1 2 F^{MN} F_{MN}$.  
When the space is curved (and therefore $\nabla_{\mu} \ve \neq 0$) we need to be
more careful.  Using~\eqref{eq:GammaMPQ} we get
\begin{equation}
  S_{FF} = \frac 1 2 F^{MN} F_{MN} + \frac 1 4 \ve \tilde \Gamma^{N} \Gamma^{M} \tilde \Gamma^{P} \Gamma^{Q} \ve
  \frac 1 3 \lb  F_{MN} F_{PQ} + F_{PN} F_{QM} + F_{QN} F_{MP} \rb.
\end{equation}

To simplify the last term, we split the indices into two groups:
$M,N,P,Q=(1,\dots,4,9,0)$ and $M,N,P,Q = (5, \dots, 8)$ describing respectively
the fields of the vectormultiplet and hypermultiplet.  The restriction
$\Gamma^{5678}\ve=\ve$ implies
that the nonvanishing terms have only $0$, $2$, or $4$  
indices in the range $(5,\dots, 8)$. 
Let us call the respective terms as vector-vector,
vector-hyper and hyper-hyper. 

Considering vector-vector
terms, we split indices to the gauge field part
$(1,\dots,4)$ and to the scalar part $(0,9)$. 
In the nonvanishing terms the indices  $M,N,P,Q$ are all distinct.
The term $F_{\mu \nu} F_{\rho
  \lambda}$  is simplified  to
\begin{multline}
  \frac 1 4 \cdot \frac 1 3 \cdot 24 \cdot \ve \Gamma^{1234} \ve (F^{21} F^{34}
  + F^{31}F^{42}+F^{41}F^{23}) = - \frac 1 2 \ve \Gamma^{1234} \ve (F,*F) =
  +\frac 1 2 \cos \theta (F, *F),
\end{multline}
where $*F$ is the Hodge dual of $F$.
All terms in which one of
the indices is $0$ vanish because $\Gamma^{MPQ}$ is antisymmetric matrix, therefore
$\ve \Gamma^{0} \Gamma^{MPQ} \ve = 0$.  The remaining vector-vector terms
have $D_{\mu}\Phi_9 F_{\nu \rho}$ structure. Integrating by parts
and using Bianchi identity we get
\begin{equation}
  - \frac 1 3 D_{\mu} (\ve \Gamma^9 \Gamma^{\mu \nu \rho} \ve) \Phi_9 F_{\nu \rho} + \text{cyclic}(\mu \nu \rho)
  = 4 (\tilde \ve \Gamma^9 \Gamma^{ \mu \nu} \ve) \Phi_9 F_{\mu \nu}.
\end{equation}
After similar manipulations we find the contribution to
the vector-hyper mixing terms in $S_{FF}$
\begin{equation}
  -8 \tilde \ve \Gamma^9 \Gamma^{ij} \ve \Phi_i [\Phi_9, \Phi_j] - 6 \tilde \ve \Gamma^{\mu} \Gamma^{ij}
  \ve \Phi_i D_{\mu} \Phi_j.
\end{equation}
Summing up all contributions to $S_{FF}$ we obtain
\begin{equation}
  S_{FF} = \frac 1 2 F_{MN} F^{MN}  + \frac 1 2 \cos \theta F_{\mu \nu} (*F)^{\mu \nu} +
  4 (\tilde \ve \Gamma^9 \Gamma^{ \mu \nu} \ve) \Phi_9 F_{\mu \nu} -8 \tilde \ve \Gamma^9 \Gamma^{ij} \ve \Phi_i [\Phi_9, \Phi_j] - 6 \tilde \ve \Gamma^{\mu} \Gamma^{ij}
  \Phi_i D_{\mu} \Phi_j.
\end{equation}

Next we consider the cross-terms between $\Phi^A$ and $F_{MN}$ in $S^{Q}|_{bos}$
\[
S_{F\Phi} = -\tilde \ve \Gamma^{A} \tilde \Gamma^{M} \Gamma^{N} \ve \Phi_A
F_{MN} - \tilde \ve \tilde \Gamma^{A} \Gamma^{M} \tilde \Gamma^{N} \ve \Phi_A
F_{MN}. \] 
We consider separately the cases when the index $A$ is in the set
$\{0,9\}$ or in the set $\{5,\dots,8\}$.  
The
terms with  $A=0$ all vanish because $\tilde \Gamma^{0} = - \Gamma^0$ and because
$\tilde \ve \Gamma^{M} \ve =0$ for our choice of $\ve$ in
$\OSp(2|4)$. The only nonvanishing terms with $A = 9$ are
\[
- 2 \tilde \ve \Gamma^{9} \Gamma^{\mu \nu} \ve \Phi_9 F_{\mu \nu} - 2 \tilde \ve
\Gamma^{9} \Gamma^{ij} \ve \Phi_9 [\Phi_i, \Phi_j],
\]
where $\mu, \nu = 1, \dots, 4$ and $i,j = 5, \dots, 8$.  Finally, we consider
 $A=5, \dots 8$. The result is
\[ 4 \tilde \ve \Gamma^{\mu } \Gamma^{ij} \ve \Phi_i D_{\mu} \Phi_j + 4 \tilde \ve
\Gamma^{9 } \Gamma^{ij} \ve \Phi_i [\Phi_9, \Phi_j]. \] Then
\[
S_{F\Phi} = - 2 \tilde \ve \Gamma^{9} \Gamma^{\mu \nu} \ve \Phi_9 F_{\mu \nu} +
4 \tilde \ve \Gamma^{\mu } \Gamma^{ij} \ve \Phi_i D_{\mu} \Phi_j + 6 \tilde \ve
\Gamma^{9 } \Gamma^{ij} \ve \Phi_i [\Phi_9, \Phi_j].
\]
The $\Phi\Phi$ term is easy
\[
S_{\Phi\Phi} = 4 \Phi^{A} \Phi^{B} \tilde \ve \Gamma^{A} \tilde \Gamma^{B}
\tilde \ve = 4 \tilde \ve \tilde \ve \Phi^A \Phi_A.
\]
Finally, we need the $\Phi K$ cross-term. Only $\Phi_0$ contributes:
\[
S_{\Phi K} = 2 K_i \Phi_0 \nu_i \tilde \Gamma^{0} \tilde \ve - 2 K_i \Phi_0
\nu_i \Gamma^{0} \tilde \ve = - 4 K_i \Phi_0 \nu_i \tilde \ve.
\]
The complete result is
\begin{multline}
  S^{Q}|_{bos} = S_{FF} + S_{F\Phi} + S_{\Phi \Phi} + S_{\Phi K} + S_{KK} = \\
  \frac 1 2 F_{MN} F^{MN} + \frac 1 2 \cos \theta F_{\mu \nu} (*F)^{\mu \nu} + 2
  \tilde \ve \Gamma^9 \Gamma^{ \mu \nu} \ve \Phi_9 F_{\mu \nu} -2 \tilde \ve
  \Gamma^9 \Gamma^{ij} \ve \Phi_i [\Phi_9, \Phi_j] - 2 \tilde \ve \Gamma^{\mu}
  \Gamma^{ij} \ve  \Phi_i D_{\mu} \Phi_j + \\
  + 4 (\tilde \ve \tilde \ve) \Phi_A \Phi^A - 4 K_i \Phi_0 \nu_i \tilde \ve -
  K_i K^i
\end{multline}
The next step is to find the critical points of
the $S^{Q}|_{bos}$.  Our strategy will be to represent $S^{Q}|_{bos}$ as a sum
of semipositive terms (full squares) and find the field configurations which
ensure vanishing all of them.

First we combine the four-dimensional curvature terms together with the
$\Phi_9$-mixing terms
\begin{multline}
  \label{eq:Q-exact-curvature}
  \frac 1 2 F^{\mu \nu} F_{\mu \nu} +\frac 1 2 \cos \theta F^{\mu \nu} (*F)_{\mu
    \nu} + 2 \tilde \ve \Gamma^9 \Gamma^{ \mu \nu} \ve \Phi_9 F_{\mu \nu} + 4
  (\tilde \ve \tilde \ve) \Phi_9 \Phi^9 = \\ = \sin^2 \frac \theta 2 ( F^-_{\mu
    \nu} + w_{\mu \nu}^- \Phi_9)^2 + {\cos^2} \frac \theta 2 (F^{+}_{\mu \nu} +
  w_{\mu \nu}^{+} \Phi_9)^2.
\end{multline}
where
\begin{equation}
  \begin{aligned}
    w_{\mu \nu}^-  = {\frac 1 {\sin^{2} \frac \theta 2 }} \tilde \ve^{L} \Gamma^{9} \Gamma_{\mu \nu} \ve^{L}  \\
    w_{\mu \nu}^+ = {\frac 1 {\cos^{2} \frac \theta 2 }} \tilde \ve^{R} \Gamma^{9}
    \Gamma_{\mu \nu} \ve^{R}.
  \end{aligned}
\end{equation}

Next we make a full square with the terms
\[D_m \Phi_{i} D^m \Phi^i -2 \tilde \ve \Gamma^9 \Gamma^{ij} \ve \Phi_i [\Phi_9,
\Phi_j] - 2 \tilde \ve \Gamma^{\mu} \Gamma^{ij} \ve \Phi_i D_{\mu} \Phi_j = (D_m
\Phi_j - \tilde \ve \Gamma_m \Gamma_{ij} \ve \Phi^i)^2 - \Phi^i \Phi_i (\tilde
\ve \tilde \ve) (\ve \ve). \] Finally we absorb the mixing term $K_i \Phi_0$ as
follows
\[ - 4 (\tilde \ve \tilde \ve) \Phi_0 \Phi_0 - 4 \Phi_0 K_{i} (\nu^{i} \tilde
\ve) - K_i K^i = -(K_i + 2 \Phi_0 (\nu_i \tilde \ve))^2. \] 
We used the following
relations in the computation
\[ (\ve \ve) = 1, \quad (\ve^L \ve^L) = {\sin^2 \frac \theta 2}, \quad (\ve^R
\ve^R) ={\cos^2 \frac \theta 2}, \quad (\tilde \ve \tilde \ve) = \frac 1 {4 r^2},
\quad w^{-}_{\mu \nu} w^{-\mu \nu} = w^{+}_{\mu \nu} w^{+\mu \nu} = \frac 1
{r^2}.
\]

The final result is
\[
S^{Q}|_{bos} = S^{Q}_{vect,bos} + S^{Q}_{hyper,bos}. \] Here
\begin{equation}
  \label{eq:N-2local}
  S^{Q}_{vect,bos} = {\sin^2} \frac \theta 2 ( F^-_{\mu \nu} + w_{\mu \nu}^- \Phi_9)^2 +
  {\cos^2 \frac \theta 2} (F^{+}_{\mu \nu} + w_{\mu \nu}^{+} \Phi_9)^2 + (D_{\mu} \Phi_a)^2  + \frac 1 2  [\Phi_a, \Phi_b] [\Phi^a, \Phi^b] + (K_i^E + w_i \Phi_0^E )^2
\end{equation}
where  $w_i = 2(\nu_i \tilde \ve)$,  the indices $a,b=0,9$ label the scalars of the vector multiplet, the
index $i=5,6,7$ labels the auxiliary fields for the vector multiplet.

One could worry that our supersymmetry transformations break the reality 
conditions on the fields in the conventions of the reduced $(10,0)$
signature.
 However, the localization argument still
holds. 
The action is  invariant under supersymmetry transformations
if we treat the action as an analytical functional in the space
of complexified fields. The path integral is understood as integration of a 
holomorphic functional of fields over a certain real half-dimensional 
subspace in the complexified space of fields. Strictly speaking, the bar 
in the formula~(\ref{eq:QPsi}) for $\overline{Q\Psi}$ literally means complex conjugation 
only if we assume that we use the specific contour of integration 
in the $(9,1)$ theory: all fields are real except,
$\Phi_0$ and $K_i$ which are imaginary. 
For a general contour of integration 
we use the functional $V$~(\ref{eq:V-definition})
where $\overline{Q\Psi}$ is~\emph{defined} by the second line of~(\ref{eq:QPsi}).
This means that the functional $V$ holomorphically 
depends on all complexified fields. The bosonic part of $QV$ is positive 
definite after restriction to the suitable contour of integration.

From any point of view, we should stress that $\delta_\ve$ squares 
to a complexified gauge transformation, whose scalar generator
is $i\Phi_0^E {- \cos \theta \Phi_9}+\sin \theta A_{v}$, where $\Phi_0^E$, $\Phi_9^E$ and $A_{v}$ take
value in the real Lie algebra of the gauge group, and where $A_{v}$ is the component of
the gauge field in the direction of the vector field $v^{\mu}$.
The theory is similar to the Donaldson-Witten \cite{Witten:1988ze}
theory near the North pole where this generator becomes ${i (\Phi_0^E + i\Phi_9)}$,
and the conjugate Donaldson-Witten theory near the South pole where this
generator becomes ${i (\Phi_0^E - i\Phi_9)}$. 

The hypermultiplet localizing action is
\[
S^{Q}_{hyper,bos} = (D_0 \Phi_i)^2 + (D_m \Phi_j - f_{m i j}\Phi^i)^2 + \frac 1
2 [\Phi_i, \Phi_j][\Phi^i,\Phi^j] + \frac 3 {4 r^2} \Phi^i \Phi_i + K^E_I K^E_I,
\]
where $m=1,\dots,5$, $i=5, \dots, 8$, $I=1, \dots 4$ and $f_{m i j} = \tilde \ve
\Gamma_m \Gamma_{ij} \ve$.  With our choice of the integration 
contour, all terms in the action $S^{Q}|_{bos}$ are
semi-positive definite.  Therefore, in the limit $t \to \infty$ we need to care
 only about the locus, where all terms vanish, and small
fluctuations in the normal directions.

For the hypermultiplet localizing action we get a simple vanishing theorem: because
of the quadratic terms $\frac 3 {4 r^2} \Phi^i \Phi_i$ and $ K_i^E K_i^E$ the functional $S^{Q}_{hyper, bos}$ vanishes
only if all fields $\Phi_i$ and $K_I$ vanish.

Next consider zeroes of $S^{Q}_{vect,bos}$. The term $(D_{\mu} \Phi_9)^2$ ensures
that the field $\Phi_9$ is covariantly constant.  Away from the North and
the South poles, requiring that the curvature terms vanish, we get the equations
\[
F_{\mu \nu} = -w_{\mu \nu} \Phi_9 \] where $w_{\mu \nu} = w^{-}_{\mu \nu} +
w^{+}_{\mu \nu}$. The curvature $F_{\mu \nu}$ satisfies Bianchi identity, hence
we get
\begin{equation}
  \label{eq:vanishing-th}
  d_{[\lambda} w_{\mu \nu]}  \Phi_9 = 0.
\end{equation}
Away from the North and the South poles, $d_{[\lambda}
w_{\mu \nu]}$ does not vanish, hence $\Phi_9$ and $F_{\mu\nu}$ must vanish. The
kinetic term $(D_\mu \Phi^E_0)^2$ ensures that $\Phi_0^E$ is covariantly
constant. Since $F_{\mu \nu} =0$,  the gauge field is trivial, thus
$\Phi_0^E$ is constant over $S^4$. We denote this zero mode of
$\Phi_0^E$ as $a_E$ and
conclude that up to a gauge transformation the only smooth solution is
\begin{equation}
  \label{eq:good-configurations}
  S^{Q}_{bos} = 0 \Rightarrow
  \begin{cases}
    A_{\mu}  = 0 \quad  \mu = 1, \dots 4 \\
    \Phi_{i} = 0 \quad i = 5, \dots, 9 \\
    \Phi_0^E = a_E \quad \text{constant over $S^4$} \\
    K_i^E = -w_i a_E \\
    K_I = 0
  \end{cases}.
\end{equation}

We have discussed the key step in the proof of the
Erickson-Semenoff-Zarembo and  Drukker-Gross conjecture using localization. The
infinite-dimensional path integral localizes to the finite dimensional
locus~\eqref{eq:good-configurations} and reduces to the integration over $a_E \in \g$ in the
resulting matrix model.

Now we evaluate the $S_{YM}$ action~\eqref{eq:actionNstar-offshell}
at~\eqref{eq:good-configurations}:
\begin{equation} S_{YM}[a] = \frac 1 {  \gym^2} \int d^4 x \sqrt {g} \lb
  \frac 2 {r^2} (\Phi^E_0)^2 + (K_i^E)^2 \rb = \frac 1 {  \gym^2} \vol(S^4)
  \frac 3 {r^2} a_E^2 = \frac {8 \pi^2 r^2 }{\gym^2} a_E^2 \end{equation}
where we used $w_i w^i = \frac 1 {r^2}$ and  $\vol(S^4) = \frac 8 3
\pi^2 r^4$. 
We have obtained precisely the Erickson-Semenoff-Zarembo/Drukker-Gross matrix model. 

Let us check that the coefficient at the quadratic action matches the
perturbation theory. The 4d YM action has the following propagators in Feynman gauge on $\BR^4$
\begin{align*}
  \langle A_{\mu} (x) A_{\nu}(x')  \rangle = \frac {\gym^2 } {8 \pi^2}  \frac {g_{\mu \nu}}{ (x-x')^2}  \\
  \langle \Phi^E_0 (x) \Phi^E_0 (x') \rangle = \frac {\gym^2} {8 \pi^2} \frac
  {1}{ (x-x')^2}.
\end{align*}
Hence, the correlation functions which appear in the perturbative expansion of
the Wilson loop operator, have the structure
\[
\langle A_{\mu}(\alpha) \dot x^{\mu} \, A_{\nu}(\alpha') \dot x^{\nu} + 
i \Phi^E_0(\alpha) i \Phi^E_0(\alpha') \rangle = -\frac {\gym^2} {8 \pi^2 r^2 }
\frac {\cos(\alpha - \alpha')-1} {4 \sin^2 \frac {\alpha - \alpha'}{2}} = -\frac
{\gym^2} {16 \pi^2 r^2},
\]
where $\alpha$ denotes the angular coordinate on the loop. This was the original
motivation for Erickson-Semenoff-Zarembo conjecture~\cite{Erickson:2000af}. 
We see that the
first order perturbation theory agrees with the matrix model action derived by
localization. The power of the localization computation is that it works
to all orders in perturbation theory. It is also possible to take into account the instanton
contributions, which we describe in section \ref{sec:Instanton corrections} after computing the fluctuation determinant
near the locus~\eqref{eq:good-configurations} and confirming the exact solution.

The same solution for the zeroes of $S^{Q}|_{bos}$ applies to the
$\CalN=2^*$ theories. To ensure that all terms are positive
definite, we take the mass parameter $M_{ij}$ in the Scherk-Schwarz reduction to be
pure imaginary antisymmetric self-dual matrix. Then the action of the mass
deformed $\CalN=2^{*}$ theory at configurations~\eqref{eq:good-configurations}
reduces to the same matrix model action. However, as we will see shortly, when
the mass parameter $M_{ij}$ is non zero, the matrix model measure for the $\CalN=2^*$ theory is
corrected by a non-trivial determinant.

\section{Determinant factor}

\subsection{Gauge-fixing complex}

Because of the infinite-dimensional gauge symmetry of the action we need to work
with the gauge-fixed theory. We use the Faddeev-Popov ghost fields and introduce
the BRST like complex with the differential $\delta$:
\begin{equation}
  \begin{aligned}
    \delta X  &= -[c,X]   & \delta c  &= -a_0 -\frac 1 2 [c,c] &      \delta \tilde c & = b&      
\delta \tilde a_0 & = \tilde c_0&    \delta b_0 & = c_0 \\
    & &\delta a_0 &= 0 & \delta b & = [a_0, \tilde c]& \delta \tilde c_0 & =
    [a_0,\bar a_0]& \delta c_0& = [a_0,b_0].
  \end{aligned}
\end{equation}
Here $X$ stands for all physical and auxiliary fields
entering~\eqref{eq:actionNstar-offshell}.  All other fields are the gauge-fixing
fields.  By $[c,X]$ we denote gauge transformation of $X$ parametrized by $c$. For gauge fields $A_{\mu}$ we take $\delta A_{\mu} =
-[c,\nabla_{\mu}]$.  The gauge transformation of $\Phi = v^{M} A_{M}$ is $\delta
\Phi = [v^{\mu} D_{\mu} + v^{A} \Phi_{A},c]=[\Phi, c]+L_{v}c$. 
The fields $c$ and $\tilde c$ are the usual
Faddeev-Popov ghost and anti-ghost.  The bosonic field $b$ is the standard
Lagrange multiplier used in $R_{\xi}$-gauge, where the gauge fixing is
performed by
adding terms like $(b,i d^* A + \frac \xi 2 b)$ and $(\tilde c, d^* \nabla_{A} c)$
to the action.  The fields $c$ and $\tilde c$  have zero modes. To treat
these zero modes  systematically we add constant fields $c_0,\tilde c_0,a_0,\tilde a_0,b_0$ to
the gauge-fixing complex.  The field $a_0$ is ghost  for
 ghost $c$.  The fields $a_0, \tilde a_0, b_0$ are bosonic, and the fields
$c_0, \tilde c_0$ are fermionic.  The  $\delta$ squares to the gauge
transformation by the constant bosonic field $a_0$
\[ \delta^2 \cdot = [a_0, \cdot].\] The gauge invariant action and gauge
invariant observables 
are $\delta$-closed
\[ \delta S_{YM}[X] = 0,\] therefore their correlation functions are
not changed when we add the $\delta$-exact gauge-fixing term.  

Combining the gauge-fixing terms with the 
physical action, we  see that the convergence of the path integral
requires the imaginary
integration contour for the constant field $a_0$.  This field $a_0$  will be identified with the zero 
mode of the physical field $\Phi_0$ integrated over imaginary contour. 
For consistent notations we set $a_0 = i a_0^E$ with $a_0^E$ being
real.

The $\delta$-exact term
\begin{multline}
  \label{eq:gauge-fix-term}
  S^{\delta}_{g.f.} = \delta( (\tilde c, i d^*A + \frac {\xi_1} 2 b + i b_0) - (c,
  \tilde a_0 - \frac {\xi_2} 2 a_0) ) = \\= (b, i d^*A + \frac {\xi_1} 2 b + i b_0) - (\tilde c, id^* \nabla_A c + ic_0 + \frac {\xi_1} 2 [a_0,\tilde c]) + (-ia_0^E + \frac 1 2 [c,c], \tilde a_0- \frac {\xi_2} 2 ia_0^E)  + (c, i\tilde c_0)
\end{multline}
properly fixes the gauge.

Assuming that all bosonic fields
are real, the bosonic part of the gauge-fixed action has strictly positive definite 
quadratic term for all fields and ghosts at $\xi_1 > 0 $ and $ \xi_2 >0$.

The partition function does not depend on the parameters $\xi_1$ and $\xi_2$. Let us fix $\xi_1=0$ and demonstrate explicitly independence on $\xi_2$ and equivalence
with the standard gauge-fixing procedure. First, integrating out 
$a_0^E$ we  get 
\[ 
 (ia_0^E + \frac 1 2 [c,c],i \tilde a_0- \frac {\xi_2} 2 ia_0^E)  \to +\frac {1} {2 \xi_2} (\tilde a_0 - \frac {\xi_2} {4} [c,c])^2.
\] 
The Gaussian integration over $\tilde a_0$ removes this term. 
The determinant coming from the Gaussian integration over $\tilde a_0$ is inverse to the determinant
coming from the Gaussian integral over $a_0$. 
Then we integrate out the zero mode of $b$ and $b_0$. Integration over non-zero modes of $b$ 
gives  the Dirac delta-functional inserted at the gauge-fixing hypersurface $d^* A=0$.
The remaining terms are 
\[
(\tilde c, id^* \nabla_A c) + i(\tilde c, c_0) + i (c,\tilde c_0).
\]
We can integrate out $c_0$ with the zero mode of $\tilde c$, and $\tilde c_0$ with the zero mode of $c$.
We are left with the integral over $c$ and $\tilde c$ with the zero modes projected out and the 
gauge-fixing term 
\[
 (\tilde c, id^* \nabla_A c).
\]
This reproduces the usual Faddeev-Popov determinant $\det'(d^{*} \nabla_A)$ which we need to insert 
into the path integral for the partition function after restricting to the gauge-fixing hypersurface $d^* A = 0$. 
The symbol $'$ means that the determinant is computed on the space without the zero modes. 

We summarize the gauge fixing procedure by the formula
\begin{multline}
  \label{eq:Z_{phys}}
  Z = \frac {1} {\vol(\CalG,\gym)} \int [DX] e^{-S_{YM}[X]} = \frac {1} {\vol(\CalG)} \int
  [DX] e^{-S_{YM}[X]}
  \int_{g \in \CalG'} [Dg] \, \delta_{Dirac}( d^*A^{g}) \det{}^{'}( d^* \nabla_{A} ) =\\
  =\frac {\vol(\CalG',\gym)} {\vol(\CalG,\gym)} \int [DX Db' Dc' D\tilde c'] e^{-S_{YM}[X] - \int_{S^4} \sqrt{g} \, d^4 x (i(b,d^*A) - (\tilde c, i d^* \nabla_A c))} =\\
  = \frac {1} {\vol{(G,\gym)}} \int [DX \, Db \, Db_0 \, Dc \, Dc_0 \, D\tilde c \,
  D\tilde c_0 \, Da_0 \, D\tilde a_0] e^{-S_{YM}[X] - S_{g.f.}^{\delta}[X,ghosts]},
\end{multline}
where $\CalG'=\CalG/G$ is the coset of the group of gauge transformations by
constant gauge transformations. In our conventions 
we take the volume of the group of gauge transformations with respect 
to the measure rescaled by a power of the coupling constant
$\gym$. That is, we take $\vol(G,\gym) = \gym^{dim G} \vol(G)$, where $\vol(G)$ 
is the volume of the gauge group computed with respect to the Haar measure 
induced by the canonical $\gym$ independent  Killing form $(,)$ on the Lie algebra.

\subsection{Supersymmetry complex}
To compute the path integral, it is convenient to present the supersymmetry
transformations in the cohomological form by making a change 
of variable with trivial Jacobian. 

Using that the conformal Killing spinor $\ve$
in~\eqref{eq:off-shell-susy} has the unit norm 
on $S^4$, we see that the set of the sixteen spinors $\{\Gamma^{M} \ve\}$,
$M=1,\dots,9$ and $\{\nu_i\}$, $i = 1,\dots, 7$ constitutes 
an orthonormal basis in the space of $\Spin(9,1)$ chiral Weyl spinors on $S^4$.  We expand $\Psi$ over this basis
\[
\Psi = \sum_{M=1}^{9} \Psi_M \Gamma^M \ve + \sum_{i=1}^{7} \Upsilon_i \nu^i.
\]
In terms of $(\Psi_M, \Upsilon_i)$, the supersymmetry
transformations~\eqref{eq:off-shell-susy} are:
\newcommand{\sdelta}{s}
\begin{equation}
  \begin{aligned}
    &\begin{cases}
      \sdelta A_{M} = \Psi_M \\
      \sdelta \Psi_{M} = -(L_v + R +M +G_{\Phi}) A_{M}
    \end{cases} \\
    &\begin{cases} \sdelta \Upsilon_i = H^i \\
      \sdelta H^i = -(L_v + R + M + G_{\Phi}) \Upsilon_i,
    \end{cases}
  \end{aligned}
\end{equation}
where
\begin{equation}
  H^i \equiv K^{i} + w_i \Phi_0 + s_i(A_M).
\end{equation}
Here $s$ denotes  $\delta_{\ve}$ to distinguish it from the differential $\delta$ of the Faddeev-Popov
complex. By $L_v$ we denote the Lie derivative in the direction of the vector field
$v^{\mu}$, $R$ denotes the $R$-symmetry transformation in $\SU^{R}_{L}(2)$, $M$
denotes the mass-term induced transformation by $M_{ij}$ in $\SU^{R}_{R}(2)$, and
$G_{\Phi}$ denotes the gauge transformation by $\Phi$.  The functionals
$s_i(A_M)$,  $i=1,\dots, 7$,
are the equations of the equivariantly cohomological field theory
\begin{equation}
  s_i(A_M) = \frac 1 2 F_{MN} \nu_i \Gamma^{MN} \ve + \frac 1 2 \Phi_A  \nu_i \Gamma^{\mu A} \nabla_{\mu} \ve \quad \text{for} \quad M,N = 1, \dots, 9 \quad A =5,\dots,9.
\end{equation}

Conceptually, the supersymmetry complex is
\begin{equation}
  \begin{aligned}
   & s X   = X'\\
   & s X' = [\phi + \ve, X]\\
   & s \phi =0 
  \end{aligned}
\end{equation}
 where $\phi = - \Phi$, $[\phi,X] = -G_{\Phi} X$ and $[\ve,X]= -(L_v +
R + M) X $.

All fields except $\Phi$ are paired in the
$s$-doublets $(X, X')$.
The fields $X$ and $X'$ are of the opposite statistics. We think about the fields $X$ as
coordinates on an infinite-dimensional supermanifold $\CalM$ acted by a
group $\CalG$.  The fields $X'$ are interpreted as de Rham differentials $X'
\equiv d X$, if we identify the operator $s$ with the differential in the Cartan
model of $\CalG$-equivariant cohomology on $\CalM$
\begin{equation}
  s = d + \phi^a i_{v^a}.
\end{equation}
Here $\phi^a$ are the coordinates on the Lie algebra of the group $\CalG$
with respect to some basis $\{e_a\}$, and $i_{v^a}$ is the contraction with a
vector field $v^a$ representing action of $e_{a}$ on $\CalM$.  The differential
$s$ squares to the Lie derivative $\CalL_{\phi}$.  For us the group $\CalG$ is the semi-direct product
\begin{equation}
  \label{eq:group_G}
  \CalG=\CalG_{gauge} \ltimes \U(1)
\end{equation}
of the infinite-dimensional group of gauge transformations $\CalG_{gauge}$ and
the $\U(1)$ Lorentz subgroup of the $\OSp(2|4)$.

In the path integral~\eqref{eq:Z_{phys}},
we integrate $s$-equivariantly closed form $e^S$ over $\CalM$ and then over
$\phi$.  See~\cite{Vafa:1994tf,Labastida:1997vq,Kapustin:2006pk}
discussing twisted $\CalN=4$, which have similar cohomological structure,
and~\cite{Witten:1992xu} where similar integration over $\phi$ 
is performed.

\subsection{The combined $Q$-complex}
We have constructed  the gauge-fixing complex with the differential $\delta$ and
the supersymmetry complex with the differential $s$:
\begin{equation}
  \begin{aligned}
    \delta a_0 &= 0  & \delta X  &= -[c,X]   & \delta c  &= -a_0 -\frac 1 2 [c,c] &      \delta \tilde c & = b&      \delta \tilde a_0 & = \tilde c_0&    \delta b_0 & = c_0 \\
    &  &  \delta X' &= -[c, X'] & \delta \phi &= -[c + \ve, \phi]  & \delta b & = [a_0, \tilde c]&       \delta \tilde c_0   & = [a_0,\tilde a_0]&       \delta c_0& = [a_0,b_0] \\
    s a_0 &= 0 & s X  &= X'   & s c  &= \phi &      s \tilde c & = 0&      s\tilde a_0 & = 0&    s b_0 & = 0 \\
    & & s X' &= [\phi+\ve, X] & s \phi &= 0 & s b & = [\ve,\tilde c]& s \tilde c_0 &
    = 0 & s c_0& = 0.
  \end{aligned}
\end{equation}
The operators $\delta$ and $s$ anticommute as follows
\begin{equation}
  \begin{aligned}
    \{\delta,\delta\} X^{(\prime)} &= [a_0, X^{(\prime)}] & \{\delta,\delta\} (ghost) &= [a_0, ghost] \\
   \{s,s\} X^{(\prime)} & = [\phi+\ve,X^{(\prime)}] &  \{s,s\} (ghost) &= 0 \\
   \{s,\delta\} X^{(\prime)} & = -[\phi,X^{(\prime)}] &    \{s,\delta\} (ghost) &= [\ve, ghost].
  \end{aligned}
\end{equation}
Here $X^{(')}$ stands for all physical and auxiliary fields $X$ and $X'$, 
and $ghost$ stands for all field of the BRST gauge fixing complex.

Combining the operators $\delta$ and $s$, we define the operator
\[
Q  = s+\delta.
\]
Then we get
\begin{equation}
  \begin{aligned}
    Q X  &= X' -[c,X]   & Q c  &= \phi -a_0 -\frac 1 2 [c,c] &      Q \tilde c & = b&      Q \tilde a_0 & = \tilde c_0&    Q b_0 & = c_0 \\
    Q X' &= [\phi+\ve, X] - [c, X']  &Q \phi &= -[c, \phi + \ve ]  &    Q b &=[a_0+\ve,\tilde c] & Q \tilde c_0   & = [a_0,\tilde c_0] &  Q c_0 & = [a_0, b_0]  \\
    Q a_0 &= 0.
\end{aligned}
\end{equation}
The $Q^2$ acts as
\[Q^2 \cdot = [a_0 + \ve, \cdot] ,\]
that is a constant gauge transformation generated by $a_0$ 
and the $\U(1)$ self-dual Lorentz rotation around the North pole generated by $\ve$. 

Now, since $sS_{phys} = 0$ and $\delta
S_{phys} = 0$ we have
\[ Q S_{phys}= 0.\] 
To  localize, we need to check that the gauge-fixing
term~\eqref{eq:gauge-fix-term} is  $Q$-closed.

We  use this $Q$-exact gauge-fixing term: 
\begin{multline} S_{g.f.}^{Q} = (\delta + s) ((\tilde c, i d^*A + \frac {\xi_1} 2 b +
  i b_0) - (c,  \tilde a_0 - \frac {\xi_2} 2 a_0) )
  = S_{g.f}^{\delta} - (\tilde c, s(i d^*A + \frac {\xi_1} 2 b + i b_0)) - (\phi, \tilde a_0) =\\
  = S_{g.f}^{\delta} - (\tilde c, d^*\psi + \frac {\xi_1} 2 [\ve, \tilde c]) - (\phi,  \tilde a_0 - \frac {\xi_2} 2 a_0).
\end{multline}
The replacement of $S_{g.f.}^{\delta}$ by $S_{g.f.}^{Q}$ does not change the
partition function $Z_{phys}$~\eqref{eq:Z_{phys}}. We check this claim at $\xi_1 = 0$. 
Integrating over $a_0$ we get
\[
 (ia_0^E + \frac 1 2 [c,c] - \phi,  \tilde a_0 - \frac {\xi_2} {2} ia_0^E) \to \frac {1}{2 \xi_2}\lb -\frac {\xi_2} {2}
( \frac 1 2 [c,c] - \phi) + i \tilde a_0\rb^2.
\]
After we integrate over $\tilde a_0$ the above term goes away completely. The determinants 
for the Gaussian integrals over $a_0$ and $\tilde a_0$ cancel. 
Then we are left with the following gauge-fixing terms
\[
 i(b, d^*A + b_0) - i(\tilde c, d^* \nabla c + c_0) + i (c,  \tilde c) - (\tilde c, d^{*} \psi),
\]
where $\psi$ is the fermionic one-form: the superpartner of the
gauge field $A$. Then we notice that the term $(\tilde c, d^{*} \psi)$ does not change the fermionic determinant arising from 
the integral over $c, \tilde c, c_0$ and $\tilde c_0$ because all modes of $c$ are coupled 
to $\tilde c$ by this quadratic action 
\[
i(\tilde c, d^* \nabla c + c_0) + i(c,  \tilde c),
\]  
and because there are no other terms in the gauge-fixed action that contain modes of $c$. 
In other words, if treat the term $(\tilde c, d^{*} \psi)$ as the perturbation to the usual 
gauge fixed action, all diagrams with insertion of  $(\tilde c, d^{*}
\psi)$  vanish because $\tilde c$ can be connected by propagator only to
$c$, 
but there are no vertices containing $c$.

Let us summarize the construction. The  standard   gauge-fixed  action 
(\ref{eq:Z_{phys}}) is $\delta$-closed, but not $Q$-closed. 
To make the action $Q$-closed, we add certain terms to the action in
such a way that the result of integration does not change.

We conclude that the total gauge-fixed action
\begin{equation}
  \tilde S_{phys} = S_{phys} + S'_{g.f.}
\end{equation}
is $Q$-closed
\begin{equation}
  Q \tilde S_{phys} = 0,
\end{equation}
and that the path integral over the space of all fields and ghosts 
with the action $\tilde S_{phys}$ is equivalent 
to the usual gauge-fixed~(\ref{eq:Z_{phys}}) partition function .

Formally, $Q$ is
the equivariant differential in the Cartan model for the $\tilde G = G \ltimes \U(1)$ 
equivariant cohomology with parameters 
by $a_0$ and $\ve$ on the space of all other fields 
in the path
integral~\eqref{eq:Z_{phys}}.  The pairs $(\tilde c, b)$, $(\tilde a_0, \tilde c_0)$
and $(b_0 , c_0)$ are the canonical multiplets.  
If we make a change variables with trivial Jacobian (therefore the path
integral is unchanged),
\begin{equation}
  \begin{aligned}
    \tilde X' = X' - [c,X] \\
    \tilde \phi = \phi - a_0 - \frac 1 2 [c,c],
  \end{aligned}
\end{equation}
in terms of the new fields the $Q$-complex
turns into canonical form. 
All fields are paired in doublets  $(Field, Field')$:
\begin{equation}
  \begin{aligned}
    \label{eq:Q_complex_long}
    Q (Field)  &= (Field') \\
    Q (Field') &= [a_0 + \ve, Field] .
  \end{aligned}
\end{equation}
Moreover, $Qa_0 = Q \ve = 0$.

Now recall Atiyah-Bott-Berline-Vergne localization formula for the integrals of $G$-equivariantly closed differential forms~\cite{MR721448,MR685019} on
a $G$-manifold $\CalM$
\begin{equation}
  \label{eq:ABBV}
  \int_{\CalM} \alpha = \int_{F \subset \CalM} \frac {i^*_{F} \alpha} { e(\CalN)}.
\end{equation}
The numerator corresponds to the physical action evaluated at the critical locus
of the $tQV$ term. The equivariant Euler class of the
normal bundle in the denominator is the determinant produced by
the  Gaussian integration of the quadratic part of $tQV$ in the normal directions $\CalN$.  
This determinant is actually the product of weights of the $G$-action on $\CalN$ defined
by~\eqref{eq:Q_complex_long}. For our purposes, we use the
straightforward generalization of the localization
formula~\eqref{eq:ABBV} for the situation when the manifold $\CalM$ is
an infinite-dimensional
 supermanifold. The equivariant Euler class is interpreted in the
super-formalism~\cite{lavaud-equiv,lavaud-superpf}. 
If we split the normal bundle to the bosonic and the fermionic subspaces, the resulting determinant is the product of
weights on the bosonic subspace divided by the product of weights on the
fermionic subspace.

Before we gauge-fixed the action, we had argued 
that the theory localizes to the zero modes of the field $\Phi_0$.
The localization principle for the gauge-fixed theory remains the same, except that 
now we identify the zero mode of the field $\Phi_0$ with $a_0$. Indeed, 
if we integrate over $\tilde a_0$ using the gauge fixing terms at $\xi_2 = 0$
\[
 (ia_0^E + \frac 1 2 [c,c] - i \phi^E, \tilde a_0),
\]
we get the constraint that the zero mode of $\phi^E$ is equal to $a_0^E$.

\subsection{Computation of the determinant using the index theory of transversally elliptic operators}
The linearized $Q$-complex is 
\begin{equation}
  \begin{aligned}
    \label{eq:eqX}
    Q X_0  &= X_0'      & Q X_1  &= X_1'  \\
    Q X_0' &= R_0 X_0 & Q X_1' &= R_1 X_1,
  \end{aligned}
\end{equation}
where all bosonic and fermionic fields in the first line
of~\eqref{eq:Q_complex_long} are denoted as $X_0$ and $X_1$ respectively, and
their $Q$-differentials are denoted as $X_0'$ and $X_1'$. The fields $X_0, X_1'$ are
bosonic, and the fields $X_0', X_1$ are fermionic.

\newcommand{\bos}{\text{b}}
\newcommand{\fer}{\text{f}}

The quadratic part of the functional $V$ is
\begin{align}
  V^{(2)} = \left(
    \begin{array}{c}
      X_0' \\ X_1
    \end{array}
  \right)^{t} \left(
    \begin{array}{cc}
      D_{00} & D_{01} \\
      D_{10} & D_{11} \\
    \end{array}
  \right) \left(
    \begin{array}{c}
      X_0 \\
      X_1' \\
    \end{array}
  \right),
\end{align}
where $D_{00}, D_{01}, D_{10}, D_{11}$ are  differential operators.  Then 
\[ QV^{(2)} = (X_{\bos}, K_{\bos} X_{\bos}) + (X_{\fer}, K_{\fer} X_{\fer}), \]
where the kinetic operators $K_{\bos}, K_{\fer}$ are expressed in terms of
$D_{00}, D_{01}, D_{10}, D_{11}$ and $R_0,R_1$ in a certain way.  The Gaussian
integration produces
\begin{equation}
\label{eq:Gauss-Z-1-loop}
 Z_{\text{1-loop}} = \left ( \frac { \det K_{\bos}}{\det K_{\fer}}
\right)^{-\frac 1 2}.
\end{equation}
 Let $E_0$ and $E_1$ be the vector bundles such that 
their sections are the fields $X_0,X_1$.  Linear algebra shows
that the ratio of the determinants in (\ref{eq:Gauss-Z-1-loop}) depends
only on the restriction of $R_0$ and $R_1$ on the kernel and cokernel
spaces, respectively,  of the operator $D_{10}: \Gamma(E_0) \to
\Gamma(E_1)$. Namely we have
\begin{equation}
  \frac { \det K_{\bos}} { \det K_{\fer}} = \frac { \det_{\ker D_{10}} R_0} { \det_{\coker D_{10}} R_1}.
\end{equation}

The operator $D_{10}$ in our problem is not elliptic, 
but transversally elliptic with respect to the $U(1)$
rotation of $S^4$ \cite{MR0482866}.

This means the following. Let $E_0$ and $E_1$ be vector bundles over a manifold $X$
and $D: \Gamma(E_0) \to \Gamma(E_1)$ be a differential operator. 
(In our problem $X = S^{4}$.) Let a compact Lie group $\tilde G$
act equivariantly on $E_i \to X$. Let $\pi: T^*X \to X$ be the cotangent bundle of $X$. 
Then the pullback $\pi^*E_i$ is a bundle
over $T^*X$.  By definition, the symbol of the differential operator $D: \Gamma(E_0) \to
\Gamma(E_1)$ is a vector bundle homomorphism $\sigma(D): \pi^* E_0 \to \pi^*
E_1$, such that in local coordinates $x_i$, the symbol is defined  replacing all partial
derivatives in the highest order component of $D$ by momenta:  $\frac {\p} {\p
  x^i} \to i p_i$, where $p_i$ are the coordinates on fibers of $T^*X$.
The operator $D$ is called elliptic if its symbol $\sigma(D)$ is invertible on 
$T^*X \setminus 0$, where $0$ denotes the zero section.  
The kernel and cokernel of an elliptic operator are finite dimensional vector spaces. 
Using the Atiyah-Singer index
theory~\cite{Atiyah-Singer:1968-ind1,Atiyah-Segal:1968-ind2,Atiyah-Singer:1968-ind3,MR0190950,MR0212836,MR0232406}
we can find the virtual difference of these spaces as a graded $\tilde G$-module.  If the operator $D_{10}$ is not elliptic, 
the ordinary Atiyah-Singer index theory does not apply. We need to use 
 generalization of Atiyah-Singer index theory for transversally elliptic
 operators. Those are the operators which are elliptic in all directions 
transversal to the $\tilde G$-orbits~\cite{MR0482866,MR0341538}.

In more details, for a point $x \in X$, let  $T_{\tilde G}^*X_x$ denote the subspace of $T^*X_x$
transversal to all $\tilde G$ orbits passing through $x$.
The family of the vector spaces $T^*_{\tilde G} X$ over $X$ is the union
of $T_{\tilde G}^*X_{x}$ for all $x \in X$. 
The operator $D$ is called transversally elliptic if its symbol
$\sigma(D)$ is invertible on $T^*_{\tilde G} X \setminus 0$.
When we compute the symbol of $D_{10}$, we will see explicitly
in~(\ref{eq:symbol-explicit}) that $D_{10}$ is  transversally elliptic
but not elliptic. The kernel and the cokernel of such  operators
usually are not finite dimensional. But if we decompose the kernel and cokernel into
 irreducible representations of $G$, each irreducible representation $\alpha$ 
appears with a finite multiplicity $m_{\alpha}$ ~\cite{MR0482866,MR0341538}:
\begin{equation}
  \begin{aligned}
    \ker D_{10} = \oplus_{\alpha} m_{\alpha}^{(0)} R_{\alpha} \\
    \coker D_{10} = \oplus_{\alpha} m_{\alpha}^{(1)} R_{\alpha}.
  \end{aligned}
\end{equation}
Therefore, 
\begin{equation}
  \label{eq:defKK}
  \frac { \det K_{\bos}} { \det K_{\fer}} = \prod_{\alpha} (\det R_{\alpha})^{m^{(0)}_{\alpha} - m^{(1)}_{\alpha}}.
\end{equation}

To compute  $m^{(0)}_{\alpha} - m^{(1)}_{\alpha}$ we use 
the Atiyah-Singer index theory~\cite{MR0482866,MR0341538} for transversally elliptic 
operators, generalizing elliptic operators~\cite{Atiyah-Singer:1968-ind1,Atiyah-Segal:1968-ind2,Atiyah-Singer:1968-ind3,MR0190950,MR0212836,MR0232406}.
In our problem, $R_{\alpha}$ is an irreducible representation of the
group $\tilde G =H \times G$, with $H \equiv \U(1)$. 
Let $t \in \BC, |t|=1$ denote an element of $\U(1)$ in the defining representation. 
Irreducible representations of $\U(1)$ are labeled by integers $n \in
\BZ$, with the character being $t^n$.  The $\U(1)$-equivariant index of
$D_{10}$ is 
\[ \ind (D_{10}) = \tr_{\ker D_{10}} R(t) - \tr_{\coker D_{10}} R(t) =
\sum_{n} ( m^{(0)}_n - m^{(1)}_n) t^{n}. \] Hence, if we compute the equivariant
index of $D_{10}$ as a formal Laurent series in $t$, we find $m^{(0)}_n
- m^{(1)}_n$ and evaluate~\eqref{eq:defKK}.

To compute the index of $D_{10}$, we need to describe concretely the bundles $E_0,
E_1$ and the symbol of the operator $D_{10}: \Gamma(E_0) \to
\Gamma(E_1)$.  
Our abstract notations  $X_0, X_0', X_1, X_1'$ correspond to the original
fields as
\begin{equation}
  \begin{aligned}
    X_0  &= (A_{M}, \tilde a_0, b_0) &    X_1 &= (\Upsilon_i, c, \tilde c) \\
    X'_1 &= (\tilde \Psi_{M}, \tilde c_0, c_0) & X_1' &= (\tilde H_i, \tilde \phi,
    b).
  \end{aligned}
\end{equation}

The linearized space of all fields splits $Q$-equivariantly~\eqref{eq:eqX} into the
direct sum of the fields
of vectormultiplet and hypermultiplet.  The vectormultiplet subspace
is
\begin{align}
  X_0^{vect} &= (\Phi_9, A_M, \tilde a_0, b_0), \quad
  M=1,\dots,4 \quad \\
  X_1^{vect} &= (\Upsilon_i, c, \tilde c ), \quad  i=5, \dots, 7
\end{align}
and the $Q$-superpartners. 
The hypermultiplet subspace is
\begin{align}
  X_0^{hyper} &= (A_M), \quad M=5,\dots,8 \quad \\
  X_1^{hyper} & = (\Upsilon_i ), \quad  i =1, \dots, 4
\end{align}
and the $Q$-superpartners. The operator $D_{10}$ commutes 
with the vector-hyper splitting. The vector bundles split as $E_0 = E_0^{vect} \oplus
E_{0}^{hyper}$, and $E_1 = E_1^{vect} \oplus E_1^{hyper}$, as well as the
operator split $D_{10} =D_{10}^{vect} + D_{10}^{hyper}$, where $D_{10}^{vect}:
\Gamma(E_0^{vect}) \to \Gamma(E_1^{vect})$ and $D_{10}^{hyper}: \Gamma(E_0^{hyper})
\to \Gamma(E_1^{hyper})$.

First we compute the index of $D_{10}^{vect}$. The constant fields $(\tilde a_0,
b_0)$ are in the kernel of $D_{10}^{vect}$ and have zero $\U(1)$ weights, hence
their contribution to the index is $2$:
\begin{equation}
  \label{eq:difference-d-dprime}
  \ind(D_{10}^{vect}) = \ind'(D_{10}^{vect}) + 2.
\end{equation}

 The remaining fields, denoted by
$X_0^{vect'}$, are identified with sections of bundle $(T^*_{S^4} \oplus \CalE_{S^4}) \otimes
\ad E$, \ where $T^*_{S^4}$ is the cotangent bundle over $S^4$, and $\CalE_{S^4}$ is the rank one trivial
bundle over $S^4$.  The fields $X_1^{vect'}$ are identified with sections of
$(\CalE^{3}_{S^4} \oplus \CalE^2_{S^4}) \otimes \ad E$,
 where $\CalE^3_{S^4}$ is the  bundle of
auxiliary scalar fields, and $\CalE^2$ is the bundle of the gauge
fixing fields $c$ and $\tilde c$.
Now we compute the symbol of the operator $D_{10}^{vect}$. The relevant terms
are
\begin{equation}
  V^{(2)} = (\tilde c, d^{*} A) - (c, \nabla_{\mu} \CalL_{v} A_{\mu}) + (\Upsilon_i, (*F_{1i}) + F_{1i} \cos \theta
  + \nabla_{i} \Phi_9 \sin \theta),
\end{equation}
where index $i$ runs over vielbein elements on $S^4$.  

We label vielbein basis elements such that  $i=1$ is directed along the $\U(1)$ vector field,
and $i=2,3,4$ are the remaining orthogonal directions.  The term $(c,
\nabla_{\mu} \CalL_{v} A_{\mu})$ comes from the term $(\psi_{\mu}, \CalL_{v}
A_{\mu})$ and the relation $\psi_{\mu} = \tilde \psi_{\mu} - \nabla_{\mu} c$.
The symbol $\sigma(D_{10}^{vect}): \pi^* E_{0}^{vect} \to \pi^{*}E_1^{vect}$, 
where $\pi$ denotes the projection of the cotangent bundle $\pi: T^*X \to X$, is
represented by the following matrix
\newcommand{\ct}{ c_{\theta} }
\newcommand{\st}{ s_{\theta} }
\begin{equation}
\label{eq:symbol-explicit}
 \begin{pmatrix}
    c \\
    \tilde c \\
    \Upsilon_2 \\
    \Upsilon_3 \\
    \Upsilon_4 \\
  \end{pmatrix}
  \leftarrow
  \begin{pmatrix}
    \ct p^2  & \st \vect{p}^2 & -\st p_2 p_1 & -\st p_3 p_1 & -\st p_4 p_1 \\
    0 & p_1 & p_2 & p_3 & p_4 \\
    \st p_2 & -\ct p_2 & \ct p_1 & -p_4 & p_3 \\
    \st p_3 & -\ct p_3 & p_4  & \ct p_1 & -p_2 \\
    \st p_4 & -\ct p_4 & -p_3 & p_2 & \ct p_1
  \end{pmatrix}
  \begin{pmatrix}
    \Phi_9 \\ A_1 \\ A_2 \\ A_3 \\ A_4
  \end{pmatrix}.
\end{equation}
Here $p_{i}$,  $i=1,\dots,4$ denotes coordinates on fibers of $T^*X$, 
$\vect{p}=(p_2,p_3,p_4)$ denotes coordinate on fibers of $T^*_H X$ (the
 momentum transversal to  the $\U(1)$ vector field), 
and $\ct, \st$ stand for $\cos \theta$, $\st \equiv \sin \theta$.
Changing the coordinates on the fibers of the bundles $E_0 \to T^*X$ and $E_1 \to T^*X$
\begin{equation}
  \begin{aligned}
    c \to c + \st p_0 \tilde c \\
    \Phi_9 \to \ct \Phi_9  + \st A_1  \\
    A_1 \to - \st \Phi_9   + \ct A_1,
  \end{aligned}
\end{equation}
we bring the matrix of the symbol of $D_{10}^{vect}$ to
\begin{equation}
\label{eq:symbol-simplified-nearly}
  \begin{pmatrix}
    p^2 & 0 & 0 & 0 & 0 \\
   - \st p_1 & \ct p_1 & p_2 & p_3 & p_4 \\
    0 & -p_2 & \ct p_1 & -p_4 & p_3 \\
    0 & -p_3 & p_4 & \ct p_1  & -p_2 \\
    0 & -p_4 & -p_3 & p_2 & \ct p_1 .
  \end{pmatrix}.
\end{equation}
The term $\st p_1$ in the second row  can be removed by
adding the first line multiplied by $\st p_1 /p^2$. 
The nontrivial part of the symbol is represented by the $4 \times 4 $ matrix
\begin{equation}
\label{eq:symbol-simplified}
\sigma=\begin{pmatrix}
     \ct p_1 & p_2 & p_3 & p_4 \\
     -p_2 & \ct p_1 & -p_4 & p_3 \\
     -p_3 & p_4 & \ct p_1  & -p_2 \\
     -p_4 & -p_3 & p_2 & \ct p_1 .
  \end{pmatrix}.
\end{equation}
The determinant of this matrix is $(\cos^2\theta p_1^2 + \vect{p}^2)^2$.
We see that the symbol is not elliptic at the equator of $S^4$, 
since for $\cos \theta = 0$ and $(p_1 \neq 0, \vect{p}=0)$ the
determinant vanishes.
But the symbol is transversally elliptic with respect to the $H=\U(1)$
action, since the determinant does not vanish if the transversal
momentum $\vect{p} \neq 0$.

Near the North pole ($\ct =1$) 
the symbol is equivalent to the elliptic symbol of the standard self-dual complex $(d,d^{+})$
\begin{equation}
\label{eq:sd-complex}
\Omega^{0} \overset{d}{\to} \Omega^{1} \overset{d^{+}}{\to} \Omega^{2+},
\end{equation}
and near the South pole ($\ct = -1$), the symbol is equivalent to the elliptic symbol of the 
standard anti-self-dual complex
$(d,d^{-})$
\begin{equation}
\label{eq:asd-complex}
\Omega^{0} \overset{d}{\to} \Omega^{1} \overset{d^{-}}{\to} \Omega^{2-}.
\end{equation}

For the elliptic complex, we can use the Atiyah-Bott formula~\cite{MR0212836,MR0232406} to
compute the index as a sum of local contributions from $H$-fixed points on $X$. 
In the transversally elliptic case the situation is more complicated. 
By definition, the index is the sum of characters
of irreducible representations
\begin{equation}
  \label{eq:index-u-1-general}   
\ind_{\U(1)} (D) = \sum_{n=-\infty}^{\infty} a_n t^n, \quad t  \in \U(1)
\end{equation}
where $a_n = m_n^{(0)} - m_n^{(1)}$ is the difference of  multiplicities
between the  kernel and cokernel of $D$.
In the elliptic case, only a finite number of $a_n$ does not vanish,
so that the index is a Laurent polynomial in $t$, and, therefore,  
 the index is a regular function on the circle $|t| = 1$.
In the transversally elliptic case, the range of summation 
in the formal Laurent series~(\ref{eq:index-u-1-general}) can be infinite,
so that index as a function of $t$ might be singular. 
Atiyah and Singer \cite{MR0482866,MR0341538} showed that in the transversally elliptic 
case, all coefficients $a_n$ are finite, 
and that the index is a well defined distribution (a generalized
function) on the group manifold.

For example, consider $X = S^1$ and consider the zero operator  $D:
C^{\infty}(S^1) \to 0$. Such trivial operator is a transversally elliptic operator with respect
to the defining  $\U(1)$ action on  $S^1$. 
The kernel of $D$ is the space of all functions on $S^1$, 
the cokernel is zero.  Then $m_n^{(0)} = 1, m_{n}^{(1)}=0$ for all $n$, 
hence the index is $\sum_{n=-\infty}^{\infty}{t^n}$, which is the Dirac delta-function 
supported at $t=1$.

The indices of transversally elliptic operators were treated
in~\cite{MR0482866,MR0341538,MR1369411,MR1288997}. We chop an $H$-manifold $X$
into a collection of small neighborhoods of $H$-fixed points
 and a manifold $Y \subset X$ such that $H$ acts freely on $Y$.
At each $H$-fixed point the symbol of the transversally elliptic operator 
is  elliptic, hence the ordinary equivariant index theory applies. 
Since $H$ acts freely on $Y$, the quotient $Y/H$ is well-defined manifold.
An $H$-transversally elliptic operator on $Y$ is pushed to an elliptic
operator on $Y/H$ under the projection map $Y \to Y/H$.  
Then we can combine the representation theory of $G$ and the usual index theory on the quotient $Y/H$ to find
the index of transversally elliptic operator on $Y$~\cite{MR0482866}.

Let $R(H)$ be the space of regular functions on $H$, equivalently, the space of finite polynomials in $t$ and $t^{-1}$.
Let $\CalD'(H)$ be the space of distributions on $H$, equivalently the
space of formal Laurent series in $t$ and $t^{-1}$.
The space of distributions  $\CalD'(H)$ is a module over the space of regular functions $R(H)$, 
since there is a well defined term by term multiplication of Laurent series
 in $t$ and $t^{-1}$ by finite polynomials in $t$ and $t^{-1}$.
Certain  generalized functions, such as the Dirac delta-function $\sum_{n=-\infty}^{\infty} t^{n}$,
are annihilated by non-zero regular functions. For example,  Dirac delta-function 
$\sum_{n=-\infty}^{\infty} t^{n} \in \CalD'(H)$ vanishes after
multiplication  by  $(1-t)$.
The elements of $\CalD'(H)$  annihilated by non-zero regular functions in $R(H)$ are called torsion
elements. 

To find the index of transversally elliptic operator up to a distribution supported 
at $t=1$, or a torsion element of $\CalD'(H)$, 
we can use the usual Atiyah-Bott
formula~\cite{MR0232406,MR0212836,MR0190950}, see
appendix~\eqref{se:trans-elliptic}. The formula gives the contribution to the index from each fixed point as a rational 
function of $t$, which might have a pole at $t=1$. For example, if $H=\U(1)$ acts 
on $\BC$ as $z \to t z$, then the Atiyah-Bott formula for the index of the $\bar \p$-operator
at the fixed point $z=0$ gives 
\begin{equation}
  \label{eq:Daulbealt-example}
  \ind ( \bar \p)|_0 = \frac {1} {1 - t^{-1}}.
\end{equation}

Expanding the series 
in $t$ and $t^{-1}$ we  get a distribution associated with this rational
function.  The expansion is not unique, the two expansions might differ 
by a distribution supported at $t=1$.  
For $H=\U(1)$, there are two natural regularizations to fix the singular part~\cite{MR0482866}.
The regularization $[f(t)]_+$ is defined by taking expansion at $t=0$,
which produces series infinite in positive
 powers of $t$. The regularization $[f(t)]_-$ is defined by taking
 expansion at $t=\infty$, which produces   infinite series in negative
 powers of $t$. The two regularizations differ by a distribution supported at $t=1$. 
For example, for the $\bar \p$-operator the difference is the Dirac 
delta-function 
$[ (1-t^{-1})^{-1}]_+ - [(1-t^{-1})^{-1}]_- = - \sum_{n=-\infty}^{n=\infty} t^{n}$.

Let $X=\BC^n$ be a $H \equiv \U(1)$ module with positive weights $m_1, \dots, m_n$,
so that $\U(1)$ acts as $z_i \to t^{m_i} z_i$. The origin $Y=\{0\}$ is the $H$-fixed point set. 
Let $v$ be the vector field generated by the $\U(1)$ 
action on $X$. Let $\sigma(D)$ be an elliptic symbol defined on  
  $T^* X |_Y$. 
Atiyah showed~\cite{MR0482866} that we can use the vector field $v$ in two
different ways, called $[\cdot ]_{+}$ and 
$[\cdot]_{-}$, to construct a transversally elliptic symbol $\tilde \sigma=[\sigma]_{\pm}$ 
on  $T_{H}^* X$ such that $\tilde \sigma$ is an isomorphism outside of $Y$. 
See appendix \ref{se:trans-elliptic} for  details. 
The index of the transversally elliptic symbol $\tilde \sigma$ is well defined as
a distribution on $H$.
Moreover, if $\ind(\sigma)$ is a rational function of $t$ associated
at the fixed point $Y$ to
 the elliptic symbol $\sigma$ by the Atiyah-Bott formula, then
\begin{equation}
  \label{eq:atiyha-trans-elliptic}
  \ind([\sigma]_\pm) = [\ind(\sigma)]_{\pm}.
\end{equation}

We apply this construction to our problem. 
Namely, we use the vector field generated by the $H \equiv \U(1)$-action on $X=S^4$ 
to trivialize the symbol $\sigma(D_{10}^{vect})$ everywhere on $T^*_H
X$ except the North and the South pole. The index is the sum of contributions from the fixed points, where each
contribution is expanded in either  positive or negative powers of $t$ according
to the~(\ref{eq:atiyha-trans-elliptic}). More concretely, we  
trivialize the transversally elliptic symbol $\sigma = \sigma(D_{10}^{vect})$ everywhere
outside the North and the South poles on $T^*_H X$ by replacing
 $\ct p_1$ by $\ct p_1 + v$ 
on the diagonal in~(\ref{eq:symbol-simplified}) with $v = \sin \theta$.
 In other words, we deform the operator 
by adding the Lie derivative in the direction of the vector field $v$. 
The resulting symbol 
\begin{equation}
\label{eq:symbol-simplified-new}
\tilde \sigma=
  \begin{pmatrix}
     \ct p_1 + \st & p_2 & p_3 & p_4 \\
     -p_2 & \ct p_1 + \st & -p_4 & p_3 \\
     -p_3 & p_4 & \ct p_1 +\st  & -p_2 \\
     -p_4 & -p_3 & p_2 & \ct p_1 + \st
  \end{pmatrix}.
\end{equation}
has determinant $(\vect{p}^2 + (\ct p_1 + \st)^2)^2$ which is non-zero everywhere
outside the North and the South poles at $T^{*}_{H} X$.
The index of $\tilde \sigma$ is equal to the index of $\sigma$, since $\tilde \sigma$ is a continuous 
deformation of $\sigma$. On the other hand, since $\tilde \sigma$ is an isomorphism outside of the 
North and the South pole,  to get the index of $\tilde \sigma$ we sum up contributions
from the North and the South pole. At the North pole $\cos \theta =
1$. Therefore, near the North pole, the transversally elliptic symbol $\tilde \sigma$ 
coincides with the symbol associated to the elliptic symbol $\tilde \sigma_{\theta =0}$
by the $[\cdot ]_{+}$ regularization. 
At the South pole $\cos \theta  = -1$. Therefore, 
near the South pole, the transversally elliptic symbol $\tilde \sigma$ 
coincides with the symbol associated to the elliptic symbol $\tilde \sigma_{\theta =\pi}$
by the $[ \cdot ]_{-}$ regularization.

Finally we obtain\footnote{This deformation is similar to~\cite{MR792703}, where 
the index of the Dirac operator is computed using the deformation $\Gamma^{\mu} D_{\mu} \to \Gamma^{\mu}D_{\mu} 
+ t \Gamma^{\mu} v_{\mu}$.  }
\begin{equation}
  \ind'(D_{10}^{vect}) = \left [ \ind (d,d^{+}) |_{\theta=0} \right ]_+ +
  \left [ \ind (d,d^{-}) |_{\theta=\pi} \right ]_- .
\end{equation}

Let $(z_1, z_2)$ be complex coordinates near the North pole, such
that the $\U(1)$ action is $z_1 \to t z_1, z_2 \to t z_2$.  
Under such action, the complexified self-dual
complex is isomorphic to the Dolbeault $\bar \p$-complex twisted by the bundle
$\CalO \oplus \Lambda^2 T^*_{1,0}$. 
Using  $\ind(\bar \p)=(1-t^{-1})^{-2}$,
we get
\begin{equation}
  \ind'(D_{10}^{vect}) = \left [ - \frac {1+t^2} {(1-t)^2}\right]_{+} +
  \left [ - \frac {1+t^2} {(1-t)^2}\right]_{-}.
\end{equation}
In our conventions $E_0$ corresponds to the
middle term of the standard (anti)-self dual
complex~(\ref{eq:sd-complex}), therefore we put the minus sign.

Finally,
\begin{multline}
  \ind (D^{vect}_{10}) = 2+\ind'(D_{10}^{vect}) =  \\= 2
  -(1+t^2)(1+2t+3t^2+\dots) - (1+t^{-2})(1+2t^{-1} + 3t^{-2} + \dots)  =
  \\ =  -\sum_{n=-\infty}^{\infty} 2 |n | t^{n}.
\end{multline}

Let us proceed to the hypermultiplet contribution to the index.  The computation
is similar to the vector multiplet. The transversally elliptic operator
$D_{10}^{hyper}: \Gamma(E_0^{hyper}) \to \Gamma(E_1^{hyper})$ can be trivialized
everywhere over $T^*_H X$ except fixed points. 
The transversally elliptic complex describing hypermultiplet 
is isomorphic to the
anti-self-dual complex at the North pole, or self-dual complex at the South
pole. For the hypermultiplet, the chirality of the complex is opposite
to the chirality of the $\U(1)$ rotation near each of the fixed points.  Then,
using that the index of the twisted Dolbeault operator is $(1+t
t^{-1})/((1-t)(1-t^{-1}))$, we get
\begin{equation}
  \ind_t (D^{hyper}_{10}) = \left [ - \frac {2} {(1-t)(1-t^{-1})}\right]_{+} +
  \left [ - \frac {2} {(1-t)(1-t^{-1})}\right]_{-},
\end{equation}
which results in
\begin{equation}
  \ind_t (D^{hyper}_{10}) = +\sum_{n=-\infty}^{\infty} |2n| t^{-n}.
\end{equation}

Above we have considered the massless hypermultiplet. Massless
adjoint hypermultiplet contribution
to the index exactly cancels the vector multiplet. Hence, the determinant factor
in the $\CalN=4$ theory is trivial.  This finishes the proof 
that the Erickson-Semenoff-Zarembo/Drukker-Gross matrix
model is exact in all orders of perturbation theory.

In the $\CalN=2^{*}$ theory  with massive adjoint  hypermultiplet,  the situation is more interesting.   In the transformations~\eqref{eq:eqX} the action of $R$
is contributed by the $\SU(2)^R_R$ generator $M_{ij}$. We normalize it as $M_{ij}
M^{ij} = 4 m^2$.  The hypermultiplet fields transform in the spin-$\frac 1 2$
representation of $\SU(2)^R_R$. Therefore, in the massive case, the index is multiplied by
the spin-$\frac 1 2$ character $\frac 1 2 (e^{im}
+ e^{-im})$.  Hence all $\U(1)$-eigenspaces split into half-dimensional
subspaces with eigenvalues shifted by $\pm m$.

Finally, all fields of $\CalN=2^*$ theory transform in the adjoint representation of gauge group. Using
a constant gauge transformation we can assume  $a_0 \in \h$, where $\h$
 is the Cartan
subalgebra $\h \subset \g$. 
The non-zero
eigenvalues of $a_0$ in the adjoint representation are $\{\alpha\cdot a_0 \}$, where
$\alpha$ runs over all roots of $\g$.  Hence, combining all contributions to the
index, we obtain for the $\CalN=2^*$ theory
\[
\left( \frac {\det K_{\bos}} { \det K_{\fer}} \right)_{\CalN=2^*} = \prod_{\text{roots }\alpha}
\prod_{n=-\infty}^{\infty} \left [ \frac { (\alpha\cdot a_0 + n \ve + m)(\alpha \cdot a_0
    + n \ve - m)} {( \alpha \cdot a_0 + n\ve)^2 } \right ]^{|n|},
\]
where $\ve = r^{-1}$. The term $n\ve$ comes from the weight $n$ representation of the $\U(1)$,
the term $\alpha \cdot a_0$ is the eigenvalue of $a_0$ acting on the eigensubspace of the adjoint 
representation corresponding to root $\alpha$. 
 
We  argued  that for the path integral  convergence 
the mass parameter $m$ and the scalar field $\Phi_0$ should be taken
 imaginary in the conventions of the $(9,1)$ reduced theory. 
The
variable $a_0$ is also imaginary, because $a_0$ is identified with the zero mode of
$\Phi_0$. Now we introduce  $m = i m^E, a_0 = i a^E \equiv i a^E_0$ with
$m^E, a^E$ being real. From (\ref{eq:Gauss-Z-1-loop}) we get
\begin{equation}
  \label{eq:Z-1-loop-N=2}
  Z_{\text{1-loop}}^{\CalN=2^*}(ia_E) = \prod_{\text{roots }\alpha} \prod_{n=1}^{\infty} \left [
    \frac {   ( (\alpha \cdot a_E)^2 + \ve^2 n^2)^2       } { ((\alpha \cdot a_E + m_E)^2 + \ve^2 n^2)
(( \alpha \cdot a_E - m_E)^2 + \ve^2 n^2 )} \right ]^{\frac n 2}.
\end{equation}
Such infinite product requires a regularization. 

We use the product formula definition of the Barnes $G$-function~\cite{Barnes} 
\begin{equation}
  G(1+z) = (2\pi)^{z/2} e^{-((1+ \gamma z^2) +z)/2} \prod_{n=1}^{\infty}  \left(1 + \frac z n \right)^n e^{-z + 
\genfrac{}{}{}{1}{ z^2}{2n}},
\end{equation}
where $\gamma$ is the Euler constant.
Let us introduce the function $H(z) = G(1+z) G(1-z)$, so that   
\begin{equation}
   H(z) = e^{-(1+\gamma) z^2} \prod_{n=1}^{\infty} \left(1 - \frac {z^2}{n^2}\right)^{n} \prod_{n=1}^{\infty} 
e^{\genfrac{}{}{}{1}{z^2} {n}}.
\end{equation}
In term of $H(z)$ we get
\begin{multline}
  \label{eq:Z-Barnes-hyper}
  Z^{\CalN=2^*}_{\text{1-loop}}(ia_E) = 
\exp\left( \frac {m_E^2}{ \ve^2} \left(
    (1+\gamma) - \sum_{n=1}^{\infty} \frac {1} {n} \right)\right) \times \\
\times \prod_{\text{roots }\alpha}  \frac { H \left(i\alpha \cdot a_E/ \ve\right)} 
{ \left[ H\left((i \alpha \cdot a_E+ i m_E) / \ve \right) H\left((i\alpha \cdot a_E-i m_E) / \ve \right)  \right]^{1/2}}. 
\end{multline}
The first exponential factor is divergent but independent of $a_E$. Therefore it cancels
when we compute expectation value of the operators normalized by the
partition function. We redefine the partition function by dropping this factor.  
The resulting product of the $G$-functions is a well defined analytic function of $a_E$. 

Our result is consistent with the usual computation of the
$\beta$-function 
is the $\CalN=2$  gauge theory. To check this we need asymptotic expansion of the
$G$-function at large $z$
\begin{equation}
  \log G(1+z) = \frac 1 {12} - \log A + \frac {z} 2 \log 2 \pi + \left( \frac {z^2} 2 - \frac 1 {12} \right) \log z
  - \frac 3 4 {z^2} + \sum_{k=1}^{\infty} \frac {B_{2k+2}} {4k(k+1)z^{2k}},
\end{equation}
where $A$ is a constant and $B_{n}$ are Bernoulli numbers.  Then
\begin{equation}
  \label{eq:asymptotic-Barnes}
  \frac 1 2 \left(\log G(1+iz_E) + \log G(1-iz_E)\right) =
  \frac 1 {12} - \log A + \left(-\frac {z_E^2} {2} - \frac {1} {12}\right) \log z_E + \frac 3 4 z_E^2 + \dots
\end{equation}

In the limit of the hypermultiplet mass $m \to \infty$, we expect to get
the minimal $\CalN=2$ theory at the energy scales much lower than $m$. 
At large $m$, we expand the denominator
in~\eqref{eq:Z-Barnes-hyper}, corresponding to the hypermultiplet contribution to
$Z_{\text{1-loop}}$, and get
\begin{equation}
  Z_{\text{1-loop}}^{hyper} = const(m_E) + \left(const + \log \frac {m_E} {\ve} \sum_{\alpha} 
\frac {(\alpha \cdot a_E)^2}{\ve^2}\right) + \CalO(\frac 1 {m^2}).
\end{equation}
The quadratic term in $a_E$ can be combined with the classical Gaussian
action in the matrix model
\begin{equation}
  \label{eq:renorm-g}
  \frac {8 \pi^2 r^2 }{\gym^2} (a_E,a_E)  \to \left( \frac {8 \pi^2 r^2 }{\gym^2} - \frac {C_2}{\ve^2} \log {\frac {m_E}\ve} \right) (a_E,a_E),
\end{equation}
where $C_2$ denotes the second Casimir constant $\tr_{\mathrm{Ad}} T_{a}
T_{b} = C_{2} \delta_{ab}$. Rewriting this as
\begin{equation}
  \label{eq:running}
  \frac 1 {\tilde \gym^2} = \frac 1 {\gym^2} - \frac{C_2}{8 \pi^2} \log \frac {m_E} \ve,
\end{equation}
we see that $\tilde \gym^2$ means the renormalized coupling
constant.  The bare microscopical constant $\gym^2$ is well defined
at the energies much greater than the UV scale $m_E$. 
At  the scales much less than $m_E$, the coupling
constant runs according to the beta-function of pure $\CalN=2$ theory. Recall that the
one-loop beta function for a gauge theory with $N_{f}$ Dirac fermions and
$N_{s}$ complex scalars in adjoint representation is
\begin{equation}
  \frac {\p \gym(\mu)}{\p \log \mu} = \beta(\gym) = -\frac {C_2
    \gym^3}{2 (4\pi)^2}\lb \frac {11}{3} - \frac 4 3 N_{f} - \frac 1 3 N_{c} \rb.
\end{equation}
Taking $N_{f} = N_{s} = 1$ for a pure $\CalN=2$ theory we get precisely the
relation~\eqref{eq:running}, which says that $\tilde \gym^2$ is the running
coupling constant at the IR scale $\ve=r^{-1}$.

We can check that the resulting integral over $a_E$ is always convergent
if  the bare coupling constant $\gym^2$ is positive.
  First of all, the Barnes function
$G(1+z)$ does not have poles or zeroes on the integration contour $\Re\,
z = 0$.   To see the nice behavior of the integrand at infinity we
use the asymptotic expansion~\eqref{eq:asymptotic-Barnes}. 

In the pure $\CalN=2$
theory the leading term in the exponent comes from the numerator of
$Z_{\text{1-loop}}$ and is equal to $-\frac 1 2 z_E^2 \log z_E$. This is a 
negative function which grows in absolute value faster than any other terms
including the renormalized quadratic term~\eqref{eq:renorm-g} even 
if $\tilde \gym^2$ formally becomes negative.  

In the $\CalN=2^*$ case we need to
take asymptotic expansion at large $z_E$ of both the numerator and denominator
of~\eqref{eq:Z-Barnes-hyper} to check convergence at infinity. 
The leading terms $(\alpha \cdot a_E)^2 \log
(\alpha \cdot a_E)$ cancel, and the next order term is proportional to $m^2_E \log
(\alpha \cdot a_E)$. This does not spoil the convergence insured by the Gaussian
classical factor $\exp(-\frac{8\pi^2 r^2}{\gym^2} (a_E,a_E))$.

To summarize, in the pure $\CalN=2$ theory we need to insert the factor
\begin{equation}
  \label{eq:Z-1-loop-N-2}
 Z_{\text{1-loop}}^{\CalN=2} = \prod_{\text{roots }\alpha} H \left(      {i\alpha \cdot a_E}/ {\ve}\right),
\end{equation}
in the matrix model integrand and 
replace $\gym$ by the renormalized coupling constant $\tilde \gym$.

At  $m=0$ we get the $\CalN=4$ theory. The one-loop determinant
of the the hypermultiplet exactly cancels the one-loop determinant of
the
vectormultiplet
\begin{equation}
  \label{eq:Z-1-loop-N-4}
  Z_{\text{1-loop}}^{\CalN=4} = 1.
\end{equation}
Most of the above computations applies to the $\CalN=2$ 
theory with a massless hypermultiplet in any 
$G$-representation $W$:
\begin{equation}
\label{eq:Z-1-loop-any-matter}
  Z_{1-loop}^{\CalN=2, W}(ia_E) = \frac{ \prod_{\alpha \in \text{roots}(
\g)} H(i\alpha \cdot a_E /\ve) }
                                    { \prod_{w \in \text{weights}(W)} H(iw\cdot a_E/\ve) }.
\end{equation}
This formula literally holds if the divergent factors are the same in 
the one-loop determinants  for the vector and hypermultiplets.
This happens for  representations $W$
such that $\sum_{\alpha} (\alpha \cdot a)^2 = \sum_{w} (w \cdot a)^2$, 
$a \in \g$, that is if the  $\beta$-function vanishes and the $\CalN=2$
theory is superconformal.
For a general matter representation, the one-loop determinant requires
regularization similar to the pure $\CalN=2$ theory.

\subsection{Example}

We give a  simple example of a non-trivial prediction 
of the formula~(\ref{eq:Z-1-loop-any-matter}), which perhaps can be checked using the traditional 
methods of the perturbation theory. 

We consider the $\SU(2)$ $\CalN=2$ theory with $N_f=4$ hypermultiplets
in the fundamental  
representation. We choose coordinate $a$ on the Cartan subalgebra of the real Lie algebra of the
gauge group $\SU(2)$, such that an element $a$ is represented by an
anti-hermitian matrix $diag(ia,-ia)$,
and our conventions for the kinetic term in the Yang-Mills is
 $-\frac {1} {2\gym^2}  \int d^4 x \sqrt{g}  \tr F_{\mu \nu},F^{\mu
   \nu} $.
In the spin-$j$ representation of dimension $2j+1$ 
we have
\[
\{w a\} = 2 a \{-j , -j + 1, \dots, j-1, j \}.
\]

Then our matrix model
 for the expectation value of the Wilson loop in the spin-$j$
 representation
gives
\begin{equation}
\label{eq:j-int}
  \langle \tr_j \Pexp (\int Adx + i\Phi_0 ds) \rangle = \frac{1}{Z} \int_{-\infty}^{\infty} da e^{ - \frac {16 \pi^2} {\gym^2} a^2 } (2a)^2 \frac { H(2ia) H(-2ia)} { (H(ia) H(-ia))^4} ( \sum_{m=-j}^{j} e^{4\pi m a} ),
\end{equation}
where $Z$ is a $j$-independent  constant.

The extra factor $(2a)^2$ is the usual Vandermonde determinant appearing 
when we reduce the integration from the Lie algebra to its Cartan subalgebra.
In  the weak coupling limit $\gym \to 0$ we evaluate the integral 
in series in $\gym$.  
The Barnes G-function at $z \to 0$ has expansion
\[
 \log G(1+z) = \frac 1 2 (\log (2 \pi) -1) z - (1+ \gamma) \frac {z^2} {2} + \sum_{n=3}^{\infty} (-1)^{n-1} \zeta(n-1)
\frac{z^n} {n}.
\]
Then our prediction is  
\begin{equation}
\label{se:first-corrections}
 \langle e^{2\pi n a} \rangle = 1 + \frac 3 {4 \cdot 2^2} n^2 \gym^2 + \frac 5 {32\cdot 2^4} n^4 \gym^4 
+ \frac {7} {8 \cdot 48 \cdot 2^6} n^6 \gym^6 + \frac{35} {8 \cdot 2^4 (4\pi)^2} c_2 n^2 \gym^6 + O(g^8),
\end{equation}
where $c_2$ is the coefficient coming from the expansion of the Barnes
$G$-function:
\[
 c_2 = -12 \zeta(3),
\]
where $\zeta$ is the Riemann zeta-function.
To get (\ref{se:first-corrections}) we expanded the determinant factor  in powers of $a$:
\[
 \log \lb \frac { H(2ia) H(-2ia)} { (H(ia) H(-ia))^4} \rb = -8 \sum_{k=2}^{\infty} \frac {\zeta(2k-1)} {k} (2^{2k-2} -1)
(-1)^k a^{2k} =: \sum_{k=2}^{\infty} c_{k} a^{2k}.
\]
For Gaussian integrals $\int da \, e^{-\frac {1} {2 \sigma^2} a^2 }$ with $\sigma^2 = \frac {\gym^2}{32 \pi^2}$
we have  

\[
\left \langle a^2 \exp \left (\sum c_{k} a^{2k} \right)  e^{wa} \right \rangle_{\text{gauss}} = 
\left (\frac{\p} {\p w} \right)^2 \exp\left( \sum c_k 
\left( \frac {\p} {\p w} \right)^k \right) e^{\frac{1}{2} w^2 \sigma^2}.
\]

 The perturbative result for the $\CalN=4$ $\SU(2)$ theory is given by the same formula but with $c_k = 0$:
 \[
  \langle e^{wa} \rangle_{\CalN=4} = (1 + \sigma^2 w^2) \exp ( \frac 1 2  \sigma^2 w^2 )=1+\frac 3 2 ( \sigma w)^2 + \frac 5 8 
 (\sigma w)^4 + \frac {7} {48} ( \sigma w)^6 + O((\sigma w)^8).
 \]

 Taking $w = 2 \pi n$ we get the result (\ref{se:first-corrections})
 for the $\CalN=4$ theory with $c_2=0$. For a superconformal $\CalN=2$ theory the Gaussian 
 matrix model action is corrected by the terms $c_{k} a^{2k}$. The first correction is quartic $c_2 a^4$,
 and at the lowest order it gives the result~(\ref{se:first-corrections}) for the $\SU(2)$ 
theory with $N_f = 4$ fundamental hypermultiplets.
The first difference for $\langle W_{R} (C) \rangle$  between the $\CalN=2$ $\SU(2)$ gauge theory 
with 4 fundamental hypermultiplets and the $\CalN=4$ $\SU(2)$ gauge theory 
appears at the order $\gym^6$. In this order
the Feynman diagrams have been computed  in $\CalN=4$ gauge theory
in \cite{Plefka:2001bu,Arutyunov:2001hs}, so the $\CalN=2$ computation seems
to be possible in this order as well.

Clearly the higher order terms can be
elementary evaluated from the matrix model integral 
(\ref{eq:j-int}), unlike the 4d gauge theory Feynman  diagrams.

\section{Instanton corrections}

\label{sec:Instanton corrections}
Assuming smooth gauge fields, we showed in~\eqref{eq:vanishing-th} that the path integral localizes to the
trivial gauge field configurations because $d_{[\lambda}w_{\mu
  \nu]}$ vanishes only at the North and the South poles. 
If we allow singular field
configurations, the gauge field strength could be non-vanishing
at the North or the South pole and still solve $QV = 0$.
From~\eqref{eq:N-2local} we see that $F^{-}$ could be non zero at the North
pole, where $\sin^2 \frac \theta 2$ vanish, while $F^{+}$ could be non zero at
the South pole, where $\cos^2 \frac \theta 2$ vanish.  Thus, if we allow
non-smooth gauge fields in the path integral, we need to account for configurations
with point instantons ($F^{+}=0$) localized at the North pole, and point
anti-instantons ($F^{-}=0$) localized at the South pole. The $Q$-complex in
our problem on $S^4$ at the North pole coincides with the
$Q$-complex in the topological gauge theory on $\BR^4$ in the
$\Omega$-background~\cite{Nekrasov:2002qd}. In this theory, the moduli
space of
instantons is considered equivariantly under the $\U(1)^2$ action on $\BR^4
 \simeq \BC^2$:  $(z_{1},z_2) \mapsto (t_1 z_1, t_2
 z_{2})$ with $t_1 = e^{i \ep_1}, t_2 = e^{i \ep_2}$.   Our $Q$-complex corresponds to the equivariant parameters 
$\ep_{1} = \ep_{2} = r^{-1}$.

In the discussion of the
instanton corrections, we focus on  $\U(N)$ gauge group.
  We define the instanton charge as the second Chern class\footnote{
For $U(N)$ bundles the total Chern class is $c = \det (1 + \frac {iF} { 2\pi}) = \prod (1 +x_i) = c_0 + c_1 + \dots$,
where $F$ is the curvature which takes value in the Lie algebra of
the gauge group, $x_i$ are the Chern roots, and $c_k$ is polynomial of degree $k$ in $x_i$. We have $c_2 = \sum_{i < j} x_i x_j = \frac 1 2 (\sum x_i)^2 - \frac 1 2 \sum x_i^2$. If $c_1 = \sum x_i$ vanishes, we get $c_2 = -\frac 1 2 \int \tr 
\frac {iF}{2\pi} \wedge \frac {iF}{2\pi}= \frac 1 {8\pi^2} \int \tr F \wedge F = -\frac 1 {8 \pi^2} \int (F,\wedge F)$, 
where the trace is taken in the fundamental representation. The parentheses $(a,b)=-\tr a b$ denote the positive
definite bilinear form on the Lie algebra which is assumed in the most of the formulas.}
\[
 k = c_2 = \frac {1} {8 \pi^2}  \tr \int  F \wedge F,
\]
and modify the action by the $\theta$-term
\[
 S_{YM} \to S_{YM} - \frac {i \theta} {8 \pi^2} \tr \int F \wedge F.
\]
At $F^{+}=0$ we have $ \sqrt{g} F_{\mu \nu}F^{\mu \nu} d^4 x = 2 F \wedge *F = -2 F \wedge F$.
Then the Yang-Mills action of instanton of charge $k$ is 
\[
S_{YM}(k) = -\frac 1 {2 \gym^2} \int \sqrt{g} d^4 x \tr F_{\mu \nu}
 F^{\mu \nu} - \tr \frac {i \theta} {8\pi^2} \int 
F \wedge F = \lb \frac {8\pi^2} { \gym^2} - {i \theta}  \rb k.
\]
The charge $k$ instanton contribution to the partition function is proportional to 
\[
 \exp (-S_{YM}(k))  = \exp \lb 2 \pi i \tau k \rb = \qq^{k},
\] 
where we introduced the complexified coupling constant
\[
 \tau = \frac {4 \pi i} {\gym^2 } + \frac {\theta}{ 2\pi},
\]
and the expansion parameter
\[
\qq=\exp(2\pi i\tau).
\]

 Near the North pole our $\CalN=2$ theory on $S^4$ is like 
the twisted topological $\CalN=2$ theory in the $\Omega$-background
localized to the instantons $F^{+}=0$.
Near the South pole our theory localizes to the anti-instantons $F^{-} =
0$.

Explicitly, the equivariant instanton
partition function on $\BR^4$ in the $\Omega$-background is~\cite{Nekrasov:2002qd,Nekrasov:2003rj,Flume:2002az,Bruzzo:2002xf,nakajima-inst,nakajima-lectures,Moore:1997dj,Losev:1997tp}
\begin{equation}
  \label{eq:Z-inst}
  Z_{\text{inst}}^{\CalN=2}(a; \ep_1,\ep_2) = \sum_{\vect{Y}} \frac {{\qq}^{|Y|}} {\prod_{\alpha, \beta=1}^{N}
    n^{\vect{Y}}_{\alpha, \beta} (\ep_1,\ep_2, \vect{a})},
\end{equation}
where the summation is over the set of Young diagrams $N$-tuples
$\{Y_\alpha\}$, $\alpha = 1\dots N$. By $|\vect{Y}|$ we denote the total
size $\sum{|Y_{\alpha}|}$ equal to the instanton number $k$. 
Each Young diagram $N$-tuple uniquely correspond to a fixed point on the instanton
moduli space. 
The factor $n_{\alpha, \beta}^{\vect{Y}}(\ep_1, \ep_2, \vect{a})$ is
the equivariant Euler class of the tangent bundle at the fixed point
described by $\vect{Y}$ 
\begin{multline}
  n^{\vect{Y}}_{\alpha, \beta}(\ep_1, \ep_2, \vect{a}) = \prod_{s \in
    Y_{\alpha}} ( - h_{Y_{\beta}}(s) \ep_1 + (v_{Y_{\alpha}}(s)+1)\ep_2 +
  a_{\beta} - a_{\alpha}) \times \\ \times \prod_{t \in Y_{\beta}}
  ((h_{Y_{\alpha}}(t) + 1 ) \ep_1 - v_{Y_{\beta}}(t) \ep_2 + a_{\beta} -
  a_{\alpha}).
\end{multline}
In our conventions  $a \in \h \subset \g$ 
is represented by the matrix $\diag (ia_{1}, \dots, ia_{N})$.  The indices $s$ and $t$ run over
squares of Young diagrams $Y_{\alpha}$ and $Y_{\beta}$.  Let $Y$ be a Young
diagram $\nu_{1} \geq \nu_{2} \dots \geq \nu_{\nu_{1}'}$, where $\nu_{i}$ is the
length of the $i$-th column, $\nu_{j}'$ is the length of the $j$-th
row. The square $s=(i,j)$ is at $i$-th column and the $j$-th row, 
and $v_{Y}(s) = \nu_{i}(Y) - j$ and $h_{Y}(s)=\nu_{j}'(Y) - i$, 
so $v_{Y}(s)$ and $h_{Y}(s)$ is respectively the vertical and horizontal distance from the
square $s$ to the edge of the diagram $Y$. We can rewrite the product in the
denominator of~\eqref{eq:Z-inst} as
\begin{equation}
  \prod_{\alpha, \beta=1}^{N}
  n^{\vect{Y}}_{\alpha, \beta} (\ep_1,\ep_2, \vect{a}) = \prod_{\alpha,\beta=1}^{N} \prod_{s \in Y_{\alpha}}
  E_{\alpha\beta}(s)
  (\ep_1 + \ep_2 - E_{\alpha\beta}(s)),
\end{equation}
where
\begin{equation}
  E_{\alpha\beta} (s)= (-h_{Y_{\beta}}(s) \ep_1 + (v_{Y_{\alpha}}(s) + 1)\ep_2 + a_{\beta} - a_{\alpha}).
\end{equation}
For illustration consider  $\U(1)$ case.  The summation is over the set of all Young
diagrams.  
For instanton charge $k=1$, there is only
one diagram $Y=(1)$ and $E_{11}=\ep_2$, hence
\begin{equation}
  Z^{\CalN=2}_{k=1}(a; \ep_1,\ep_2) = \frac {1}{ \ep_2 \ep_1}.
\end{equation}
For instanton charge $k=2$, there are two diagrams $Y=(2,0)$ and
$Y=(1,1)$.  Their contribution is
\begin{equation}
  Z^{\CalN=2}_{k=2}(\ep_1,\ep_2,a_1) = \frac{1}{(2\ep_2)(\ep_1 - \ep_2)(\ep_2)(\ep_1)} +
  \frac {1} {(-\ep_1 + \ep_2)(2 \ep_1) (\ep_2) (\ep_1)} = \frac {1}{2(\ep_1 \ep_2)^2},
\end{equation}
and for any $k$ one gets 
\begin{equation}
  Z^{\CalN=2}_{k}(\ep_1,\ep_2,a) = \frac {1} { k! (\ep_1 \ep_2)^k},
\end{equation}
hence
\begin{equation}
  Z^{{\CalN=2}}_{\U(1)}(a; \ep_1,\ep_2) = \sum_{k=1}^{\infty} \frac {\qq^k}{k! (\ep_1\ep_2)^k} =
  \exp\left(\frac  {\qq}{\ep_1 \ep_2}\right).
\end{equation}
Similarly, for the $\U(2)$ gauge group, 
at  $k=1$ we have two colored Young diagrams $((1),0)$ and $(0,(1))$ contributing
\begin{multline}
  Z^{\CalN=2}_{k=1}(a_1,a_2; \ep_1,\ep_2) = \frac {1} { \ep_1 \ep_2 (a_2 - a_1 +
    \ep_1 + \ep_2)(a_1 - a_2) }
  + \frac {1} { (a_1 - a_2 + \ep_1 + \ep_2)(a_2 - a_1) \ep_1 \ep_2} = \\
  =\frac {2} {\ep_1 \ep_2 ( (\ep_1+\ep_2)^2 - a^2)},
\end{multline}
and 
at  $k=2$, substituting $a_1 = i a_E , a_2 = - i a_E$,  we get 
 \begin{equation}
   Z^{\CalN=2}_{k=2}(ia_E, -i a_E; \ep_1,\ep_2) = \frac{ (2a_E^2 + 8 \ep_1^2 + 8 \ep_2^2 + 17 \ep_1 \ep_2) }
   { ((\ep_1+2 \ep_2)^2 + a_E^2)((2\ep_1+ \ep_2)^2 + a_E^2)((\ep_1 + \ep_2)^2 + a_E^2)\ep_1^2 \ep_2^2 }.
 \end{equation}

For any $k$, the instanton contribution is a rational functions of $a_i$
and $\ep_i$.  

Often the literature on the gauge theory in the
$\Omega$-background specializes to the case $\ep_1 = - \ep_2$, relating
the gauge theory to the topological 
string~\cite{Nekrasov:2002qd,Nekrasov:2003rj,nakajima-lectures}. 
On the other hand, the physical $\CalN=2^*$ gauge theory on $S^4$
corresponds to the values $\ep_1=\ep_2$ of the $\Omega$-background
parameters.

We notice that  at $\ep_{1} =
\ep_{2}$ the instanton partition function does not have poles at the integration
contour  $a_E \in \BR$. Indeed, the denominator is the product of factors
like $n_1 \ep_1 + n_2 \ep_2 + a$.   The integration contour passes
through a pole only if $n_1 \ep_1 
+ n_2 \ep_2=0$.  This is possible at $\ep_1 = -\ep_2$, but not 
at $\ep_1 = \ep_2$.\footnote{   
We checked this up to $k=5$ for $N=2$, and a general technical proof is
possible. The function $Z_{\text{inst}}(a; \ep_1, \ep_2)$ has simple
poles at $a_{\alpha} - a_{\beta} = n_1 \ep_1 + n_2 \ep_2$ for all positive 
integers $n_1, n_2$. The author
thanks H.~Nakajima for a discussion.} Hence, 
the integrand in~\eqref{eq:main-result} is non-singular 
function on the integration contour rapidly decreasing at infinity, 
and the integral
is well defined.

When the hypermultiplets are introduced to the theory, the instanton
contributions are multiplied by extra factors. In the $\CalN=2^*$
theory, for each fixed point, the factor is the same as for the tangent
bundle of the moduli space, but all weights are shifted by the
hypermultiplet mass.
From \cite{Nekrasov:2002qd,Nekrasov:2003rj,Bruzzo:2002xf} 
we get 
\begin{equation}
\label{eq:Z-inst-N-star}
  Z^{\CalN=2^{*}}_{\text{inst}}(a,  \tilde m; \ep_1, \ep_2) = \sum_{\vect{Y}} \qq^{|\vect{Y}|} \prod_{\alpha,\beta=1}^{N} \prod_{s \in Y_{\alpha}}
  \frac { (E_{\alpha \beta}(s) -\tilde m )(\ep_1 + \ep_2 - E_{\alpha \beta}(s) -\tilde m)} { E_{\alpha \beta}(s)(\ep_1 + \ep_2 - E_{\alpha \beta}(s))  }.
\end{equation}
Here  $\tilde m$ is the equivariant mass parameter
for the gauge theory in the  $\Omega$-background
in conventions of \cite{Nekrasov:2002qd}. The $\tilde m$ is related to
 the hypermultiplet mass $m$ in the present work as $\tilde m = m + (\ep_1 + \ep_2)/2$ 
(see \cite{Okuda:2010ke} for details)\footnote{The correction $\tilde m
  = m  + (\ep_1 + \ep_2)/2$ appeared in the v2 of this preprint on
  arXiv, while in v1 it was erroneously assumed that $\tilde m = m$. The
  author thanks Takuya Okuda for pointing out this issue.}.
For example,
\begin{multline}
  Z^{\CalN=2^*}_{k=1}( a_1,a_2, \tilde m; \ep_1,\ep_2) = \frac { (\ep_1- \tilde m) (\ep_2-\tilde m) (a_2 -
    a_1 + \ep_1 + \ep_2- \tilde m)(a_1 - a_2- \tilde m) } { \ep_1 \ep_2
    (a_2 - a_1 + \ep_1 + \ep_2)(a_1 - a_2) } + \\
  + \frac { (a_1 - a_2 + \ep_1 + \ep_2- \tilde m)(a_2 - a_1- \tilde m) (\ep_1-\tilde m) (\ep_2-\tilde m) } { (a_1 - a_2 + \ep_1 + \ep_2)(a_2 - a_1) \ep_1 \ep_2} = \\
  =\frac{ 2(\tilde m-\ep_2)(\tilde m-\ep_1)( \tilde m^2-a^2- \tilde m(\ep_1 + \ep_2)+(\ep_1+\ep_2)^2) } {
    ((\ep_1 + \ep_2)^2 - a^2) \ep_1 \ep_2 }
\end{multline}
In the $\CalN=2^*$ case, the integrand is again a smooth function on the
integration contour rapidly decreasing  at infinity.

We conclude,  the matrix model integral including all instanton
corrections is well defined in the $\CalN=2,2^*,4$ theories. 

For generic $m$ in the $\CalN=2^{*}$ theory, the one-loop determinant factor $Z_{\text{1-loop}}$ and
the instanton factor $Z_{\text{inst}}$ are nontrivial. 
However, for $m=0$, when $\CalN=4$ symmetry is recovered, $Z_{\text{1-loop}} = 1$,
as well as $Z_{\text{inst}} = 1$ \cite{Okuda:2010ke}.
We conclude that in the $\CalN=4$ theory there are no instanton corrections,
and the the Gaussian matrix model conjecture~(\ref{eq:main-result-op}) is exact.

Another interesting case is $\tilde m = 0$. It is easy to evaluate $Z_{\text{1-loop}}$
and $Z_{\text{inst}}$. The numerator and denominator cancel in 
each of the fixed point instanton contribution to $Z_{\text{inst}}$,
hence in the $\U(N)$ theory
\begin{equation}
  \label{eq:Z-N-4-inst}
  Z_{\U(N), \text{inst}}^{\tilde m = 0 } = \sum_{\vect{Y}} \qq^{|\vect{Y}|}  = \prod_{k=1}^{\infty} \frac 1 {(1-\qq^k)^N}
\end{equation}
is the generating function for the number of $N$-colored partitions. 

Using the definition of the Dedekind eta-function $\eta(\tau) =
\qq^{1/24}\prod_{k=1}^{\infty}(1-\qq^k)$ we  rewrite
\begin{equation}
  \label{eq:Z-N-4-eta}
    Z_{\U(N), \text{inst}}^{\tilde m = 0} = \lb \frac {1} {\qq^{-1/24} \eta(\tau)} \rb^{N}.
\end{equation}

At $\tilde m = 0$ most of the factors in the infinite product (\ref{eq:Z-1-loop-N-2}) cancel each other, 
and we are left with 
\begin{equation}
  Z_{\U(N), \text{1-loop}}^{\tilde m = 0}(ia_E) = \prod_{\text{roots }\alpha} \frac{1}  { |(\alpha \cdot a_E)|}
\end{equation}
We see that the 1-loop contribution at $\tilde m  = 0$ gives exactly the
inverse Weyl measure (the Vandermonde determinant).
Therefore, the total partition function at $\tilde m = 0$ for $\U(N)$
theory at $r  = 1$ is
\begin{equation}
  Z_{\U(N)}^{\tilde m = 0} =   |Z_{\U(N), \text{inst}}^{\tilde m = 0 }|^2 \int d^N a \, e^{- \frac { 8 \pi^2 }{\gym^2} a_E^2}  =  \lb \frac {1} { ( \qq \bar \qq)^{-1/24} \eta(\tau) \bar \eta(\tau) \sqrt{ 2 \tau_2}} \rb^{N} .
\end{equation}

This function does not transform well under the $S$-duality $\tau \to -1/\tau$. However, 
we can deform the action by $c$-number gravitational
curvature terms~\cite{Vafa:1994tf}, for example the  $R^2$-term:
\begin{equation}
   S_{YM} \to S_{YM} - 2 \pi \tau_2 \frac {1} {24} \frac {N} {32 \pi^2} \int_{S^4} R_{\mu \nu \rho \lambda}R^{\mu \nu \rho \lambda}.
\end{equation}
Such $R^2$ terms are usually generated as  gravitational corrections to 
the effective action on branes in string theory~\cite{Bachas:1999um}. The $R^2$ term cancels $\qq^{-1/24}$ in the partition function, and we get 
\begin{equation}
  \label{eq:z-final-1}
  Z^{\tilde m = 0 }_{\U(N), \text{$R^2$ background}} =  \frac {1} { ( \eta(\tau) \bar \eta (\tau) \sqrt{2 \tau_2})^{N}}.
\end{equation}

Now we consider instanton corrections to the Wilson loop operator $W_R(C)$.
One can show that $W_{R}(C)$  is in the same $\delta_{\ve}$ cohomology 
class as the operator $\tr_{R} \exp (\frac {2\pi} {\ep} i \Phi)$ inserted
at the North pole with  $\Phi =  \Phi_0^E + i \Phi_9$. Instanton corrections to the operator $\exp ( \beta \Phi)$ 
in the gauge theory in the $\Omega$-background were computed in~\cite{Losev:2003py,Flume:2004rp,nakajima-lectures,Nekrasov:2003rj}.
Using these results we see that if $\beta = \frac {2 \pi i n}{\ep}$,
$n \in \BZ$,  there are no instanton corrections to the operator $\tr_{R} \exp(\beta \Phi)$.
In other words, the operator $\tr_R \exp( \beta \Phi)$ in the field
theory
exactly corresponds to  the operator $\tr_R \exp(2 \pi i r a)$ in the matrix model. 

Still, the expectation value $\la W(C) \ra$ is deformed by 
instanton corrections because the measure in the matrix
integral~(\ref{eq:main-result}) is deformed
by the instanton factor $|Z_{\text{inst}}(ia_E; \ep,\ep, q)|^2$.

It would be interesting to integrate over $a_E$ and check the
$S$-duality predictions for generic $\CalN=2$ superconformal theories (see e.g.\cite{Argyres:2007cn,Kapustin:2006hi}).

\appendix

\section{Clifford algebra}
\label{sec:Octonionic-gamma-matrices}
Our notations for symmetrized and antisymmetrized tensors are:
\begin{equation}\label{eq:symmetr-antisymmetr}
  \begin{aligned}
    a_{[i} b_{j]}= \frac 1 2 (a_{i} b_{j} - a_{j} b_{i}) \\
    a_{\{i} b_{j\}}= \frac 1 2 (a_{i} b_{j} + a_{j} b_{i}),
  \end{aligned}
\end{equation}
where $a$ and $b$ are any indexed variables.

In the $(9,1)$ signature we use the metric 
\[ds^2 = -dx_0^2 + dx_1^2 + \dots dx_9^2 = g_{MN}
dx^{M} dx^{N}.\] We 
use capital letters
from the middle of the Latin alphabet $M,N,P,Q = 0,\dots, 9$  to denote the ten-dimensional 
space-time indices.  
Let $\gamma^{M}$ for $M=0,\dots,9$ be $32 \times 32$ matrices representing 
the Clifford algebra $\Cl(9,1)$,  satisfying the standard anticommutation relations
\begin{equation}\label{eq:Gamma-anti-comm}
  \gamma^{\{M} \gamma^{N\}} = g^{MN}.
\end{equation}

The corresponding spin representation of $\Spin(9,1)$  can be decomposed into
irreducible spin representations $\CalS^+$ and $\CalS^-$ of rank 16 each.
The chirality operator   
\[
  \gamma^{11} = \gamma^{1}\gamma^{2}\dots \gamma^{9} \gamma^{0}
\]
acts on $\CalS^{+}$ and $\CalS^{-}$ with eigenvalues $1$ and $-1$, respectively. 
The gamma-matrices reverse
chirality
\begin{equation}
  \Gamma^{M}: \CalS^{\pm} \to \CalS^{\mp}.
\end{equation}
Representing the Dirac spin representation of $\Spin(9,1)$ in the form 
\begin{equation}
\begin{pmatrix}
  \CalS^{+} \\
  \CalS^{-}
\end{pmatrix},
\end{equation}
the matrices $\gamma^{M}$ have the block structure
\begin{equation}\label{eq:gamma-ident}
  \gamma^{M} = \begin{pmatrix}
    0 & \tilde \Gamma^{M} \\
    \Gamma^{M} & 0 \\
  \end{pmatrix},
\end{equation}
with
\begin{equation}\label{eq:gamma-anti-comm}
  \tilde \Gamma^{\{M} \Gamma^{N\}} = g^{MN}, \quad
  \Gamma^{\{M} \tilde \Gamma^{N\}} = g^{MN}.
\end{equation}
We define $\gamma^{MN}, \Gamma^{MN}$ and $\tilde \Gamma^{MN}$ as follows
\begin{equation}\label{eq:gamma-commutator}
  \gamma^{MN} = \gamma^{[M} \gamma^{N]} = \begin{pmatrix}
    \tilde \Gamma^{[M} \Gamma^{N]} & 0 \\
    0 & \Gamma^{[M} \tilde \Gamma^{N]} \\
  \end{pmatrix} =:
  \begin{pmatrix}
    \Gamma^{MN} &  0\\
    0 & \tilde \Gamma^{MN} \\
  \end{pmatrix}.
\end{equation}
A useful commutation relation is
\begin{equation}\label{eq:gamma-tri}
  \Gamma^{M} \Gamma^{PQ} = 4 g^{M[P} \Gamma^{Q]} + \tilde \Gamma^{PQ} \Gamma^{M}.
\end{equation}
For computations in the four-dimensional theory, we split the ten-dimensional 
space-time indices into two groups. The first group, for which we use
Greek letters in the middle of the alphabet $\mu, \nu, \lambda, \rho$, 
 denotes
space-time directions in the range $1,\dots,4$. 
The second group, for which we use  capital letters from the beginning
of the Latin alphabet
 $A,B,C,D$,  denotes the scalar field directions $5,\dots,9,0$,

The following identities are useful
\begin{equation}\label{eq:gamma-tri-spec}
  \begin{aligned}
    & \Gamma_{\mu A} \tilde \Gamma^{\mu} = - 4 \tilde \Gamma_{A} \\
    & \Gamma^{\mu} \Gamma_{\nu \rho} \tilde \Gamma_{\mu} = 0 \\
    & \Gamma^{\mu} \Gamma_{\nu A} \tilde \Gamma_{\mu} = 2 \tilde \Gamma_{\nu A} \\
    & \Gamma^{\mu} \Gamma_{A B} \tilde \Gamma_{\mu} = 4 \tilde \Gamma_{AB}.
  \end{aligned}
\end{equation}
We choose matrices $\Gamma_{M}$ and $\tilde \Gamma^{M}$ to be symmetric
\[
(\Gamma^{M})^{T} = \Gamma_{M} \quad (\tilde \Gamma^{M})^{T} = \tilde \Gamma^{M},
\]
then $(\Gamma^{MN})^{T} = - \tilde \Gamma^{MN}$, and 
the representations $\CalS^{+}$ and $\CalS^{-}$ are dual to each other. 

The important triality identity ensures supersymmetry of $\CalN=1$
$d =10$ Yang-Mills 
\begin{equation}\label{eq:triality}
  (\Gamma_{M})_{ \alpha_1 \{\alpha_2} (\Gamma^{M})_{\alpha_3 \alpha_4\}} = 0,
\end{equation}
where $\alpha_1,\alpha_2,\alpha_3,\alpha_4=1,\dots,16$ are the matrix indices of $\Gamma^{M}$.

 In the $(10,0)$  signature we use  matrices $\Gamma^{i}_{E} = \{i\Gamma^{0}, \Gamma^{1}, \dots, \Gamma^{9}\}$.
Hence, in our conventions, all Euclidean gamma-matrices are real except
the pure imaginary matrix $\Gamma^{0}_E = i \Gamma^{0}$. 
In the $(10,0)$  signature the representation $\CalS^{+}$ and $\CalS^{-}$
are unitary, dual and complex conjugate to each other.

It is convenient to use octonions to explicitly represent $\Gamma^{M}$.
In the $(9,1)$ signature we choose
\begin{equation}
  \begin{aligned}
    &\Gamma^{i} = \left(
      \begin{array}{cc}
        0 & E_{i}^{T} \\
        E_{i} & 0 \\
      \end{array}
    \right), \quad  i=1\dots 7 \\
    &\Gamma^9 = \left(
      \begin{array}{cc}
        1_{8\times8} & 0 \\
        0 & -1_{8\times 8} \\
      \end{array}
    \right), \\
    &\Gamma^0 = \left(
      \begin{array}{cc}
        1_{8\times8} & 0 \\
        0 & 1_{8\times 8} \\
      \end{array}
    \right),
  \end{aligned}
\end{equation}
where $E_{i}$, $i = 1 \dots 8$, are $8\times8$ matrices representing left
 multiplication of the octonions.  

Let $e_{i}$  be the
generators of the octonion algebra with the octonionic structure constants
$c^{k}_{ij}$ defined by the multiplication table $e_{i} \cdot e_{j} = c^{k}_{ij}
e_{k}$, with $e_1  = 1$.  We set  $(E_i)^{k}_{j} = c^{k}_{ij}$. 
Concretely, the multiplication can be encoded by the list of quaternionic 
triples:
$ (234),(256),(357),(458),(836),(647),(728)$, which means $e_2 e_3 =
e_4$, and so on. 
Then $E_{i}$ take the following form
\begin{equation}
  \begin{aligned}
    & E_{\mu} = \left(
      \begin{array}{cc}
        J_{\mu} & 0 \\
        0 & \bar{J_\mu} \\
      \end{array} \right), \quad \mu=1\dots 4 \\
    & E_{A} = \left(
      \begin{array}{cc}
        0 & -J_{A}^{T}\\
        J_{A} & 0 \\
      \end{array}
    \right), \quad A = 5 \dots 8,
  \end{aligned}
\end{equation}
where $J_{\mu}$ for $\mu = 1 \dots 4$ are the $4 \times 4$ matrices representing
generators of the quaternionic left action, and $\bar{J_{\mu}}$ 
are the $4\times 4$ matrices representing generators  quaternionic right action.
Concretely,
\begin{equation}
  (J_{1},J_{2},J_{3},J_{4}) =\tiny  \left( \left(
      \begin{array}{cccc}
        1 & 0 & 0 & 0 \\
        0 & 1 & 0 & 0 \\
        0 & 0 & 1 & 0 \\
        0 & 0 & 0 & 1 \\
      \end{array}
    \right),
    \left(
      \begin{array}{cccc}
        0 & -1& 0 & 0 \\
        1 & 0 & 0 & 0 \\
        0 & 0 & 0 & -1 \\
        0 & 0 & 1& 0 \\
      \end{array}
    \right),
    \left(
      \begin{array}{cccc}
        0 & 0 & -1& 0 \\
        0 & 0 & 0& 1 \\
        1 & 0 & 0 & 0 \\
        0 & -1& 0 & 0 \\
      \end{array}
    \right),
    \left(
      \begin{array}{cccc}
        0 & 0 & 0 & -1\\
        0 & 0 & -1& 0 \\
        0 & 1 & 0 & 0 \\
        1 & 0 & 0 & 0 \\
      \end{array}
    \right) \right),
\end{equation}
satisfy
\[
J_{i} J_{j} = \ve_{ijk} J_{k}, \quad i,j,k=2 \dots 4,
\]
and
\begin{equation}
  (\bar {J_{1}},\bar {J_{2}},\bar {J_{3}},\bar {J_{4}}) =\tiny  \left( \left(
      \begin{array}{cccc}
        1 & 0 & 0 & 0 \\
        0 & 1 & 0 & 0 \\
        0 & 0 & 1 & 0 \\
        0 & 0 & 0 & 1 \\
      \end{array}
    \right),
    \left(
      \begin{array}{cccc}
        0 & -1& 0 & 0 \\
        1 & 0 & 0 & 0 \\
        0 & 0 & 0 & 1 \\
        0 & 0 &-1& 0 \\
      \end{array}
    \right),
    \left(
      \begin{array}{cccc}
        0 & 0 & -1& 0 \\
        0 & 0 & 0& -1\\
        1 & 0 & 0 & 0 \\
        0 & 1& 0 & 0 \\
      \end{array}
    \right),
    \left(
      \begin{array}{cccc}
        0 & 0 & 0 & -1\\
        0 & 0 & 1& 0 \\
        0 & -1& 0 & 0 \\
        1 & 0 & 0 & 0 \\
      \end{array}
    \right) \right),
\end{equation}
satisfy
\[
\bar J_{i} \bar J_{j} = -\ve_{ijk} \bar J_{k}, \quad i,j,k=2 \dots 4.
\]
Similarly,
\begin{equation}
  ({J_{5}},{J_{6}},{J_{7}},{J_{8}}) =\tiny  \left( \left(
      \begin{array}{cccc}
        1 & 0 & 0 & 0 \\
        0 & -1 & 0 & 0 \\
        0 & 0 & -1 & 0 \\
        0 & 0 & 0 & -1 \\
      \end{array}
    \right),
    \left(
      \begin{array}{cccc}
        0 & 1& 0 & 0 \\
        1 & 0 & 0 & 0 \\
        0 & 0 & 0 & 1 \\
        0 & 0 &-1& 0 \\
      \end{array}
    \right),
    \left(
      \begin{array}{cccc}
        0 & 0 & 1& 0 \\
        0 & 0 & 0& -1\\
        1 & 0 & 0 & 0 \\
        0 & 1& 0 & 0 \\
      \end{array}
    \right),
    \left(
      \begin{array}{cccc}
        0 & 0 & 0 & 1\\
        0 & 0 & 1& 0 \\
        0 & -1& 0 & 0 \\
        1 & 0 & 0 & 0 \\
      \end{array}
    \right) \right).
\end{equation}

We choose orientation in the $(1\dots 4)$-plane and the $(5\dots 8)$-plane by
taking $1234$ and $5678$ to be the positive cycles.

Then  $\Gamma^{\mu \nu}$,  $\mu,\nu=1 \dots 4$, and $\Gamma^{i j}$,
 $i,j=5 \dots 8$ decompose as 
\begin{equation}
  \begin{aligned}
    \Gamma_{\mu \nu} = \left(
      \begin{array}{cc}
        E_{[\mu}^{T} E_{\nu]} & 0 \\
        0 & E_{[\mu} E_{\nu]}^{T}\\
      \end{array}
    \right) = \left(
      \begin{array}{cccc}
        J_{\mu \nu}^{-} & 0 & 0 & 0 \\
        0 & \bar J_{\mu \nu}^{+} & 0 & 0 \\
        0 & 0 & -J_{\mu \nu}^{+} & 0 \\
        0 & 0 & 0 & -\bar J_{\mu \nu}^{-} \\
      \end{array}
    \right), \\
    \Gamma_{i j} = \left(
      \begin{array}{cc}
        E_{[i}^{T} E_{i]} & 0 \\
        0 & E_{[i} E_{i]}^{T}\\
      \end{array}
    \right) = \left(
      \begin{array}{cccc}
        - \bar J_{ i j}^{-} & 0 & 0 & 0 \\
        0 & -  J_{ij}^{+} & 0 & 0 \\
        0 & 0 & -\bar J_{ij}^{-} & 0 \\
        0 & 0 & 0 & - J_{ij}^{+} \\
      \end{array}
    \right), \\
  \end{aligned}
\end{equation}
where the $\pm$-superscript denotes the self-dual and anti-self-dual tensors;
$J_{12} = J_1^{T} J_2 = J_2$, etc.  

The four-dimensional chirality
operator is 
\[
\gfive=\Gamma_1 \Gamma_2 \Gamma_3 \Gamma_4,
\]
explicitly 
\begin{equation}
  \gfive = \left(
    \begin{array}{cccc}
      1_{4\times 4} & 0 & 0 & 0 \\
      0 & -1_{4\times 4} & 0 & 0 \\
      0 & 0 & -1_{4\times 4} & 0 \\
      0 & 0 & 0 & 1_{4\times 4} \\
    \end{array}
  \right).
\end{equation}

The chirality operator defining the $\CalN=2$ subalgebra of $\CalN=4$ is  
\[
\geight=\Gamma_5 \Gamma_6 \Gamma_7 \Gamma_8,\]
explicitly
\begin{equation}
  \geight = \left(
    \begin{array}{cccc}
      1_{4\times 4} & 0 & 0 & 0 \\
      0 & -1_{4\times 4} & 0 & 0 \\
      0 & 0 & 1_{4\times 4} & 0 \\
      0 & 0 & 0 & -1_{4\times 4} \\
    \end{array}
  \right).
\end{equation}
Notice that 
\begin{equation}
 \Gamma^{9} = \gfive \geight = \left(
    \begin{array}{cccc}
      1_{4\times 4} & 0 & 0 & 0 \\
      0 & 1_{4\times 4} & 0 & 0 \\
      0 & 0 & -1_{4\times 4} & 0 \\
      0 & 0 & 0 & -1_{4\times 4} \\
    \end{array}
  \right).
\end{equation}

The representation $\bf{16} = \CalS^{+}$ (Majorana-Weyl fermion of $\Spin(9,1)$)
 splits as $\bf{16} = \bf{8} + \bf{8'}$ with respect to the $\Spin(8) \subset \Spin(9,1)$ acting 
in the directions $M=1,\dots, 8$.
Then we brake $\Spin(8)$ into
 $ \Spin(4) \times \Spin(4)^R  \hookrightarrow   \Spin(8)$
where the  $\Spin(4)$ acts in the directions $M=1, \dots, 4$, 
and the $\Spin(4)^R$ acts in the directions $M = 5, \dots, 8$.
Next we represent the $\Spin(4)$ as $\SU(2)_L \times \SU(2)_R$ 
and the $\Spin(4)^R$ as $\Spin(4)^R = \SU(2)_L^R \times \SU(2)_R^R$.
With respect to these $\SU(2)$-subgroups, the representation $\bf{16} = \CalS^{+}$  of $\Spin(9,1)$ 
transforms as 
\[
{\bf 16 = (2,1,2,1) + (1,2,1,2) + (1,2,2,1) + (2,1,1,2)}.
\]

The $(10,0)$ gamma-matrices are 
\begin{equation}
  \begin{aligned}
    &\Gamma^{M}_E = \left(
      \begin{array}{cc}
        0 & E_{M}^{T} \\
        E_{M} & 0 \\
      \end{array}
    \right), \quad  M=1\dots 7 \\
    &\Gamma^9_E = \left(
      \begin{array}{cc}
        1_{8\times8} & 0 \\
        0 & -1_{8\times 8} \\
      \end{array}
    \right), \\
    &\Gamma^0_E = \left(
      \begin{array}{cc}
        i1_{8\times8} & 0 \\
        0 & i1_{8\times 8} \\
      \end{array}
    \right).
  \end{aligned}
\end{equation}

\section{Conformal killing spinors on $S^4$ \label{se:Killing-spinors}}

The explicit form of conformal Killing spinors on $S^4$ depends on the choice 
of the vielbein.  For
solution in spherical coordinates see~\cite{Lu:1998nu}.  In stereographic
coordinates the solution has simpler form and is easily related to the flat
limit.

We pick two opposite points on $S^4$ and denote them the North pole and
the South pole. Let $x^{\mu}$ be the stereographic coordinates in the
neighborhood of the North pole. The metric is
\begin{equation}\label{eq:metric-S4-stereo-app}
  g_{\mu \nu} = \delta_{\mu \nu} e^{2\Omega},
  \quad \text{where} \quad  e^{2 \Omega} := \frac 1 { (1 + \frac {x^2} {4 r^2})^2 }.
\end{equation}
Let $\theta$ be the polar angle in spherical coordinates measured from
the North pole, so that $\theta=\frac {\pi} {2}$ is the equator, and
 $\theta = \pi$ is the South pole. We have  $|x| = 2 r \tan \frac \theta 2$
and  $e^{\Omega} = \cos^2 \frac \theta 2$.  Take the
vielbein\footnote{In this section we use the indices $\hat \mu ,\hat
  \nu=1,\dots,4 $ to enumerate the vielbein basis elements, that is $e^{\hat \mu
  }_{\lambda} e^{\hat \nu}_{\nu} = \delta^{\hat \mu \hat \nu}$ where
  $\delta^{\hat \mu \hat \nu}$ is the four-dimensional Kronecker symbol. Then
  $\Gamma^{\hat \mu}$ are the four-dimensional gamma-matrices normalized as
  $\Gamma^{(\hat \mu} \Gamma^{\hat \nu)}= \delta^{\hat \mu \hat \nu}$, }
$e^{\hat \mu}_{\lambda} = \delta^{\hat \mu}_{\lambda} e^{\Omega}$.  The spin
connection $\omega^{\hat \mu}_{\hat \nu \lambda}$ induced by the Levi-Civita
connection can be computed using the Weyl transformation of the flat metric
$\delta_{\mu \nu} \mapsto e^{2 \Omega} \delta_{\mu \nu}$. Under such
transformation $\omega^{\hat \mu}_{\hat \nu \mu} \mapsto \omega^{\hat \mu}_{\hat
  \nu \lambda} + (e^{\hat \mu}_{\lambda} e^{\nu}_{\hat \nu} \Omega_\nu - e_{\hat
  \nu \lambda} e^{\hat \mu \nu} \Omega_{\nu})$.  Since in the flat case
$\omega^{\hat \mu}_{\hat \nu \lambda} = 0$, we get
\begin{equation}
  \omega^{\hat \mu}_{\hat \nu \lambda} = (e^{\hat \mu}_{\lambda} e^{\nu}_{\hat \nu} \Omega_\nu  - e_{\hat \nu \lambda} e^{\hat \mu \nu} \Omega_{\nu}),
\end{equation}
where $\Omega_{\nu} := \p_{\nu} \Omega$.

The conformal Killing spinor satisfies 
\begin{equation}\label{eq:Killing-equation-ster}
  \begin{aligned}
    &(\p_{\lambda} + \frac 1 4 \omega_{\hat \mu \hat \nu \lambda} \Gamma^{\hat \mu \hat \nu} ) \ve  = \Gamma_{\lambda} \tilde \ve\\
    &(\p_{\lambda} + \frac 1 4 \omega_{\hat \mu \hat \nu \lambda} \Gamma^{\hat
      \mu \hat \nu} ) \tilde \ve = - \frac 1 {4 r^2} \Gamma_{\lambda} \ve.
  \end{aligned}
\end{equation}
In the limit $r \to \infty$, the equations simplify as $\p_{\lambda} \ve =
\Gamma_{\lambda} \tilde \ve$ and $\p_{\lambda} \tilde \ve = 0$; hence the flat
space solution is
\begin{equation}
  \begin{aligned}
    \label{eq:conformal_susy_in_R^4-sol}
    &\ve = \hat \ve_{s} + x^{\hat \mu} \Gamma_{\hat \mu} \hat{\ve_c} \\\
    &\tilde \ve = \hat \ve_c,
  \end{aligned}
\end{equation}
where $\hat \ve_{s}, \hat \ve_{c}$ are constant spinors on $\BR^4$.  The spinor
$\hat \ve_{s}$ generates the usual Poincare supersymmetry transformations, the spinor $\hat \ve_c$ generates 
special superconformal transformations.

For finite $r$, the solution is
\begin{align}
  \label{eq:Killing-solution-in-S^4-stereo-app}
  \ve =   \frac 1 {\sqrt{1+ \frac {x^2} {4 r^2}}} (\hat \ve_{s} + x^{\hat \mu} \Gamma_{\hat \mu} \hat \ve_{c})  \\
  \tilde \ve = \frac 1 {\sqrt{1+ \frac {x^2} {4 r^2}}} (\hat \ve_{c} - \frac
  {x^{\hat \mu} \Gamma_{\hat \mu} }{4 r^2} \hat \ve_{s}),
\end{align}
where $\hat \ve_{s}$ and $\hat \ve_{c}$ are arbitrary spinor parameters.

\newcommand{\gfiv}{\Gamma^{9}}

Now consider conformal Killing spinors generating $\OSp(2|4)$ subgroup. We take chiral $\hat \ve_{s}$ and
$\hat \ve_{c}$, such that $\gfiv \hat \ve_s = -\hat \ve_s$ and $\gfiv \hat
\ve_{c} = -\hat \ve_{c}$, so
\begin{align}
  \label{eq:Killing-solution-in-S^4-stereo2}
  \ve = \frac 1 {\sqrt{1+ \frac {x^2} {4 r^2}}} (\hat \ve_{s} \GREEN{-} x^{\hat \mu}
  \Gamma_{\hat \mu} \gfiv \hat \ve_{c}).
\end{align}

Moreover, for such spinor $\ve$ we have $\hat \ve_{c} = \frac 1 {2r} \frac 1 4
\omega_{\hat \mu \hat \nu} \Gamma^{\hat \mu \hat \nu} \hat \ve_{s}$, where
$\omega_{\hat \mu \hat \nu}$ is a self-dual generator of $SO(4)$
normalized $\omega_{\hat \mu \hat \nu} \omega^{\hat \mu \hat \nu} = 4$. Therefore, $\delta_{\ve}$ squares to a rotation around the North pole
generated by $\omega$.  Then $(\hat \ve_{c}, \hat \ve_{c}) = \frac 1 {4r^2} (\hat \ve_{s}, \hat \ve_{s})$,
and thus $(\ve,\ve)$ is constant over $S^4$.  

Assume $(\hat \ve_{s}, \hat \ve_{s}) =1$. Then we get the vector field $v_{\hat \nu} = \ve \Gamma_{\hat
  \nu} \ve = 2 \hat \ve_{s} \Gamma_{\hat \nu} \Gamma_{\hat \mu} x^{\hat \mu}
\hat \ve^{c} = 2 \hat \ve_{s} \Gamma_{\hat \nu} \Gamma_{\hat \mu} x^{\hat \mu}
\frac 1 {2r} \frac 1 4 \omega_{\hat \rho \hat \lambda}\Gamma^{\hat \rho \hat
  \lambda} \hat \ve_{s} = \frac 1 r x^{\hat \mu} \omega_{\hat \mu \hat \nu}
(\hat \ve_{s} \hat \ve_{s}) = \frac 1 r x^{\hat \mu} \omega_{\hat \mu \hat \nu}
$. Using this equation we rewrite conformal Killing spinor
$\ve\equiv\ve(x)$ as  a  $\Spin(5)$ rotation of $\ve(0)$
\begin{align}
  \label{eq:Killing-solution-in-S^4-stereo3}
 \ve(x) = \frac 1 {\sqrt{1+ \frac {x^2} {4 r^2}}} (\hat \ve_{s} - \frac 1 {2r}
  \frac 1 4 x^{\hat \mu} \Gamma_{\hat \mu} \omega_{\hat \rho \hat \lambda}
  \Gamma^{\hat \rho \hat \lambda} \gfiv \hat \ve_{s}) =
  \frac 1 {\sqrt{1+ \frac {x^2} {4 r^2}}} (\hat \ve_{s} - \frac {1} {2r} x^{\hat \rho} \Gamma^{\hat \lambda} \omega_{\hat \rho \hat \lambda} \gfiv \hat \ve_{s}) = \\
  =\frac 1 {\sqrt{1+ \frac {x^2} {4 r^2}}} (\hat \ve_{s} - \frac 1 2 v_{\hat
    \lambda} \Gamma^{\hat \lambda} \gfiv \hat \ve_{s}) = \frac 1 {\sqrt{1+ \frac
      {x^2} {4 r^2}}} (\hat \ve_{s} - \frac {|x|}{2r} n_{\hat \lambda}(x)
  \Gamma^{\hat \lambda} \gfiv \hat \ve_{s}) = \\= \lb \cos \frac \theta 2 - \sin
  \frac \theta 2 (n_{\hat \lambda}(x) \Gamma^{\hat \lambda} \gfiv )\rb \hat
  \ve_{s} = \exp  \lb -\frac \theta 2 n_{\hat \lambda}(x) \Gamma^{\hat \lambda}
  \gfiv \rb \hat \ve_{s},
\end{align}
where $n_{\hat \lambda}$ is the unit vector in the direction of the vector field
$v_{\hat \lambda}$.

\section{Off-shell supersymmetry\label{se:off-shell susy}}
 
Let $\delta_{\ve}$ be the supersymmetry transformation generated by a conformal Killing spinor
$\ve$.  Then  $\delta_{\ve}^2$ is represented as 
\begin{equation}
  \delta_{\ve}^2 A_{M} =\delta_{\ve} (\ve \Gamma_{M} \Psi) = \ve \Gamma_{M} ( \frac 1 2 \Gamma^{PQ} \ve F_{PQ} + \frac 1 2 \Gamma^{\mu A} \Phi_{A} D_{\mu} \ve ).
\end{equation}
Since
\[ \ve \Gamma_{M} \Gamma_{PQ} \ve = \ve \Gamma_{PQ}^T \Gamma_M \ve = -\ve \tilde
\Gamma_{PQ} \Gamma_{M} \ve = \frac 1 2 \ve (\Gamma_{M} \Gamma_{PQ} - \tilde
\Gamma_{PQ} \Gamma_{M}) = 2 g_{M[P} \ve \Gamma_{Q]} \ve ,\] the first term for
$\delta_{\ve}^2 A_{M}$ gives $ -\ve \Gamma^{N} \ve F_{NM}$.  The second term is
\[ \frac 1 2 \ve \Gamma_{M} \Gamma^{\mu A } \Phi_{A} D_{\mu} \ve = -2 \ve
\Gamma_M \tilde \Gamma_{A} \ve \Phi^{A}. \] Then
\begin{equation}
  \label{eq:delta2onA}
  \delta_{\ve}^2 A_{M} = -(\ve \Gamma^{N} \ve) F_{NM} -
  2 \ve \Gamma_{M} \tilde \Gamma_{A} \ve \Phi^{A}.
\end{equation}
Restricting the index $m$ to the range labeling gauge fields or scalar fields, we get respectively
\begin{equation}
  \begin{aligned}
    & \delta_{\ve}^2 A_{\mu} = - v^{\nu} F_{\nu \mu} - [v^{B} \Phi_B, D_{\mu}] \\
    & \delta_{\ve}^2 \Phi_A = - v^{\nu} D_{\nu} \Phi_{A} - [v^{B} \Phi_B, \Phi_A] - 2
    \ve \tilde \Gamma_{AB} \tilde \ve \Phi^{B} - 2\ve \tilde \ve \Phi_A,
  \end{aligned}
\end{equation}
where we introduced the vector field $v$
\begin{equation}\label{eq:v-in-terms-eps-2}
  v^\mu \equiv \ve \Gamma^{\mu} \ve, \quad v^{A} \equiv \ve \Gamma^{A} \ve.
\end{equation}
Therefore
\begin{equation}
  \delta_{\ve}^2 = -L_{v} - G_{v^M A_{M}} - R - \Omega.
\end{equation}
Here $L_{v}$ is the Lie derivative in the direction of the vector field
$v^{\mu}$.  The transformation $G_{v^{M}A_{M}}$ is the gauge transformation
generated by the parameter $v^{M}A_{M}$.  On matter fields $G$ acts as $G_{u}
\cdot \Phi \equiv [u, \Phi]$, on gauge fields $G$ acts as $G_{u} \cdot A_{\mu} =
- D_{\mu} u$.  The transformation $R$ is the rotation of the scalar fields $(R
\cdot \Phi)_{A} = R_{AB} \Phi^{B}$ with the generator $R_{AB} = 2 \ve \tilde
\Gamma_{AB} \tilde \ve$. Finally, the transformation $\Omega$ is the dilation
transformation with the parameter $2(\ve \tilde \ve)$.

The $\delta_{\ve}^2$ acts on the fermions as follows
\begin{multline}\label{eq:delta-fermion}
  \delta_{\ve}^2 \Psi = D_{M} (\ve \Gamma_N \Psi) \Gamma^{MN} \ve +
  \frac 1 2 \Gamma^{\mu A} (\ve \Gamma_{A} \Psi) D_{\mu} \ve = \\
  = (\ve \Gamma_N D_{M} \Psi) \Gamma^{MN} \ve + ((D_{\mu} \ve) \Gamma_{N} \Psi)
  \Gamma^{\mu N} \ve + \frac 1 2 \Gamma^{\mu A} (\ve \Gamma_{A} \Psi) D_{\mu}
  \ve.
\end{multline}
From the triality identity, we have $\Gamma_{N \alpha_2 (\alpha_1
}\Gamma^{N}_{\alpha_3) \xi } = -\frac 1 2 \Gamma^{N}_{\alpha_2 \xi} \Gamma_{N
  \alpha_1 \alpha_3}.$ Then the first term gives
\begin{multline}
  \label{eq:delta-fermion-on-shell}
  (\ve \Gamma_N D_{M} \Psi) (\Gamma^{MN} \ve)_{\alpha_4} =
  (\ve \Gamma_N D_{M} \Psi) ((\tilde \Gamma^{M} \Gamma^{N} \ve)_{\alpha_4} - g^{MN}\ve_{\alpha_4}) = \\
  = \ve^{\alpha_1} \Gamma_{N \alpha_1 \alpha_2 } D_{M} \Psi^{\alpha_2} \tilde
  \Gamma^{M}_{\alpha_4 \xi} \Gamma^{N}_{\xi \alpha_3} \ve^{\alpha_3}
  - (\ve \Gamma^{N} D_{N} \Psi) \ve_{\alpha_4}=\\
  =-\frac 1 2 (\ve^{\alpha_1} \Gamma_{N \alpha_1 \alpha_3} \ve^{\alpha_3})
  (\tilde \Gamma^{M}_{\alpha_4 \xi} \Gamma^{N}_{\alpha_2 \xi} D_{M} \Psi^{\alpha_2}) - (\ve \Gamma^{N} D_{N} \Psi) \ve_{\alpha_4}=\\
  = -\frac 1 2 (\ve \Gamma_N \ve) (\tilde \Gamma^{M} \Gamma^{N} D_{M}
  \Psi)_{\alpha_4}
  -(\ve \Gamma^{N} D_{N} \Psi) \ve_{\alpha_4} = \\
  = -\frac 1 2 (\ve \Gamma_N \ve) ( - \tilde \Gamma^{N} \Gamma^{M} D_{M} \Psi +
  2 D_N \Psi)_{\alpha_4} - (\ve \Gamma^N D_{N} \Psi) \ve_{\alpha_4} = \\
  = \frac 1 2 (\ve \Gamma_N \ve) \tilde \Gamma^{N} (\Dslash \Psi)_{\alpha_4} -
  (\ve \Gamma^{N} \ve) (D_{N} \Psi)_{\alpha_4} - (\ve \Dslash \Psi)
  \ve_{\alpha_4}.
\end{multline}
The first and the third term in the last line vanish on-shell.  When we add
auxiliary fields, they cancel the first and the third term explicitly.
Then we get
\begin{equation}
  \delta_{\ve}^{2} \Psi = -(\ve \Gamma^{N} \ve) D_{N} \Psi
  +  ( \Psi \Gamma_{N}  D_{\mu} \ve) \Gamma^{\mu N} \ve +
  \frac 1 2 \Gamma^{\mu A} (\ve \Gamma_{A} \Psi) D_{\mu} \ve + \text{eom}[\Psi], \\
\end{equation}
where $\text{eom}[\Psi]$ stands for the terms proportional to the Dirac equation
of motion for $\Psi$. 
Then we rewrite the last two terms as follows 
\begin{multline}
  \label{eq:begin-trans}
  (\Psi \Gamma_N \Gamma_{\mu} \tilde \ve) \Gamma^{\mu N} \ve + \frac 1 2
  \Gamma^{\mu A} (\ve \Gamma_{A} \Psi) \Gamma_{\mu}
  \tilde \ve = \\
  = (\Psi \Gamma_N \Gamma_{\mu} \tilde \ve) (\tilde \Gamma^{\mu} \Gamma^{N} - g^{\mu
    N}) \ve - 2 (\ve \Gamma_{A} \Psi) \tilde \Gamma^{A} \tilde \ve = (\Psi
  \Gamma_N \Gamma_\mu \tilde \ve) \tilde \Gamma^{\mu} \Gamma^{N} \ve
  - 4(\Psi \tilde \ve) \ve  - 2 (\ve \Gamma_{A} \Psi) \tilde \Gamma^{A} \tilde \ve \\
  \overset{triality}{=} -(\tilde \ve \tilde \Gamma_{\mu} \Gamma_N \ve) \tilde
  \Gamma^{\mu} \Gamma^{N} \Psi -(\ve \Gamma_N \Psi) \tilde \Gamma^{\mu}
  \Gamma^{N} \Gamma_{\mu} \tilde \ve
  -4 (\Psi \tilde \ve) \ve - 2 (\ve \Gamma_{A} \Psi) \tilde \Gamma^{A} \tilde \ve = \\
  = -(\tilde \ve \tilde \Gamma_{\mu} \Gamma_\nu \ve) \tilde \Gamma^{\mu} \Gamma^{\nu}
  \Psi - (\tilde \ve \tilde \Gamma_{\mu} \Gamma_A \ve) \tilde \Gamma^{\mu} \Gamma^{A}
  \Psi + 2(\ve \Gamma_\nu \Psi) \Gamma^{\nu} \tilde \ve + 4 (\ve \Gamma_{A} \Psi)
  \tilde \Gamma^{A}
  \tilde \ve - 4 (\Psi \tilde \ve) \ve - 2 (\ve \Gamma_{A} \Psi) \tilde \Gamma^{A} \tilde \ve =\\
  -(\tilde \ve \Gamma_{\mu \nu} \ve) \Gamma^{\mu \nu} \ve - 4 (\ve \tilde \ve)
  \Psi\ -(\tilde \ve \Gamma_{\mu A} \ve) \Gamma^{\mu A} \ve + 2( \ve
  \Gamma_{\nu} \Psi) \tilde \Gamma^{\nu} \tilde \ve +
  2(\ve \Gamma_{A} \Psi) \tilde \Gamma^{A} \tilde \ve - 4( \Psi \tilde \ve) \ve =\\
  = -\frac 1 2 (\tilde \ve \Gamma_{\mu \nu} \ve ) \Gamma^{\mu \nu} \Psi - \frac 1
  2 (\tilde \ve \Gamma_{\mu \nu} \ve) \Gamma^{\mu \nu} \Psi -4 (\ve \tilde \ve)
  \Psi - (\tilde \ve \Gamma_{\mu A} \ve) \Gamma^{\mu A} \ve - \frac 1 2 (\tilde
  \ve \Gamma_{AB} \ve) \Gamma^{AB} \Psi + \\ + \frac 1 2 (\tilde \ve \Gamma_{AB} \ve)
  \Gamma^{AB} \Psi  +
  2 (\ve \Gamma_{N} \Psi) \tilde \Gamma^{N} \tilde \ve - 4(\Psi \tilde \ve) \ve = \\
  = \left ( -\frac 1 2 (\tilde \ve \Gamma_{\mu \nu} \ve) \Gamma^{\mu \nu} \Psi +
    \frac 1 2 (\tilde \ve \Gamma_{AB} \ve) \Gamma^{AB} \Psi \right)+\\ +  \left (
    -\frac 1 2 (\tilde \ve \Gamma_{MN} \ve) \Gamma^{MN} \Psi - 4( \ve \tilde
    \ve)\Psi - 4(\Psi \tilde \ve) \ve + 2 (\ve \Gamma_{N} \Psi) \tilde
    \Gamma^{N} \tilde \ve \right)
\end{multline}
The first term is a part of the Lie derivative along the vector field $v^{\mu} =
(\ve \Gamma^{\mu} \ve)$ acting on $\Psi$. The second term corresponds to the
rotations of the scalar fields $\Phi^{A}$ by the generator $R_{AB}$ and the
properly induced action on the fermions.

In the $\CalN=4$ case we use Fierz identity for $\Gamma^{MN}_{\alpha_1 \alpha_2}
\Gamma_{MN \, \alpha_3 \alpha_4}$ in the last line of~\eqref{eq:begin-trans} to
see that all terms in the second pair of parentheses are canceled except for
$-3(\ve \tilde \ve)\Psi$, so that
\begin{equation}\label{eq:delta4fermions}
  \delta_{\ve}^{2} \Psi = -(\ve \Gamma^{N} \ve) D_{N} \Psi -
  \frac 1 2 (\tilde \ve \Gamma_{\mu \nu} \ve) \Gamma^{\mu \nu} \Psi
  - \frac 1 2 (\ve \tilde \Gamma_{AB} \tilde  \ve) \Gamma^{AB} \Psi -
  3(\tilde \ve \ve) \Psi + \text{eom}[\Psi].
\end{equation}

To close off-shell the supersymmetry transformation of the $\CalN=4$ theory, we add  auxiliary
fields $K_i$, $i=1, \dots, 7$ and modify the transformations 
\begin{equation}
  \begin{aligned}
    &  \delta_{\ve} \Psi = \frac 1 2 \Gamma^{MN} F_{MN} + \frac 1 2 \Gamma^{\mu A} \Phi_{A} D_{\mu} \ve + K^i \nu_i \\
    & \delta_{\ve} K_i = -\nu_i \Gamma^{M} D_{M} \Psi.
  \end{aligned}
\end{equation}
Here we have introduced seven spinors $\nu_i$. They depend on $\ve$ and are required to satisfy the following relations:
\begin{align}
  \label{eq:nu-relations1}
  &\ve \Gamma^M \nu_i = 0 \\
  \label{eq:nu-relations2}
  &\frac 1 2 (\ve \Gamma_N \ve) \tilde \Gamma^{N}_{\alpha \beta} =
  \nu^i_{\alpha} \nu^i_{\beta} + \ve_{\alpha} \ve_{\beta} \\
  \label{eq:nu-relations3}
  &\nu_i \Gamma^M \nu_j = \delta_{ij} \ve \Gamma_M \ve.
\end{align}
The equation~\eqref{eq:nu-relations1} ensures 
closure on $A_M$, the equation~\eqref{eq:nu-relations2} ensures closure on $\Psi$. 

The new term in the transformations for $\Psi$ modifies the last line of~\eqref{eq:delta-fermion-on-shell}
as 
\[ \delta_{\ve} (K^i \nu_i ) = -(\nu_i \Dslash \Psi) \nu_i. \] 
Then the terms in $\delta_{\ve}^2
\Psi$, which we have not taken into an account in~\eqref{eq:remain}, are
\begin{equation}
  -(\nu_i \Dslash \Psi) \nu_i + \frac 1 2 (\ve \Gamma_N \ve) \tilde \Gamma^{N} \Dslash \Psi -
  (\ve \Dslash \Psi) \ve.
\end{equation}
This expression is identically zero because of (\ref{eq:nu-relations2}).  
Hence, after inclusion of the auxiliary
fields $K_i$, the $\delta_{\ve}^2 $ \eqref{eq:delta4fermions} closes
 off-shell on $\Psi$. 

For $\delta_{\ve}^2 K_i$ we get
\begin{equation}
  \delta_{\ve}^2  K_i = -\nu_i \Gamma^{M} [ (\ve \Gamma_M \Psi), \Psi ]
  - \nu_i \Gamma^{M} D_{M} (\frac 1 2 \Gamma^{PQ} F_{PQ} \ve + \frac 1 2 \Gamma^{\mu A}
  \Phi_A D_{\mu} \ve + K^{i} \nu_i).
\end{equation}

Using the gamma matrices triality identity, the first term is transformed to
$\frac 1 2 (\nu_i \Gamma^M \ve) [(\Psi, \Gamma^{M} \Psi)]$, which vanishes
because of~\eqref{eq:nu-relations1}.  The second term with derivative acting on
$F$ is equal by Bianchi identity to $(\nu_i \Gamma_N \ve) D_{M} F^{MN}$ and
vanishes because of~\eqref{eq:nu-relations1}. Then we
use~\eqref{eq:gamma-tri-spec} to simplify the remaining terms
\begin{multline}
  \delta_{\ve}^2 K_i = - \frac 1 2 \nu_i \Gamma^{\mu} \Gamma^{PQ} \Gamma_{\mu} \tilde
  \ve F_{PQ} - \frac 1 2 (\nu_i \Gamma^{M} \Gamma_{\mu A} \Gamma^{\mu} \tilde
  \ve) D_{M} \Phi_A - \frac 1 2 (-\frac 1 {4 r^2}) \Phi_A \nu_i \Gamma^{\nu}
  \Gamma^{\mu A}
  \Gamma_{\mu} \Gamma_{\nu} \ve -\\
  - \nu_i \Gamma^{M} (D_{M} K^{j}) \nu_j- (\nu_i \Gamma^{\mu} D_{\mu} \nu_j)
  K^{j} = -\frac 1 2 (4) \nu_i \tilde \Gamma^{M B} \tilde \ve D_{M} \Phi_B
  -\frac 1 2 (-4) \nu_i \tilde \Gamma^{M B} \tilde \ve D_{M} \Phi_B
  + (\frac 2 {r^2}) \nu_i \Gamma^{A} \ve \Phi_A + \\
  - (\nu_i \Gamma^{M} \nu_j) D_{M} K^{j} - (\nu_i \Gamma^{\mu} D_{\mu} \nu_j)
  K^{j} =
  - (\ve \Gamma^{M} \ve) D_{M} K^{j} -  (\nu_i \Gamma^{M} D_{M} \nu_j) K^{j} =\\
  = - (\ve \Gamma^{M} \ve) D_{M} K^{i} - (\nu_{[i} \Gamma^{\mu} D_{\mu}
  \nu_{j]}) K^{j} - 4(\tilde \ve \ve) K_{i}.
\end{multline}
To get the last line we use the differential of~\eqref{eq:nu-relations3},
i.e. $\nu_{(i} \Dslash \nu_{j)} = 4 (\ve \tilde \ve) \delta_{ij}$.

Now we consider  pure $\CalN=2$ Yang-Mills. 
First we rewrite the last terms in~(\ref{eq:begin-trans}), where $d=6$
is
the space-time dimension of the $\CalN=1$ SYM generating $\CalN=2$ by the
dimensional reduction
\begin{multline}
  \label{eq:remain}
  (\tilde \ve \Gamma_{M N} \ve) \Gamma^{MN} \Psi = (\tilde \ve \tilde
  \Gamma_{M} \Gamma_N \ve) \Gamma^{MN} \Psi = (\tilde \ve \tilde \Gamma_{M}
  \Gamma_N \ve)
  \tilde \Gamma^{M} \Gamma^{N} \Psi - d(\tilde \ve \ve) \Psi \overset{triality}{=} \\
  - (\ve \Gamma_N \Psi) \tilde \Gamma^{M} \Gamma^{N} \tilde \Gamma_{M} \tilde
  \ve - (\Psi \Gamma_N \tilde \Gamma_M \tilde \ve) \tilde \Gamma^{M} \Gamma^{N}
  \ve - d(\tilde \ve \ve) \Psi = (d-2) (\ve \Gamma_N \Psi) \tilde \Gamma^{N}
  \tilde \ve -(\Psi \Gamma_N \tilde \Gamma_M \tilde \ve) \tilde \Gamma^{M}
  \Gamma^{N} \ve - d(\tilde \ve \ve) \Psi.
\end{multline}
Then 
\begin{multline}
  \label{eq:tolambda}
  \left ( -\frac 1 2 (\tilde \ve \Gamma_{MN} \ve) \Gamma^{MN} \Psi - 4( \ve
    \tilde \ve)\Psi - 4(\Psi \tilde \ve) \ve + 2 (\ve \Gamma_{N} \Psi) \tilde
    \Gamma^{N}
    \tilde \ve \right) = \\
  -\frac 1 2 \left ( 4 (\ve \Gamma_N \Psi) \tilde \Gamma^{N} \tilde \ve -(\Psi
    \Gamma_N \tilde \Gamma_M \tilde \ve) \tilde \Gamma^{M} \Gamma^{N} \ve -
    6(\tilde \ve \ve) \Psi \right ) -\\ - 4( \ve \tilde \ve)\Psi - 4(\Psi \tilde
  \ve) \ve + 2 (\ve \Gamma_{N} \Psi) \tilde \Gamma^{N} \tilde \ve = \frac 1 2
  (\Psi \Gamma_N \tilde \Gamma_M \tilde \ve) \tilde \Gamma^{M} \Gamma^{N} \ve
  -(\ve \tilde \ve) \Psi - 4(\Psi \tilde \ve) \ve = \\ = \frac 1 2
  (\Psi(-\Gamma_M \tilde \Gamma_{N} + 2 g_{MN}) \tilde \ve) \tilde \Gamma^{M}
  \Gamma^N \ve -(\ve \tilde \ve) \Psi - 4(\Psi \tilde \ve) \ve = \\ = -\frac 1 2
  (\Psi \Gamma_M \tilde \Gamma_{N} \tilde \ve) \tilde \Gamma^{M} \Gamma^{N} + 6
  (\Psi \tilde \ve) \ve -(\ve \tilde \ve) \Psi - 4(\Psi \tilde \ve) \ve = -\frac
  1 2 (\Psi \Gamma_M \tilde \Gamma_{N} \tilde \ve) \tilde \Gamma^{M} \Gamma^{N}
  \ve + 2 (\Psi \tilde \ve) \ve -(\ve \tilde \ve) \Psi.
\end{multline}
We express the first term using  the triplet of antisymmetric matrices
$\Lambda^{i}$ such that
\begin{align}\label{eq:Lambda-relations}
  &\Lambda^{i}_{\alpha_1 \alpha_3} \Lambda^{j}_{\alpha_2 \alpha_3} =
  \epsilon^{ijk} \Lambda^{k}_{\alpha_1 \alpha_2} + \delta^{ij} 1_{\alpha_1 \alpha_2},
  \quad i,j,k = 1,\dots, 3.\\
  &[\Lambda_i, \Gamma^{M}] = 0 \\
 &\label{eq:Lambda-ident}
  \frac 1 2  \Gamma^{M}_{\alpha_1 \alpha_2 } \tilde \Gamma_{M \, \alpha_3 \alpha_4 } =
  \delta_{\alpha_2 (\alpha_1 } \delta_{\alpha_3) \alpha_4} -
 \Lambda^i_{ \alpha_2  (\alpha_1 } \Lambda^i_{\alpha_3) \alpha_4}.
\end{align}
Then
\begin{equation}
  (\Psi \Gamma_M \tilde \Gamma_{N} \tilde \ve) \tilde \Gamma^{M} \Gamma^{N} \ve =
  4 (\Psi \tilde \ve) \ve + 4 (\ve \tilde \ve)\Psi  +
  4 (\ve \Lambda^{i} \tilde \ve) \Lambda^i \Psi,
\end{equation}
and finally the equation~\eqref{eq:tolambda} turns into
\begin{equation}
  -2  (\Psi \tilde \ve) \ve - 2 (\ve \tilde \ve)\Psi -
  2(\ve \Lambda^{i} \tilde \ve) \Lambda^i \Psi +
  2 (\Psi \tilde \ve) \ve -(\ve \tilde \ve) \Psi =
  - 2(\ve \Lambda^{i} \tilde \ve) \Lambda^i \Psi - 3(\tilde \ve \ve) \Psi.
\end{equation}
Finally,  $\delta_{\ve}^{2}$  on fermions is
\begin{equation}\label{eq:delta2fermions}
  \delta_{\ve}^{2} \Psi = -(\ve \Gamma^{N} \ve) D_{N} \Psi -
  \frac 1 2 (\tilde \ve \Gamma_{\mu \nu} \ve) \Gamma^{\mu \nu} \Psi
  - \frac 1 2 (\ve \tilde \Gamma_{AB} \tilde  \ve) \Gamma^{AB} \Psi -
  2(\ve \Lambda^{i} \tilde \ve) \Lambda^i \Psi - 3(\tilde \ve \ve) \Psi.
\end{equation}
In terms of the supergroup generators we can write the above as
\begin{equation}
  \delta_{\ve}^{2} \Psi = -L_{v} \Psi  - G_{v^{N} A_{N}} \Psi - R \Psi - R' \Psi - \Omega \Psi,
\end{equation}
where the notations for the generators are the same as in the bosonic case. The
only new generator here is $R'$, corresponding to the term $\delta_{\ve}^2 \Psi = -
2(\ve \Lambda^{i} \tilde \ve) \Lambda^i \Psi$.  It generates an $\SU(2)_L$
R-symmetry transformation of $\CalN=2$ which acts trivially on the bosonic fields of the
theory, and as $\Psi \mapsto e^{r_i \Lambda_i} \Psi$ on fermionic fields.

To close off-shell the supersymmetry transformation for  $\CalN=2$
theory, we add the triplet of auxiliary
fields $K_i$ and modify the transformations as
\begin{equation}
  \begin{aligned}
    &  \delta_{\ve} \Psi = \frac 1 2 \Gamma^{MN} F_{MN} + \frac 1 2 \Gamma^{\mu A} \Phi_{A} D_{\mu} \ve + K^i \Lambda_i \ve \\
    & \delta_{\ve} K_i = \ve \Lambda_i \Gamma^{M} D_{M} \Psi,
  \end{aligned}
\end{equation}
The new term in the transformations for $\Psi$ modifies the last line of~\eqref{eq:delta-fermion-on-shell}
as 
\[ \delta_{\ve}(K^i \Lambda_i \ve) = (\ve \Lambda_i \Dslash \Psi) \Lambda_i \ve. \] 
Then the terms in $\delta_{\ve}^2
\Psi$ which were not considered in~\eqref{eq:remain} are
\begin{equation}
  (\ve \Lambda_i \Dslash \Psi) \Lambda_i \ve + \frac 1 2 (\ve \Gamma_N \ve) \tilde \Gamma^{N} \Dslash \Psi -
  (\ve \Dslash \Psi) \ve.
\end{equation}
This expression is identically zero because of the
relation~\eqref{eq:Lambda-relations}.  Hence, after inclusion of the auxiliary
fields $K_i$, the formula~\eqref{eq:delta4fermions} for $\delta_{\ve}^2 \Psi$ is valid
off-shell.

\emph{Remark.} The second equation~\eqref{eq:nu-relations2} follows from the
first equation~\eqref{eq:nu-relations1} and the third
equation~\eqref{eq:nu-relations3} as follows.  Let
\[ M_{\alpha \beta} = \nu^i_{\alpha} \nu^i_{\beta} + \ve_{\alpha} \ve_{\beta}.\]
We want to show that $M_{\alpha \beta} = \frac 1 2 v_N \tilde \Gamma^{N}_{\alpha
  \beta}$, that is the matrix $M_{\alpha \beta}$ can be expanded over the
matrices $\tilde \Gamma^{N}_{\alpha \beta}$ with the coefficients $\frac 1 2
v_N$. Fix the positive definite metric on the space $\BR^{16 \times 16}$ of $16
\times 16$ matrices as $(M,M): = M_{\alpha \beta} M_{\alpha \beta}$.  Since
$\tilde \Gamma^N = \Gamma_N$ and $\Gamma^{\alpha \beta}_M \tilde
\Gamma^N_{\alpha \beta} = 16 \delta^N_M$, the set of 10 matrices $\frac 1 4
\Gamma_N$ is orthonormal in $\BR^{16 \times 16}$.  Complete this set to the
basis of $\BR^{16 \times 16}$.  Then the coefficient $m_N$ of $\frac 1 4
\Gamma_N$ in the expansion of $M$ over this basis is given by the scalar product
\[ m_N = (M, \frac 1 4 \Gamma_N) = \frac 1 4(\nu^i \Gamma_N \nu^i + \ve \Gamma_N
\ve)= 2 v_N .\] Therefore we have $M = 2 v_N (\frac 1 4 \Gamma_N) + (\dots)$,
where $(\dots)$ stand for possible other terms in the expansion over the
completion of the set $\{\frac 1 4 \Gamma_N\}$ to the basis of $\BR^{16 \times
  16}$.  To prove that all other terms vanish, compare the norm of $M$
\[ (M,M) = (\ve \ve)(\ve \ve) + (v_i v_j)(v_iv_j) = (\ve\ve) + \delta_{ij}
(\ve\ve) \delta_{ij} (\ve\ve) = 8 (\ve\ve)(\ve \ve) \] with the $\sum_{N} m_N^2$
\[ \sum_{N} m_N^2 = 4 v_N v_N = 4 (\ve \Gamma_N \ve) (\ve \tilde \Gamma^N \ve) =
4 ( ( \ve \Gamma_N \ve) (\ve \Gamma^N \ve) + 2 (\ve \ve) (\ve \ve)) = 8 (\ve
\ve) (\ve \ve). \] Since the norms are the same, $(M,M) = \sum_{N} m_N^2$, and
the metric is positive definite, we conclude that all other coefficients vanish.

\section{Index of transversally elliptic operators \label{se:trans-elliptic}}
Here we review the index theory for  transversally elliptic operators
mostly following Atiyah~\cite{MR0482866} and Singer~\cite{MR0341538}.

Let $\dots \to E^{i} \overset{D_i}{\to} E^{i+1} \to \dots $ be an elliptic complex of vector bundles over a manifold $X$. Let a Lie group $G$
act on $X$ and bundles $E^i$. This means that for
any transformation $g: X \to X$, which sends a point $x \in X$ to $g(x)$, we are
given a vector bundle homomorphisms $\gamma^i: g^* E^i \to E^i$. Then we have
natural linear maps $\hat \gamma^i: \Gamma(E^i) \to \Gamma(E^i)$ defined by
$\hat \gamma^i = \gamma^i \circ g^*$. On any section $s(x) \in \Gamma(E^i)$ the
map $\hat \gamma^i$ acts by the formula $(\hat \gamma s)(x) = \gamma_x s
(g(x))$. We assume that $\hat \gamma$ commutes with the differential operators $D_{i}$ of
the complex $E$. Then $\hat \gamma$ descends to a well-defined action on the cohomology groups $H^{i}(E)$.  

The $G$-equivariant index is a complex valued function on $G$
defined as the $G$-character of $\oplus H^{i}(E)$ viewed as a graded
$G$-module
\begin{equation}
  \label{eq:equiv-index-def}
  \ind_g(E) = \sum_{i} (-1)^i \tr_{H^i} \hat \gamma^i.
\end{equation}

If the set of $G$-fixed points is discrete and the action of $G$ is nice in 
a neighborhood of each of the fixed point, the Atiyah-Bott fixed point formula gives~\cite{MR0232406,MR0212836,MR0190950}
\begin{equation}
  \label{eq:Atiyah-Bott-formula}
  \ind_g(E) = \sum_{x \in \text{fixed point set } } \frac {\sum (-1)^i \tr \gamma_x^i} { |\det (1 - dg(x))|}.
\end{equation}

This formula can be easily argued in the following way (see~\cite{Goodman:1985bw} for a derivation
using supersymmetric quantum mechanics). For  illustration we
consider $E$ to have only  two terms:  $E^0 \overset{D}{\to} E^1$, 
and we assume that the bundles $E_i$ are equipped with a hermitian $G$-invariant
metric, and $D: \Gamma(E^0) \to \Gamma(E^1)$ is the differential. Then we consider the
Laplacian $\Delta = D D^* + D^*D$. The zero modes of the Laplacian are
identified with the cohomology groups of $E$, which are in this case: $H^{0}(E) = \ker D$ and 
$H^1 (E) = \coker D$. 
Hence, the index can be computed  as
\[ \ind_{g} (E) = \lim_{\beta \to \infty} \str_{\Gamma(E)} \hat \gamma e^{-\beta
  \Delta}. \] Here the supertrace for operators acting on $\Gamma(E)$ is defined
assuming even parity on $\Gamma(E^0)$ and  odd parity on $\Gamma(E^1)$.
However, the expression under the limit does not depend on $\beta$
because $[\Delta,\hat \gamma]=0$. 
Taking the limit  $\beta \to 0$ we get  supertrace of $\hat
\gamma$.  The trace can be easily computed in the coordinate representation. By definition,
the operator $\hat \gamma$ has  kernel $\hat \gamma(x,y) = \gamma_x \delta
(g(x) - y)$ if we write $(\hat \gamma s)(x) = \int_{X} \hat \gamma(x,y) s(y)$.
Here $\delta(x)$ is the Dirac delta-function. Computing the trace we get
Atiyah-Bott formula
\begin{multline}
  \label{eq:index-derivation}
  \ind_{g} (E) = \lim_{\beta \to 0} \str_{\Gamma(E)} \hat \gamma e^{-\beta \Delta} = \int dx \str_{E_x} \hat \gamma(x,x) = \int dx \str_{E_x} \gamma_{x} \delta (g(x) - x) = \\
  = \sum_{g(x) = x} \frac { \str_{E_{x}} \gamma_x } { | \det (1 - dg(x))| }.
\end{multline}

Let $X$ be a complex $n$-dimensional manifold. Consider the Dolbeault complex of
$(0,p)$-forms with the differential $\bar \p$. Let $G= \U(1)$ act on $X$
holomorphically. Near a fixed point we choose
such coordinates $(z^1, \dots, z^n)$ that $t \in \U(1)$ acts by $z^i \to t_i
z^i$. If $z^i$ transforms with the $\U(1)$  weight $m_i \in \BZ$, then $t_i =
t^{m_i}$.  The one-forms $f_{\bar i}$ transform as $f_{\bar i}
\to \bar t_i^{-1} f_{\bar i}$.  Since $|t|=1$ we have $f_{\bar i} \to t_i
f_{\bar i}$. Computing the supertrace for the numerator on external powers of
the anti-holomorphic subspace of the fiber of the cotangent bundle at the
origin, we get $\str_{\Omega^{0,\bullet}} t = \prod_{i=1}^{n} (1 - t_i)$.  The
denominator is $\prod_{i=1}^{n} (1-t_i)(1-t_i^{-1})$. Then  contribution of
a fixed point with weights $\{t_1,\dots, t_n\}$ to the index of $\bar \p$ is
\[
\ind_t (\bar \p)|_0 = \frac {1} {\prod_{i=1}^{n} (1 - t_i^{-1})}.
\]

Let $\pi: T^*X \to X$ be the cotangent bundle. Then $\pi^*E_i$ are the bundles
over $T^*X$.  The symbol of the differential operator $D: \Gamma(E_0) \to
\Gamma(E_1)$ is a vector bundle homomorphism $\sigma(D): \pi^* E_0 \to \pi^*
E_1$.  In local coordinates $x_i$ it is defined by replacing all partial
derivatives in the highest order part of $D$ by momenta symbols $\frac {\p} {\p
  x^i} \to i p_i$, and then taking $p_i$ to be coordinates on fibers of $T^*X$.
Let the family of the vector spaces $T^*_{G} X$ be a union
of vector spaces $T^{*}_G X_x$ over all points $x\in X$, where
$T^{*}_G X_x$ denotes a subspace of $T^* X$ transversal to the $G$-orbit 
through $x$. The operator $D$ is transversally elliptic if its symbol
$\sigma(D)$ is invertible on $T^*_G X \setminus 0$, where $0$ denotes the zero
section.

\newcommand{\Vect}{\mathrm{Vect}} We need a few notions of the $K$-theory~\cite{MR0224083}. 
Let
$\Vect(X)$ be the set of isomorphism classes of vector bundles on $X$. It is an
abelian semigroup where the addition is defined as the direct sum of vector
bundles. For any abelian semigroup $A$ we can associate an abelian group $K(A)$
by taking all equivalence classes of pairs $(a,b) \sim (a+c,b+c)$, where $a,b,c
\in A$. Taking $\Vect(X)$ as $A$ we define the $K$-theory group $K(X)$. Its
elements are pairs of isomorphism classes of vector bundles $(E_0,E_1)$ over $X$
up to the equivalence relation $(E_0,E_1) \sim (E_0 \oplus H, E_1 \oplus H)$ for
all vector bundles $H$ over $X$.  If $X$ is a space with a base point $x_0$, then
we define $\tilde K(X)$ as a kernel of the map $i^*: K(X) \to K(x_0)$ where $i:
x_0 \to X$ is the inclusion map.  Next we define relative $K$-theory group
$K(X,Y)$ for a compact pair of spaces $(X,Y)$.  Let $X/Y$ be the space obtained
by considering all points in $Y$ to be equivalent and taking this equivalence
class as a base point. Then $K(X,Y)$ is defined as $\tilde K(X/Y)$.
Equivalently, $K(X,Y)$ consists of pairs of vector bundles $(E_0,E_1)$ over $X$
such that $E_0$ is isomorphic to $E_1$ over $Y$, and considered up to the
equivalence relation $(E_0,E_1) \sim (E_0 \oplus H, E_1 \oplus H)$ for all
vector bundles $H$ over $X$. For a non-compact space, such as a total space of
vector bundle $V \to X$, we define $K(V)$ as $\tilde K (X^V)$, where $X^V$ is a
one-point compactification of $V$, or equivalently $B(V)/S(V)$, where $B(V)$
and $S(V)$ is respectively a unit ball and unit sphere on $V$.  

If a group $G$ acts on $X$ we can
consider the set of isomorphism classes of $G$-vector bundles over $X$. It is an
abelian semi-group, to which we associate an abelian group $K_G(X)$. All
constructions above can be done in $G$-equivariant fashion.

Since the  symbol of a transversally elliptic operator is an isomorphism $\sigma(D):
\pi^* E \to \pi^* F $ of vector bundles over $T^*_{G} X$ outside of zero
section, by definition it represents an element of $K_{G}(T^*_G X)$. One can
show that the index of transversally elliptic operator does not depend on
continuous deformations of its symbol, hence it depends only on the homotopy type
of the symbol. The index vanishes for a symbol which is induced by an
isomorphism of vector bundles $E$ and $F$. Therefore the index of $D$ depends only on
an element of $K_{G}(T^*_G X)$ which represents symbol $\sigma(D)$.

The equivariant index of a complex of vector bundles $E$ is defined as a $G$-character of the cohomology
groups $\oplus H^i(E)$ treated as a graded $G$-module. 
  One can show that for transversally
elliptic operators $\oplus H^i(E)$ can be decomposed into a direct sum
of irreducible representations where each irreducible representation enters with
a finite multiplicity. For elliptic complex the number of different irreducible
representations and multiplicities are both finite since cohomology groups $H^i$ have
finite dimensions. The index of transversally elliptic operator is
$\sum_{\alpha} m_{\alpha} \chi_{\alpha}$ where $m_{\alpha}$ are finite integer
multiplicities, and  $\chi_\alpha$ is a character of an irreducible
representation $\alpha$.  The index can be viewed as a distribution on
$G$,
and the  multiplicities $m_{\alpha}$ are coefficients in
the corresponding  Fourier series
expansion. Let $\CalD'(G)$ be the space of distributions on $G$. For
transversally elliptic complex the summation generally runs over an infinite set
of irreducible representations $\alpha$, but each multiplicity
$m_{\alpha}$ is finite \cite{MR0482866}.

We learned that the index is a map from the $K$-theory group of $T^*_{G} X$ to
distributions on $G$
\[
\ind: K_{G} (T^*_G X ) \to \CalD'(G).
\]
The index is a group homomorphism with respect to the abelian group
structure on $K_{G}(T^*_G X)$ and the addition operation on $\CalD'(G)$. The
abelian groups $\CalD'(G)$ and $K_{G}(T^*_G X)$ are modules over the character
ring $R(G)$. Indeed, $K_{G}(pt) = R(G)$ since elements of $R(G)$ are formal
linear combinations of irreducible representations of $G$, and $K_{G}(X)$ has a
module structure over $K_{G}(pt)$, since we can take tensor products of vector
bundles representing $K_{G}(X)$ with trivial vector bundles representing
$K_{G}(pt)$. The module $\CalD'(G)$ has a torsion submodule. For example, the
Dirac delta-function on the circle $|t| = 1$ supported at $t=1$ is a torsion element of
$\CalD'(\U(1))$, because it is annihilated by $t-1$. One can show that the
support of the index is a subset of points $g \in G$ for which $X^{g} \neq
\varnothing$, where $X^{g} \subset X$ is the $g$-fixed set. If $G$ acts freely
on $X$ then the index is supported at the identity of $G$, hence 
the index is a pure torsion element.

From now we consider the case $G=\U(1)$. We can find the torsion free part of the index
if we know the index  as a function $\chi(t)$ defined for  $ t \in G$,  $t \neq 1$. If $X^g$
consists of non-degenerate points, we can repeat the argument used in the
elliptic case and obtain the formula~\eqref{eq:index-derivation}.  In the
elliptic case, the separate contributions from fixed points are not well defined at
$t=1$, but the total sum is well defined, since the index is a finite polynomial in
$t$ and $t^{-1}$.  In the transversally elliptic case, if we add contributions of
fixed points formally defined by the
formula~\eqref{eq:index-derivation}, we get  the index 
 up to a torsion (a singular distribution supported at $t=1$).

To fix the torsion part, we should specify how we associate distributions to rational
functions~\eqref{eq:index-derivation}. This procedure is explained
in details in~\cite{MR0482866}. 
For example, the equivariant index
of the Dolbeault operator $\bar \p$ on $\BC$ under the defining $\U(1)$
action on $\BC$, computed at the fixed point $z = 0$, is 
\begin{equation}
  \ind_t(\bar \p)|_0 = \frac {1} {1-t^{-1}}.
\end{equation}
Expanding in positive or negative powers of $t$, we can construct the  distribution associated 
with the singular function 
\begin{align}
  &\left [ \frac 1 {1-t^{-1}} \right]_+ = - \frac {t} {1-t} = - \sum_{n=1}^{\infty} t^n \\
  &\left [ \frac 1 {1-t^{-1}} \right]_- =  \sum_{n=0}^{\infty} t^{-n}.
\end{align}
The two regularizations differ by a torsion element:
\[ \left [ \frac 1 {1-t^{-1}} \right]_+ - \left [ \frac 1 {1-t^{-1}} \right]_- =
- \sum_{n=-\infty}^{\infty} t^n  = - 2\pi i \delta (t-1).\] 

The decomposition of $K_{G} (T^*_G X) $ to
the torsion part and the torsion free part can be described by the exact sequence
\begin{equation}
  \label{eq:exact-sequence}
  0 \to K_G(T^*_G(X \setminus Y)) \to K_G (T^*_G X) \to K_G (T^*X |_Y) \to 0,
\end{equation}
where $Y$ is the $G$-fixed point. Since $G$ acts freely on $X \setminus
Y$, the image of $K_{G}(T^*_G(X \setminus Y))$ under the index homomorphism is a
torsion submodule of $\CalD'(G)$.  The last term of the sequence is the torsion
free quotient determined completely by  $Y$.  Using a vector
field $v$ on $X$ generated by action of $G$, it is possible to construct two
homomorphisms
\[ \theta^{\pm}: K_G (T^*X|_Y) \to K_G (T^*_G X), \] where $\pm$ signs
correspond to a choice of the direction of the vector field.  First, given a
symbol $\sigma: \pi^*E_0 \to \pi^*E_1$, representing an element of
$K_G(T^*X|_Y)$, we extend it to an open neighborhood $U$ of $Y$. Such
extension is  an
isomorphism outside of the zero section.  Second, we define a symbol  $\tilde
\sigma: \pi^*E_0 \to \pi^*E_1$ as a  deformation of $\sigma$ 
using the vector field $v$
\[ \tilde \sigma(x,p) = \sigma (x, p + v e^{-p^2}), \] where $(x,p)$ are local
coordinates on $T^*X$ in a neighborhood of $Y$. Outside of $Y$ the symbol
$\tilde \sigma$ is an isomorphism for all points on fibers of $T^*_GX$, 
and not only outside of the zero section. In other words, $\tilde \sigma$ is an isomorphism
everywhere in the neighborhood $U$ outside of $Y$.  Hence
$\tilde \sigma$ represents an element of $K_G(T^*_G U)$. Since $U$ is open in
$X$, using the natural homomorphism $K_G( T^*_GU) \to K_G (T^*_G X)$ we get an
element of $K_G (T^*_G X)$.

Applying this construction to the space $X=\BC^n$ on which $\U(1)$ acts with
positive weights $m_1,\dots, m_n$, and taking generator of $K(T^* \BC^n|_0)$
associated with $\bar \p$ operator, we get its images under $\theta^{\pm}$ in
$K_G(T^*_G \BC^n)$:
\[ \ind \theta^{\pm} [\bar \p] = \left [ \frac 1 { \prod(1-t^{-m_i})}
\right]_{\pm} .\]

Now assume that using the vector field $v$ we can trivialize a
transversally elliptic operator everywhere on $T^*_GX$ outside of the fixed
point set $Y$, and that near the fixed point set the
trivialization is isomorphic to either $\theta^{+}$ or $\theta^{-}$ 
at each fixed point. Then the index is computed by adding contributions
 from the fixed points regularized correspondingly by $\theta^{+}$ or
 $\theta^{-}$ .

For example, consider $X = \CP^{1}$ and the natural $\U(1)$ action on
$X$. Consider the operator 
\begin{equation}
  D =  \cos^2(\tfrac{\theta}{2}) \bar \p + \sin^2(\tfrac{\theta}{2}) \p
\end{equation}
where   $\theta$ is the polar angle
on $\CP^1$ measured from the North pole. The $D$ is
approximately $\bar \p $ at the North pole and $\p$ at the South
pole.  
This operator $D$ is not  elliptic at the equator, but is transversally elliptic with respect to
the $\U(1)$ action.
We need to take the $\theta^{+}$ regularization of
the North pole contribution, and the  $\theta^{-}$
regularization  of  the South pole contribution 
\begin{equation}
  \label{eq:cp1-trans-ellipt}
 \ind (D) =
\left[ \frac {1} {1-t^{-1}} \right]_+ + \left[ \frac {1} {1-t^{-1}} \right]_-.
\end{equation}

The operator $D$ (\ref{eq:cp1-trans-ellipt}) is the two-dimensional
analogue of the four-dimensional transversally elliptic
 operator  (\ref{eq:symbol-simplified}) that we used
for localization of the $\CalN=2$ theory on $S^4$.

\bibliography{lib}
\end{document}